\documentstyle[11pt]{article}
\advance \topmargin by -\headheight
\advance \topmargin by -\headsep
     
\evensidemargin \oddsidemargin
\begin{document}
\def\rf#1{(\ref{eq:#1})}
\def\lab#1{\label{eq:#1}}
\def\nonu{\nonumber}
\def\br{\begin{eqnarray}}
\def\er{\end{eqnarray}}
\def\be{\begin{equation}}
\def\ee{\end{equation}}
\def\eq{\!\!\!\! &=& \!\!\!\! }
\def\foot#1{\footnotemark\footnotetext{#1}}
\def\lb{\lbrack}
\def\rb{\rbrack}
\def\llangle{\left\langle}
\def\rrangle{\right\rangle}
\def\blangle{\Bigl\langle}
\def\brangle{\Bigr\rangle}
\def\llb{\left\lbrack}
\def\rrb{\right\rbrack}
\def\Blb{\Bigl\lbrack}
\def\Brb{\Bigr\rbrack}
\def\lcurl{\left\{}
\def\rcurl{\right\}}
\def\({\left(}
\def\){\right)}
\def\v{\vert}                     
\def\bv{\bigm\vert}               
\def\Bgv{\;\Bigg\vert}            
\def\bgv{\bigg\vert}              
\def\lskip{\vskip\baselineskip\vskip-\parskip\noindent}
\def\mskp{\par\vskip 0.3cm \par\noindent}
\def\sskp{\par\vskip 0.15cm \par\noindent}
\def\bc{\begin{center}}
\def\ec{\end{center}}
\def\Lbf#1{{\Large {\bf {#1}}}}
\def\lbf#1{{\large {\bf {#1}}}}

\newcommand{\sect}[1]{\setcounter{equation}{0}\section{#1}}
\renewcommand{\theequation}{\thesection.\arabic{equation}}
\relax


\def\tr{\mathop{\rm tr}}                  
\def\Tr{\mathop{\rm Tr}}                  
\newcommand\partder[2]{{{\partial {#1}}\over{\partial {#2}}}}
\newcommand\partderd[2]{{{\partial^2 {#1}}\over{{\partial {#2}}^2}}}
\newcommand\partderh[3]{{{\partial^{#3} {#1}}\over{{\partial {#2}}^{#3} }}}
\newcommand\partderm[3]{{{\partial^2 {#1}}\over{\partial {#2} \partial {#3} }}}
\newcommand\partderM[6]{{{\partial^{#2} {#1}}\over{{\partial {#3}}^{#4}
{\partial {#5}}^{#6} }}}          
\newcommand\funcder[2]{{{\delta {#1}}\over{\delta {#2}}}}
\newcommand\Bil[2]{\Bigl\langle {#1} \Bigg\vert {#2} \Bigr\rangle}  
\newcommand\bil[2]{\left\langle {#1} \bigg\vert {#2} \right\rangle} 
\newcommand\me[2]{\left\langle {#1}\right|\left. {#2} \right\rangle} 

\newcommand\sbr[2]{\left\lbrack\,{#1}\, ,\,{#2}\,\right\rbrack} 
\newcommand\Sbr[2]{\Bigl\lbrack\,{#1}\, ,\,{#2}\,\Bigr\rbrack} 
\newcommand\pbr[2]{\{\,{#1}\, ,\,{#2}\,\}}       
\newcommand\Pbr[2]{\Bigl\{ \,{#1}\, ,\,{#2}\,\Bigr\}}  
\newcommand\pbbr[2]{\lcurl\,{#1}\, ,\,{#2}\,\rcurl}  
\def\a{\alpha}
\def\b{\beta}
\def\c{\chi}
\def\d{\delta}
\def\D{\Delta}
\def\eps{\epsilon}
\def\vareps{\varepsilon}
\def\g{\gamma}
\def\G{\Gamma}
\def\grad{\nabla}
\def\h{{1\over 2}}
\def\l{\lambda}
\def\L{\Lambda}
\def\m{\mu}
\def\n{\nu}
\def\ov{\over}
\def\om{\omega}
\def\O{\Omega}
\def\p{\phi}
\def\P{\Phi}
\def\pa{\partial}
\def\tpa{{\tilde \partial}}
\def\pr{\prime}
\def\ra{\rightarrow}
\def\lra{\longrightarrow}
\def\s{\sigma}
\def\S{\Sigma}
\def\t{\tau}
\def\th{\theta}
\def\Th{\Theta}
\def\z{\zeta}
\def\ti{\tilde}
\def\wti{\widetilde}
\newcommand\sumi[1]{\sum_{#1}^{\infty}}   
\newcommand\fourmat[4]{\left(\begin{array}{cc}  
{#1} & {#2} \\ {#3} & {#4} \end{array} \right)}
\newcommand\twocol[2]{\left(\begin{array}{cc}  
{#1} \\ {#2} \end{array} \right)}
\def\cA{{\cal A}}
\def\cB{{\cal B}}
\def\cC{{\cal C}}
\def\cD{{\cal D}}
\def\cE{{\cal E}}
\def\cF{{\cal F}}
\def\cG{{\cal G}}
\def\cH{{\cal H}}
\def\cI{{\cal I}}
\def\cJ{{\cal J}}
\def\cK{{\cal K}}
\def\cL{{\cal L}}
\def\cM{{\cal M}}
\def\cN{{\cal N}}
\def\cO{{\cal O}}
\def\cP{{\cal P}}
\def\cQ{{\cal Q}}
\def\cR{{\cal R}}
\def\cS{{\cal S}}
\def\cT{{\cal T}}
\def\cU{{\cal U}}
\def\cV{{\cal V}}
\def\cW{{\cal W}}
\def\cY{{\cal Y}}
\def\cZ{{\cal Z}}

\def\phanta{\phantom{aaaaaaaaaaaaaaa}}
\def\phantb{\phantom{aaaaaaaaaaaaaaaaaaaaaaaaa}}
\def\phantc{\phantom{aaaaaaaaaaaaaaaaaaaaaaaaaaaaaaaaaaa}}

\def\lie{{\cal G}}
\def\dlie{{\cal G}^{\ast}}
\def\elie{{\widetilde \lie}}
\def\edlie{{\elie}^{\ast}}
\def\hlie{{\cal H}}
\def\wlie{{\widetilde \lie}}
\def\ulie{{\cal U}\( {\cal G}\)}           
\def\f#1#2#3{f^{{#1}{#2}}_{#3}}            
\font\numbers=cmss12
\font\upright=cmu10 scaled\magstep1
\def\stroke{\vrule height8pt width0.4pt depth-0.1pt}
\def\topfleck{\vrule height8pt width0.5pt depth-5.9pt}
\def\botfleck{\vrule height2pt width0.5pt depth0.1pt}
\def\Zmath{\vcenter{\hbox{\numbers\rlap{\rlap{Z}\kern 0.8pt\topfleck}\kern
2.2pt
                   \rlap Z\kern 6pt\botfleck\kern 1pt}}}
\def\Qmath{\vcenter{\hbox{\upright\rlap{\rlap{Q}\kern
                   3.8pt\stroke}\phantom{Q}}}}
\def\Nmath{\vcenter{\hbox{\upright\rlap{I}\kern 1.7pt N}}}
\def\Cmath{\vcenter{\hbox{\upright\rlap{\rlap{C}\kern
                   3.8pt\stroke}\phantom{C}}}}
\def\Rmath{\vcenter{\hbox{\upright\rlap{I}\kern 1.7pt R}}}
\def\IZ{\ifmmode\Zmath\else$\Zmath$\fi}
\def\IQ{\ifmmode\Qmath\else$\Qmath$\fi}
\def\IN{\ifmmode\Nmath\else$\Nmath$\fi}
\def\IC{\ifmmode\Cmath\else$\Cmath$\fi}
\def\IR{\ifmmode\Rmath\else$\Rmath$\fi}

\def\one{\hbox{{1}\kern-.25em\hbox{l}}}
\def\0#1{\relax\ifmmode\mathaccent"7017{#1}%
        \else\accent23#1\relax\fi}
\def\omz{\0 \omega}

\def\mark{\noindent{\bf Remark.}\quad}
\def\prop{\noindent{\bf Proposition.}\quad}
\def\exam{\noindent{\bf Example.}\quad}

\newtheorem{definition}{Definition}[section]
\newtheorem{proposition}{Proposition}[section]
\newtheorem{theorem}{Theorem}[section]
\newtheorem{lemma}{Lemma}[section]
\newtheorem{corollary}{Corollary}[section]
\def\proof{\par{\it Proof}. \ignorespaces} \def\endproof{{$\Box$}\par}
\newenvironment{Proof}{\proof}{\endproof} 
\def\Winf{{\bf W_\infty}}               
\def\Win1{{\bf W_{1+\infty}}}           
\def\nWinf{{\bf {\hat W}_\infty}}       
\def\Winft#1{{\bf W_\infty^{\geq {#1}}}}    
\def\winf{{\bf w_\infty}}
\def\win1{{\bf w_{1+\infty}}}
\def\hWinf{{\bf {\hat W}_{\infty}}}        
\def\stc{\stackrel{\circ}{,}}              
\def\sto{\stackrel{\otimes}{,}}              
\def\sta{\, ,\,}
\def\xx{(\xi , x)}
\def\yy{(\zeta , y)}
\def\xxt{(\xi , x ; t )}
\def\intres{\int dx\, {\rm Res}_\xi \; }
\def\Intres{\int dx\, {\rm Res} \; }
\def\intrest{\int dt\, dx\, {\rm Res}_\xi \;}
\def\Intrest{\int dt\, dx\, {\rm Res} \;}
\def\Res{{\rm Res}_\xi}
\def\pexx{e^{\pa_x \pa_\xi}}
\def\mexx{e^{-\pa_x \pa_\xi}}
\def\SLinf{SL (\infty ; \IR )}             
\def\slinf{sl (\infty ; \IR )}               
\def\sumlm{\sum_{l=1}^{\infty} \sum_{\v m\v \leq l}}
\def\WDO#1{W_{DOP (S^1 )} \lb #1\rb}               

\def\PsDAS{\Psi{\cal DO} (S^1 )}
\def\PsDA{\Psi{\cal DO}}
\def\ePsDA{{\widetilde {\Psi{\cal DO}}}}
\def\dPsDA{\Psi{\cal DO}^{\ast}}
\def\PsDOS{\Psi {\rm DO} (S^1 )}
\def\PsDO{\Psi {\rm DO}}
\def\ePsDO{{\widetilde {\Psi {\rm DO}}}}
\def\Volt{\Bigl( \Psi{\cal DO} \Bigr)_{-}}
\def\eVolt{\Bigl( {\widetilde {\Psi{\cal DO}}} \Bigr)_{-}}
\def\dVolt{\Bigl( \Psi{\cal DO}^{\ast} \Bigr)_{-}}
\def\VOLT{\Bigl( \Psi {\rm DO} \Bigr)_{-}}
\def\eVOLT{\Bigl( {\widetilde {\Psi {\rm DO}}} \Bigr)_{-}}
\def\KP{${\bf \, KP\,}$}                 
\def\KPl{${\bf \,KP_{\ell}\,}$}         
\def\KPo{${\bf \,KP_{\ell = 0}\,}$}         
\def\mKPa{${\bf \,KP_{\ell = 1}\,}$}    
\def\mKPb{${\bf \,KP_{\ell = 2}\,}$}    
\def\bKP{${\bf \, KP\,}$}                 
\def\bKPl{${\bf \,KP_{\ell}\,}$}         
\def\bKPo{${\bf \,KP_{\ell = 0}\,}$}         
\def\bmKPa{${\bf \,KP_{\ell = 1}\,}$}    
\def\bmKPb{${\bf \,KP_{\ell = 2}\,}$}    
\def\faa{Fa\'a di Bruno~}
\def\bb{{\bar  B}}
\def\bom{{\bar \om}}
\newcommand\ttmat[9]{\left(\begin{array}{ccc}  
{#1} & {#2} & {#3} \\ {#4} & {#5} & {#6} \\
{#7} & {#8} & {#9} \end{array} \right)}
\newcommand\thrcol[3]{\left(\begin{array}{c}  
{#1} \\ {#2} \\ {#3} \end{array} \right)}
\def\cKP{{\sf cKP}~}
\def\cKPrm{${\sf cKP}_{r,m}$~}
\def\scKP{{\sf scKP}}
\newcommand\Back{{B\"{a}cklund}~}
\newcommand\DB{{Darboux-B\"{a}cklund}~}
\def\BH{{Burgers-Hopf}~}
\def\tQ{{\widetilde Q}}
\def\tit{{\tilde t}}
\def\hQ{{\widehat Q}}
\def\hb{{\widehat b}}
\def\hR{{\widehat R}}
\def\htt{{\hat t}}
\def\tv{{\tilde v}}
\newcommand{\nit}{\noindent}
\newcommand{\ct}[1]{\cite{#1}}
\newcommand{\bi}[1]{\bibitem{#1}}
\newcommand\PRL[3]{{\sl Phys. Rev. Lett.} {\bf#1} (#2) #3}
\newcommand\NPB[3]{{\sl Nucl. Phys.} {\bf B#1} (#2) #3}
\newcommand\NPBFS[4]{{\sl Nucl. Phys.} {\bf B#2} [FS#1] (#3) #4}
\newcommand\CMP[3]{{\sl Commun. Math. Phys.} {\bf #1} (#2) #3}
\newcommand\PRD[3]{{\sl Phys. Rev.} {\bf D#1} (#2) #3}
\newcommand\PLA[3]{{\sl Phys. Lett.} {\bf #1A} (#2) #3}
\newcommand\PLB[3]{{\sl Phys. Lett.} {\bf #1B} (#2) #3}
\newcommand\JMP[3]{{\sl J. Math. Phys.} {\bf #1} (#2) #3}
\newcommand\PTP[3]{{\sl Prog. Theor. Phys.} {\bf #1} (#2) #3}
\newcommand\SPTP[3]{{\sl Suppl. Prog. Theor. Phys.} {\bf #1} (#2) #3}
\newcommand\AoP[3]{{\sl Ann. of Phys.} {\bf #1} (#2) #3}
\newcommand\RMP[3]{{\sl Rev. Mod. Phys.} {\bf #1} (#2) #3}
\newcommand\PR[3]{{\sl Phys. Reports} {\bf #1} (#2) #3}
\newcommand\FAP[3]{{\sl Funkt. Anal. Prilozheniya} {\bf #1} (#2) #3}
\newcommand\FAaIA[3]{{\sl Functional Analysis and Its Application} {\bf #1}
(#2) #3}
\def\TAMS#1#2#3{{\sl Trans. Am. Math. Soc.} {\bf #1} (#2) #3}
\def\InvM#1#2#3{{\sl Invent. Math.} {\bf #1} (#2) #3}
\def\AdM#1#2#3{{\sl Advances in Math.} {\bf #1} (#2) #3}
\def\PNAS#1#2#3{{\sl Proc. Natl. Acad. Sci. USA} {\bf #1} (#2) #3}
\newcommand\LMP[3]{{\sl Letters in Math. Phys.} {\bf #1} (#2) #3}
\newcommand\IJMPA[3]{{\sl Int. J. Mod. Phys.} {\bf A#1} (#2) #3}
\newcommand\TMP[3]{{\sl Theor. Mat. Phys.} {\bf #1} (#2) #3}
\newcommand\JPA[3]{{\sl J. Physics} {\bf A#1} (#2) #3}
\newcommand\JSM[3]{{\sl J. Soviet Math.} {\bf #1} (#2) #3}
\newcommand\MPLA[3]{{\sl Mod. Phys. Lett.} {\bf A#1} (#2) #3}
\newcommand\JETP[3]{{\sl Sov. Phys. JETP} {\bf #1} (#2) #3}
\newcommand\JETPL[3]{{\sl  Sov. Phys. JETP Lett.} {\bf #1} (#2) #3}
\newcommand\PHSA[3]{{\sl Physica} {\bf A#1} (#2) #3}
\newcommand\PHSD[3]{{\sl Physica} {\bf D#1} (#2) #3}
\newcommand\JPSJ[3]{{\sl J. Phys. Soc. Jpn.} {\bf #1} (#2) #3}
\newcommand\JGP[3]{{\sl J. Geom. Phys.} {\bf #1} (#2) #3}

\begin{titlepage}
\vspace*{-1.5cm}
\noindent
{\sl hep-th/9607234} \hfill{BGU-96/19/July-PH}\\
\phantom{bla}
\hfill{UICHEP-TH/96-12}\\
\begin{center}
{\large {\bf Constrained KP Hierarchies: \\
Additional Symmetries, \DB Solutions and \\
Relations to Multi-Matrix Models}}
\end{center}
\vskip .15in
\begin{center}
{ H. Aratyn\footnotemark
\footnotetext{Work supported in part by the U.S. Department of Energy
under contract DE-FG02-84ER40173}}
\par \vskip .1in \noindent
Department of Physics \\
University of Illinois at Chicago\\
845 W. Taylor St., Chicago, IL 60607-7059, U.S.A.\\
{\em e-mail}: aratyn@uic.edu \\
\par \vskip .15in
{ E. Nissimov$^{\,\ast ,\,\ast\ast ,\, 2}$  and
S. Pacheva$^{\,\ast ,\,\ast\ast ,}$\foot{Supported in part by Bulgarian 
NSF grant {\em Ph-401}}}
\par \vskip .1in \noindent
$^{\ast )}$ Institute of Nuclear Research and Nuclear Energy \\
Boul. Tsarigradsko Chausee 72, BG-1784 $\;$Sofia, Bulgaria \\
{\em e-mail}: nissimov@inrne.acad.bg, svetlana@inrne.acad.bg \\
${}$ \\
$^{\ast\ast )}$ Department of Physics, Ben-Gurion University of the Negev \\
Box 653, IL-84105 $\;$Beer Sheva, Israel \\
{\em e-mail}: emil@bgumail.bgu.ac.il, svetlana@bgumail.bgu.ac.il
\par \vskip .1in
\end{center}

\begin{abstract}  
This paper provides a systematic description of the interplay between
a specific class of reductions denoted as \cKPrm ($r,m \geq 1$) 
of the primary continuum integrable system -- the Kadomtsev-Petviashvili 
({\sf KP}) hierarchy and
discrete multi-matrix models.
The relevant integrable \cKPrm structure is a generalization of the
familiar $r$-reduction of the full {\sf KP}
hierarchy to the $SL(r)$ generalized KdV hierarchy ${\sf cKP}_{r,0}$.
The important feature of \cKPrm hierarchies is the presence of a
discrete symmetry structure generated by successive \DB (DB) transformations.
This symmetry allows for expressing the relevant tau-functions as Wronskians
within a formalism which realizes the tau-functions as DB orbits
of simple initial solutions. In particular, it is shown that any DB orbit of a
${\sf cKP}_{r,1}$ defines a generalized 2-dimensional Toda lattice structure.
Furthermore, we consider the class of truncated {\sf KP}
hierarchies ({\sl i.e.}, those defined via Wilson-Sato dressing operator
with a finite truncated pseudo-differential series) and establish explicitly
their close relationship with DB orbits of \cKPrm hierarchies.
This construction is relevant for finding partition functions of the discrete
multi-matrix models.

The next important step involves the reformulation of the familiar
non-isospectral additional symmetries of the full {\sf KP} hierarchy
so that their action on \cKPrm hierarchies becomes consistent with the
constraints of the reduction. Moreover, we show that the correct modified
additional symmetries are compatible with the discrete DB symmetry on the
\cKPrm DB orbits.

The above technical arsenal is subsequently applied to obtain 
complete solutions of the discrete multi-matrix models.
The key ingredient is our identification of
$q$-matrix models as DB orbits of ${\sf cKP}_{r,1}$ integrable hierarchies
where $r= \( p_q - 1\) \ldots \( p_2 - 1\)$ with $p_1 ,\ldots ,p_q$
indicating the orders of the corresponding random matrix potentials.
Applying the notions of additional symmetry structure and the technique of 
equivalent hierarchies turns out to be instrumental in implementing 
the string equation and finding closed expressions for 
the partition functions of the discrete multi-matrix models.
As a byproduct, we obtain a representation of the $\t$-function of the most
general DB orbit of ${\sf cKP}_{1,1}$ hierarchy in terms of a new generalized
matrix model.

The present formalism is of direct relevance 
to the study of various random
matrix problems in condensed matter physics and other related areas. In
particular, we obtain a new type of joint distribution function with an
additional attractive two-body and a genuine many-body potentials.
\end{abstract}

\end{titlepage}

\tableofcontents
\newpage
\sect{Introduction}
\label{section:noak-int}
\subsection{Motivation and Strategy}

The main purpose of this paper is to present a conceptually consistent and 
complete description of a special class of integrable hierarchies of 
non-linear differential equations -- truncated and constrained 
Kadomtsev-Petviashvili ({\sf KP}) hierarchies, identified as specific 
reductions of the standard full-fledged {\sf KP} hierarchy.
Another purpose is to apply this structure towards the goal of
obtaining closed form solutions to the discrete multi-matrix models.

There are many equivalent formulations of the theory of integrable models 
and, clearly, any account will have to be selective and focus on one or few
approaches which best fit the general strategy.
Our presentation follows a path which exhibits the connection between 
integrable models and multi-matrix models.
In our case, the strategy dictates that we start with the pseudo-differential 
operator calculus of Lax operators and then establish the links  
to dressing operators and Baker-Akhiezer (wave) functions of the underlying 
linear problem. This in turn, allows us to formulate the relevant 
integrable models in terms of their tau-functions.
It is conceptually important to understand that the discrete structure
of matrix models is being captured by the canonical \DB symmetry 
in the setting of the continuous integrable models.
In this spirit, the tau-functions are constructed explicitly as \DB
orbits of certain simple initial solutions.
This part of the construction is directly relevant for finding  
the partition functions of multi-matrix models.
The story is, however, not as simple as it was briefly sketched above
due to the existence of the so called ``string'' equation constraining 
the form of corresponding partition functions. 
Accordingly, the complete understanding of the connection between 
multi-matrix models and constrained {\sf KP} hierarchies requires 
several important additional concepts.

First of all, we need to formulate the notion of {\em additional symmetries}
for constrained KP hierarchies and describe its interplay with the discrete
symmetry structure provided by the \DB transformations.
Secondly, one discovers that, within the formalism of  multi-matrix models,
the fruitful notion of {\em equivalent hierarchies} facilitates 
implementation of the string equation before the final identification
with the relevant Lax structure of the constrained KP hierarchy. 
Let us now briefly describe the key ingredients of our construction.

\subsection{Integrable Systems and KP Hierarchy }

Integrable systems constitute an outstanding branch of theoretical physics
and mathematics (for the basics, see 
refs.\ct{Faddeev,QISM,Zakh,Russ-other,integr-other,ldickey}). 
The list of their applications to physics is enormous:
they have been found essential for the
description of a vast variety of fundamental non-perturbative phenomena
ranging from the hydrodynamics of $D=2$ (space-time dimensional)
nonlinear soliton waves and planar statistical mechanics to
string and membrane theories in high-energy elementary particle physics.
Furthermore, a variety of physically
interesting theories in higher space-time dimensions can be reformulated
(under plausible assumptions keeping the essential properties of the
dynamics) as lower-dimensional ($D=2$) integrable models which in the same 
time possess {\em infinite-dimensional symmetries} 
(see, especially, the recent developments \ct{Seiberg-Witten}
related with integrability of effective low-energy theory of
(extended) supersymmetric gauge theories).
On a more formal level, the theory of integrable systems is a corner-stone of 
various basic disciplines in mathematics itself, such as algebraic geometry 
(Schottky problem), group theory (quantum groups) and topology 
(link polynomials in knot theory). For a recent account of the latter 
developments we refer to ref.\ct{Fokas-Zakharov}.

The primary integrable system, based on $\PsDA$ (the algebra of all
pseudo-differential operators on the circle) as symmetry algebra, is
the {\sf KP} integrable hierarchy \ct{Zakh,ldickey}
of soliton nonlinear evolution equations, which plays a r\^{o}le
of a universal {\em arch-type} integrable system. 
Its universality is underlined by the fact  that the 
{\sf KP} hierarchy (and multi-component generalizations thereof) is intimately
related to: the discrete generalized Toda lattice hierarchy \ct{U-T} 
(which similarly plays a r\^{o}le of primary lattice integrable system);
integrable systems of point particles \ct{Krichever-Calogero,Zakh}; 
theory of Riemann surfaces \ct{Krichever-Novikov,Zakh}; theory of
representations of infinite-dimensional Lie algebras \ct{Kac}. 

Since the seminal works of Sato's school \ct{KP} (and also of Segal-Wilson 
\ct{Segal-Wilson})
it is well-known that any solution of {\sf KP} hierarchy can be expressed in
terms of a single function of all evolution parameters -- 
the so called $\t$-function, satisfying fundamental bilinear Hirota equations.
The space of all $\t$-functions is known as the universal Sato Grassmannian
(see ref.\ct{OSTT} for a non-technical introduction to Sato theory).

In the last few years, the main interest towards {\sf KP} hierarchy 
originates from its deep connection 
\ct{itep-fian,integr-matrix,BX-2} with the 
statistical-mechanical models of random matrices
((multi-)matrix models) providing non-perturbative discretized formulation
of string theory (for a comprehensive review, see refs.\ct{Morozov}).

One of the most remarkable achievements in the theory of (multi-)matrix
models of string theory is the realization that the (multi-)matrix
partition functions are $\t$-functions of certain reductions of the full
{\sf KP} hierarchy. Therefore, the problem of {\em explicit identification}
of the relevant reduced {\sf KP}
hierarchy in terms of its Lax operator description turns out to be of 
definite interest since the Lax formalism provides the most succinct 
formulation of integrable systems. In the case of discrete 
(multi-)matrix models ({\sl i.e.}, without taking the double-scaling limit),
the pertinent reduced {\sf KP} hierarchies belong to the class described
in that which follows. 

\subsection{Constrained KP Hierarchies}
The fundamental character of the {\sf KP} hierarchy shows up, among other
things, in the fact that via various reductions schemes it 
provides a unifying description of a number of basic
soliton equations like (generalized) Korteweg-de-Vries (KdV), nonlinear
Schr{\"o}dinger (NLS, or more generally - the AKNS hierarchy), Yajima-Oikawa,
coupled Boussinesq-type equations etc.. The latter are
embedded in different ways in various reduced {\sf KP} 
hierarchies comprising the class of the so called \cKPrm {\em constrained}
{\sf KP} hierarchies.

This large and important class of integrable
hierarchies of non-linear differential equations, appearing 
as reduced {\sf KP} hierarchies, is the principal object of the present paper.
The relevant reduction scheme we are going to describe here generalizes 
(and is less restrictive that) the well-known $r$-reduction of the full 
{\sf KP} hierarchy to the $SL(r)$ Gelfand-Dickey (generalized KdV) hierarchy. 

The \cKPrm integrable hierarchies appear in different disguises from various
parallel developments:
\begin{itemize}
\item
Symmetry reductions \ct{symm-constr,oevela,chengs} of the full {\sf KP} 
hierarchy 
-- the ``eigenfunction'' \ct{Dickey-cKP,avoda}
form of \cKPrm (eqs.\rf{f-5},\rf{f-5-1} below).
\item
Abelianization, {\sl i.e.}, free-field realizations, in terms of finite number
of fields\foot{Recall that the full {\sf KP} system contains an infinite 
number of coefficient fields.} of both compatible first and second {\sf KP}
Hamiltonian structures -- ``$m$-generalized two-boson'' form 
\ct{multikp,m-2bose} and the ``ratio'' (or $SL(r+m,m)$) form
\ct{Yu,no2rabn1,office} of \cKPrm , respectively, generalizing $SL(r)$ KdV
hierarchy (eq.\rf{f-5-1} and eqs.\rf{laxmk}--\rf{i-lmk-a} below);
(see also refs.\ct{Dickey-cKP,cortona}).
\item
A method of extracting continuum integrable hierarchies (by taking 
fields associated with one fixed lattice site) from the 
generalized Toda-like lattice hierarchies \ct{BX93-94} underlying
(multi-)matrix models -- ``multi-boson'' KP reductions (eq.\rf{iss-8b} below) 
generalizing the familiar two-boson {\sf KP} reduction \ct{2bose,discrete}.
\item
Purely algebraic approach in terms of the zero-curvature equations 
for the affine Kac-Moody algebras with non-standard gradations \ct{lastyear}
(interpolating between homogeneous and principal gradations of the 
Drinfeld-Sokolov approach).
\end{itemize}
There exist explicit Miura-like transformations connecting the above 
mentioned different forms of \cKPrm hierarchies 
(see eqs.\rf{iss-8c},\rf{iss-8d} and \rf{Miura-A-v} below).

\subsection{Darboux-B\"{a}cklund Transformations, As\-so\-ci\-ated Lat\-tice
Mo\-del Structures
and Discrete Multi-Matrix Models}
The recent surprising discoveries linking the discrete 
multi-matrix models to continuum integrable systems have their 
roots in the formulation of the constrained {\sf KP} systems possessing 
extra discrete symmetry -- \DB (DB) symmetry, which mimics lattice shifts
between neighboring sites. 

Originally, the classical Darboux transformation was introduced in the
context of the linear spectral problem for purely differential operators of
Schr{\"o}dinger type. Namely, using a known eigenfunction one can transform
the initial Schr{\"o}dinger operator into a new one with modified potential
and express the spectral data of the transformed operator explicitly in
terms of the transformation-generating eigenfunction. Since auxiliary linear
spectral problems are inherently associated with the nonlinear evolution
equations of integrable systems, Darboux transformations play an important
r\^{o}le in generating multi-soliton solutions of the latter (for an extensive
overview, see ref.\ct{DT}). 

Recently, the notion of Darboux transformations was general\-ized to the 
case of pseudo-\-diffe\-rential operator formulation of integrable 
hierarchies -- the so called Darboux-B\"{a}cklund transformations 
(see \ct{oevela} and references therein).
Especially, we will be interested in {\em auto}-B{\"a}cklund transformations
applied to the \cKPrm hierarchies --
{\sl i.e.}, such DB transformations which map solutions of a given initial
\cKPrm integrable hierarchy into other solutions of this same hierarchy.
Let us stress that in our approach it is very important that all Lax operator 
eigenfunctions, generating the DB transformations, are {\em not}
Baker-Akhiezer (wave) functions (unlike the case in ref.\ct{oevela}): 
they are rather ``spectral'' superpositions of the latter. 

As demonstrated below in section \ref{section:noak2-2}, the associated
\DB orbits, {\sl i.e.}, the set of all solutions of the subclass of
${\sf cKP}_{r,1}$ hierarchies obtained upon
successive applications of (auto-)\DB transformations on a fixed initial
solution, give rise to generalized two-dimensional Toda-lattice-like 
structures which, in turn,
naturally arise in the context of (multi-)matrix string models.
This generalizes the known connection between ordinary (one-dimensional)
Toda lattice models and B{\"a}cklund transformations of 
refs.\ct{chudnovsky,discrete,leznov-shabat}.

The eigenfunctions and $\t$-functions of all members of any \cKPrm DB orbit
can be explicitly expressed in Wronskian form entirely in terms of the data in
the initial \cKPrm hierarchy (see eqs.\rf{pchi-a}--\rf{tauok} and 
\rf{pchi-a-1}--\rf{tauok-1} below).
There exists alternative ``spectral'' representation for the $\t$-functions
of \cKPrm DB orbits (cf. eq.\rf{tau-recur-r} below) which turns out to be
very instrumental in interpreting of such $\t$-functions as partition
functions of (multi-)matrix models generalizing the familiar ones with
standard polynomial matrix potentials.

\subsection{Additional Symmetries of Integrable Hie\-rar\-chies and the String
Equation}
Any integrable hierarchy is characterized by the presence of an infinite
number of mutually {\em commuting} flows of the associated evolution
parameters (``times'') -- vector fields acting on the 
space of the corresponding Lax operators. 
They form an infinite-dimensional Abelian algebra
of {\em isospectral} symmetries of the associated nonlinear evolution 
equations. However, the space of all symmetries of the nonlinear evolution 
equations turns out to be much broader \ct{Fuchs-Chen}. The extra symmetries,
which are explicitly time-dependent but commute with the ordinary isospectral
ones, are called ``non-isospectral'' or ``additional'' symmetries and,
furthermore, they form a {\em non-commutative} algebra. A particularly
efficient approach to deal with additional symmetries was developed by
Orlov and Schulman \ct{Orlovetal}. For systematic reviews we refer to
\ct{cortona,moerbeke}; see also the discussions \ct{addsym-models}
of additional symmetries in the context of specific integrable models
(AKNS, {\em truncated} KP, and generalized matrix hierarchies, respectively).

In the Orlov-Schulman approach, the additional symmetries are constructed as
flows (vector fields) acting on the space of Lax operators of the integrable
hierarchy. As proved explicitly in refs.\ct{ASvM,Dickey-addsym}, there
exists an equivalent definition of additional symmetries as vector fields
acting on the space of $\t$-functions (Sato Grassmannian) of the
corresponding hierarchy. This latter formulation allows to provide a simple
interpretation of the Virasoro (and $\Win1$) constraints on
partition functions of (multi-)matrix models in the double-scaling limit
where the underlying integrable hierarchies are of the generalized $SL(r)$ KdV
type. Namely, the Virasoro constraints express invariance of the
$\t$-functions ({\sl i.e.}, the partition functions) under the Borel
subalgebra of the Virasoro algebra of additional non-isospectral symmetries
(similarly for the $\Win1$ constraints).
In particular, the so called ``string'' equation in (multi-)matrix models is
nothing but condition for invariance of the partition function under the 
lowest additional symmetry flow.

There is, however, a problem in applying the Orlov-Schulman formulation of
additional symmetries in the case of the \cKPrm hierarchies. Namely,
the standard Orlov-Schulman additional symmetries for the full unconstrained 
{\sf KP} hierarchy are broken by the constraints when restricting their 
action to the \cKPrm Lax operators.
Therefore, they need to be suitably modified in order for the
corresponding additional symmetry flows to preserve the space of \cKPrm Lax 
operators for any fixed $r,m$ . The solution to the problem is 
adding to the Orlov-Schulman flows appropriate linear combinations of 
the so called ``ghost'' symmetry flows generated by pairs of eigenfunctions
and adjoint eigenfunctions of the \cKPrm Lax operator; see ref.\ct{addsym}
and section \ref{section:addsym} below.

For \DB orbits of \cKPrm hierarchies we have the following important property.
The ``string'' equation ({\sl i.e.}, the condition for invariance under the
lowest additional symmetry flow) on the initial hierarchy {\em alone} turns 
out to be sufficient to uniquely determine the $\t$-functions of the whole
\cKPrm DB orbit which, thus, automatically satisfy the pertinent higher 
constraints (under higher additional symmetry flows).

Implementing the string equation constraint in the setting of the discrete 
multi-matrix models and its corresponding linear problem taking the form 
of Toda-like lattice system turns out to be greatly simplified 
by application of the techniques of the equivalent hierarchy method.
As shown in \ct{office}, the constrained two-matrix Toda lattice structure 
augmented by the string condition can be re-expressed, by a simple  
change of variables in the set of coupling parameters  
as a single set of flow equations for only one independent matrix.
This technique is extended here to the multi-matrix system and
the resulting simplification opens the way for identification of the  
constrained Toda-like lattice system with the corresponding continuum
generalized reduced 
KP hierarchy with the string equation constraint imposed on it.

The method of equivalent hierarchies appeared also in the setting of the Toda 
lattice hierarchy \ct{Tak} and the Generalized Kontsevich Model \ct{KMMM}.
\subsection{Plan of Exposition}

In section \ref{section:noak2-1} we start with a brief review of the basics 
of Sato formulation of the standard unconstrained {\sf KP} hierarchy.
Among the covered topics are: pseudo-differential
Lax operator formalism, Baker-Akhiezer (wave) functions, $\t$-functions,
bi-Hamiltonian structure. Then we go on with introducing the class of reduced
(constrained) \cKPrm hierarchies in various equivalent forms and provide
explicit Miura-like transformations connecting them. Also, as a byproduct we
describe series of explicit free-field realizations of $\Win1$ and $\hWinf$
algebras -- the first and the second fundamental {\sf KP} Poisson bracket 
algebras.

Section \ref{section:noak2-2} is devoted to the explicit construction of
\DB solutions for arbitrary \cKPrm hierarchies.
First, after recalling the basic definition and properties of
DB transformations, we give the derivation of the two-dimensional
Toda lattice structure corresponding to any DB orbit of \cKPrm hierarchies.
Further, we explicitly construct, in terms of simple Wronskian determinants,
the eigenfunctions and $\t$-functions of these general DB orbits. In
particular, we provide a ``spectral'' representation of the \cKPrm
DB-generated $\t$-functions which will become useful for the subsequent
identification of the latter as ``eigenvalue'' integral representations of
the (multi-)matrix model partition functions. 

In the second half of section
\ref{section:noak2-2} we discuss a special class of constrained {\sf KP}
hierarchies -- the so called {\em truncated} {\sf KP} hierarchies, 
defined in terms of dressing operators with finite (truncated) 
pseudo-differential series.
This class includes especially the subset called generalized \BH hierarchies,
constructed in terms of the DB transformations of the  most simple,
non-trivial dressing operator built of one function only.
We identify the truncated {\sf KP} hierarchies as particular members 
of certain \cKPrm DB orbits and present
the explicit solutions for their BA and $\t$-functions which, furthermore, are
shown to contain the well-known multi-soliton {\sf KP} solutions.

In section \ref{section:addsym} we study and solve the problem of modifying
the usual Orlov-Schulman additional symmetries in order for them to preserve
the space of Lax operators of any given constrained \cKPrm hierarchy 
(for fixed $r,m$).
Explicit solution is given for the Virasoro algebra of additional symmetries,
and the generalization of the latter construction for the full $\Win1$ algebra
of additional symmetries is indicated.

In section \ref{section:noak3} the techniques developed in the previous 
sections are
applied to the treatment of the simplest model of random matrices -- the
Hermitian one-matrix model. Its underlying integrability structure is
identified with the above mentioned special case of truncated {\sf KP}
hierarchy -- the generalized \BH system, subject to the additional condition
of invariance under the lowest additional symmetry flow (the ``string''
equation constraint). Also, it is explicitly demonstrated how this last
single constraint uniquely fixes the \BH $\t$-function ({\sl i.e.}, the
one-matrix model partition function) such that the latter automatically
satisfy all higher Virasoro additional-symmetry constraints. 
Also, as a byproduct, we obtain a representation of the $\t$-function of the
most general DB orbit of ${\sf cKP}_{1,1}$ hierarchy in terms of a new 
generalized matrix model (cf. eq.\rf{ZN-1M-wti} below). 

Section \ref{section:noak4} is devoted to a detailed study of the integrable
structure underlying the discrete (Hermitian) two-matrix model. After
recalling the basic properties of the corresponding generalized Toda-like
lattice system, we employ a method similar to the method of equivalent
hierarchies to map the latter lattice system into an equivalent, 
but simpler one, where part of the pertinent ``string'' equation constraints 
are automatically fulfilled. Furthermore, by taking quantities associated 
with a single lattice site 
we explicitly construct, out of the modified Toda-like lattice system,
the relevant continuum integrable hierarchy which is identified as constrained
{\sf KP} hierarchy of ${\sf cKP}_{r,m=1}$ type  
subject to the subsidiary condition of invariance under the lowest
additional (non-isospectral) symmetry flow. The latter identification  
allows us to
apply the \DB techniques from section \ref{section:noak2-2} in order to obtain
the explicit Wronskian solution for the two-matrix model partition function.

In section \ref{section:noak5} we extend the treatment from section 
\ref{section:noak4} to the case of arbitrary multi-matrix models.

The concluding section \ref{section:noak-out} contains brief discussion of
some further developments, in particular, 
the relevance of the present formalism for
studying random matrix problems in condensed matter physics.  

Some useful formulas and identities are collected in the Appendix.

Preliminary short expositions of part of the present results were previously
reported in refs.\ct{office,avoda,symposio,pirin,addsym}.

\sect{KP Hierarchy and its Reductions}
\label{section:noak2-1}
\subsection{Background on KP Hierarchy}
We use the calculus of the pseudo-differential operators (see e.g.
\ct{KP,ldickey}) to describe the
KP hierarchy of integrable nonlinear evolution equations.
In what follows the operator $D$ is such that $ \sbr{D}{f} =
f^{\pr}$
with $f^{\pr}= \pa f = \pa f /\pa x$ and the
generalized Leibniz rule holds:
\be
D^n f  = \sumi{j=0} {n \choose j} (\pa^j f) D^ {n-j}  \quad , \quad
n \in \IZ
\lab{gleib}
\ee
In order to avoid confusion we shall employ the following notations:
for any (pseudo-)\-differential operator $A$ and a function $f$, the symbol
$\, A(f)\,$ will indicate application (action) of $A$ on $f$, whereas the
symbol $Af$ will denote just operator product of $A$ with the zero-order
(multiplication) operator $f$.

In this approach the main object is the pseudo-differential operator:
\be
L = D^r + \sum_{j=0}^{r-2} v_j D^j + \sum_{i=1}^{\infty} u_i D^{-i} 
\lab{lax-op}
\ee
called (generalized) Lax operator.
The Lax equations of motion:
\be
\partder{L}{t_n} = \lb L^{n\over r}_{+} \, , \, L \rb \quad , \; \;
n = 1, 2,
\ldots 
\lab{lax-eq}
\ee
describe isospectral deformations of $L$. In \rf{lax-eq} and in what follows
the subscripts $(\pm )$ of any pseudo-differential
operator $A = \sum_j a_j D^j$ denote its purely differential part
($A_{+} = \sum_{j\geq 0} a_j D^j$) or its purely pseudo-differential part
($A_{-} = \sum_{j \geq 1} a_{-j} D^{-j}$), respectively. Further,
$(t) \equiv (t_1 \equiv x, t_2 ,\ldots )$ collectively  denotes the infinite
set of evolution parameters from \rf{lax-eq}.

One can also represent the Lax operator \rf{lax-op} in terms of the
dressing operator:
\be
W= 1 + \sum_{n=1}^{\infty} w_n D^{-n} \qquad ; \qquad 
L = W D^r\,W^{-1}
\lab{dress-1}
\ee
In this framework eqs.\rf{lax-eq} are equivalent to the so called Wilson-Sato
equations:
\be 
\pa_n W \equiv \frac{\pa W}{\pa t_n} = L^{n\over r}_{+} W - W D^n
= - L^{n\over r}_{-} W
\lab{sato-a}
\ee
(in what follows we shall use frequently the short-hand notation
$\pa_n \equiv \partder{}{t_n}$).
Define next the Baker-Akhiezer (BA) function via:
\be
\psi (t,\l ) = W\bigl( e^{\xi (t,\l )}\bigr) = w(t,\l )e^{\xi  (t,\l )}
\quad ; \quad
w(t,\l ) = 1 + \sum_{i=1}^{\infty} w_i(t)\l^{-i} \ , 
\lab{BA}
\ee
where
\be
\xi(t,\lambda) \equiv  \sum_{n=1}^\infty t_n\lambda^n \qquad; \quad
t_1 = x
\lab{xidef}
\ee
Accordingly, there is also an adjoint BA function:
\be
\psi^{*} = W^{*-1}\bigl( e^{-\xi (t,\l )}\bigr) = 
w^{*}(t,\lambda) e^{-\xi (t,\l )}     \quad ;\quad 
w^{*}(t,\lambda) = 1 + \sum_{i=1}^{\infty} w_i^{*}(t)\lambda^{-i}
\lab{BA-adjoint} 
\ee
and one has the following linear system:
\br 
L\bigl( \psi (t,\l )\bigr) = \l^r \psi (t,\l ) \quad , \quad 
\pa_n\psi = L^{n\over r}_{+}\bigl(\psi (t,\l )\bigr)
\lab{linsys} \\ 
L^{*}\bigl( \psi^{*} (t,\l )\bigr) = \l^r\psi^{*} (t,\l )  \quad , \quad
\pa_n\psi^{*}(t,\l ) = 
- \( L^{\ast}\)^{n\over r}_{+}\bigl(\psi^{*}(t,\l )\bigr)      \nonu
\er
Note that eqs.\rf{lax-eq} for the KP hierarchy flows can be regarded as
compatibility conditions for the system \rf{linsys}.

There exists a quite natural way of developing the KP hierarchy based on one
single function of all evolution parameters -- the so-called tau function
$\tau(t)$ \ct{KP} . This approach
is an alternative to using the Lax operator and the calculus of the
pseudo-differential operators. The $\t$-function is related to the BA functions
\rf{BA}--\rf{linsys} via:
\br 
\psi(t,\l) \eq 
\frac{\tau \bigl( t - [\l^{-1}]\bigr)}{\tau (t)} e^{\xi(t,\l)} 
= e^{\xi(t,\l)} \sumi{n=0} \frac{p_n \( - [\pa]\)\tau (t)}{\tau (t)} \l^{-n} 
\lab{psi-main} \\
\psi^{*}(t,\l) \eq
\frac{\tau \bigl(t + [\l^{-1}]\bigr)}{\tau (t)} e^{-\xi(t,\l)} 
= e^{-\xi(t,\l)} \sumi{n=0} \frac{p_n \( [\pa]\)\tau (t)}{\tau (t)}
\l^{-n} 
\lab{psi-mainc}
\er
where: 
\be
[\l^{-1}] \equiv \( \l^{-1}, \h \l^{-2}, {1\over 3} \l^{-3},\dots\)
\qquad ; \qquad
[\pa] \equiv \(\pa_1, \h \pa_2, {1\over 3} \pa_3, \ldots\)
\lab{tau-short-hand}
\ee
and the Schur polynomials $p_n (t)$ are defined through:
\be
e^{\sum_{l \geq 1} \l^l t_l} = 
\sumi{n=0} \l^n \, p_n (t_1, t_2, \ldots ) 
\lab{Schur}
\ee
Taking into account \rf{psi-main} and \rf{BA}, the expansion for the
dressing operator \rf{dress-1} becomes:
\be
W = \sum_{n=0}^{\infty} \frac{p_n \( - [\pa]\)\tau (t)}{\tau (t)} D^{-n}
\lab{W-main}
\ee

The BA functions enter the fundamental bilinear identity 
\be
\oint_{\infty}\psi(t,\l) \psi^{*}(t',\l) d\l = 0 
\lab{bilide}
\ee 
which generates the entire KP hierarchy. In \rf{bilide}
$\oint_{\infty}(\cdot)d\l$ is the residue integral about $\infty$. It is
possible to rewrite the above identity in terms of the tau-functions
obtaining:
\be
\oint_{\infty} \tau \bigl( t - [\l^{-1}]\bigr) 
\tau \bigl( t' + [\l^{-1}]\bigr) e^{\xi(t,\l)-\xi(t',\l)}d\l = 0 
\lab{biltau} 
\ee
Taylor-expanding \rf{biltau} in $y$ ($t\to t-y,\,\,t'\to t+y$) leads to:
\be
\(\sumi{0} p_n(-2y) p_{n+1} \bigl( [D]\bigr) e^{\sum_1^{\infty} y_i D_i}\)
\tau\,\cdot\,\tau = 0 
\lab{hir1} 
\ee 
with $[D] \equiv (D_1,(1/2) D_2,(1/3)D_3,\ldots)$, where $D_i$ is the Hirota
derivative defined as:
\be 
D^m_j a\,\cdot\,b = (\pa^m/\pa s_j^m) a(t_j+s_j)b(t_j-s_j)\v_{s=0}
\lab{Hirota-deriv}
\ee 
The coefficients in front of $y_n$ in \rf{hir1} yield:
\be 
\( \h D_1D_n -  p_{n+1} \bigl( [D]\bigr) \) \tau \cdot \tau = 0
\lab{hir2}
\ee
Eqs.\rf{hir1} and \rf{hir2} are called Hirota bilinear equations. 

Concluding this subsection let us also recall that the space of (generalized)
KP Lax operators \rf{lax-op} is endowed with bi-Hamiltonian Poisson bracket
structures (another expression of its integrability) which result from the two
compatible Hamiltonian structures on the algebra of pseudo-differential
operators $\PsDA$ \ct{STS83}. The latter are given by:
\br
{\pbbr{\me{L}{X}}{\me{L}{Y}}}_1 \eq
- \llangle L \bv \left\lb X,\, Y \right\rb \rrangle
\lab{first-KP}\\
{\pbbr{\me{L}{X}}{\me{L}{Y}}}_2 \eq {\Tr}_A \( \( LX\)_{+} LY -
\( XL\)_{+} YL \)      \nonu  \\
&+& \!\! {1\over r}\int dx \, {\rm Res}\Bigl( \sbr{L}{X}\Bigr)
\pa^{-1}
{\rm Res}\Bigl( \sbr{L}{Y}\Bigr) \qquad
\lab{second-KP}
\er
where the following notations are used. $<\cdot \v \cdot >$ denotes the
standard bilinear pairing in $\PsDA$ via the Adler trace 
$\me{L}{X} = {\Tr}_A \( LX\)$ with ${\Tr}_A X = \int {\rm Res} X $. Here
$X,Y$ are arbitrary elements of the algebra of pseudo-differential
operators of the form $X = \sum_{k \geq - \infty} D^k X_k $ and similarly for
$Y$. The second term on the r.h.s. of \rf{second-KP} is a Dirac bracket term
originating from the second-class Hamiltonian constraint $v_{r-1}=0$ on $L$
\rf{lax-op}.

In terms of the Lax coefficient functions
$v_{r-2},\ldots ,v_0 ,u_1 ,u_2 ,\ldots ,$ the first Poisson bracket structure
\rf{first-KP} takes the form of an infinite-dimensional Lie algebra which is
a direct sum of two subalgebras spanned by $\{ v_j\}$ and
$\{ u_i\}$, respectively. The latter is called $\Win1$-algebra \ct{W-inf}.
Its Cartan subalgebra contains the infinite set of (Poisson-)commuting
KP integrals of motion ~$H_{l-1} = {1\over l} {\Tr}_A
L^{l\over r}$
whose densities are expressed in terms of the $\t$-function
\rf{psi-main} as:
\be
\pa_x \partder{}{t_l} \ln \t = {\rm Res} L^{l\over r}
\lab{tau-L}
\ee

In turn, the second Poisson bracket structure \rf{second-KP} spans a nonlinear
(quadratic) algebra called $\hWinf (r)$ \ct{W-h-inf}, which is an
infinite-dimensional generalization of Zamolodchikov's $W_N$ conformal
algebras \ct{Zam}.
\subsection{Constrained KP Hierarchy}
\begin{definition}
The function $\Phi$ ($ \Psi$) is called \underbar{(adjoint) eigenfunction} of
the Lax operator $L$ satisfying Sato's flow equation \rf{lax-eq} if its  
flows are given by the expressions:
\be
\partder{\Phi}{t_k} = L^{k\over r}_{+}\bigl( \Phi\bigr) \qquad; \qquad
\partder{\Psi}{t_k} = - \( L^{*} \)^{k\over r}_{+}\bigl( \Psi\bigr)
\lab{eigenlax}
\ee
for the infinite many times $t_k$.
\label{definition:eigen-def}
\end{definition}
Of course, an eigenfunction function, which in addition also
satisfies the spectral equations from \rf{linsys}, is a BA function.
Arbitrary (non-BA) eigenfunctions can be represented in terms of 
the BA function:
\be
\Phi (t) = \int_{\Gamma} d\l \, \p (\l ) \psi (t,\l )
\lab{Phi-psi}
\ee
with an appropriate ``density'' $\p (\l )$ and a contour of
integration $\Gamma$ such that the integral \rf{Phi-psi} exists.

Consider now the Lax operator $Q=D+ \sumi{i=0} v_i D^{-i-1}$
satisfying the 
flow equations as in \rf{lax-eq} with $r=1$.
Let furthermore $\P_i$ and $\Psi_i$ ($i=1, \ldots,m$) be, respectively, 
eigenfunctions and adjoint eigenfunctions of $Q$, according to \rf{eigenlax}.
It is easy to prove the following lemma.
\begin{lemma}
The time evolution of the pseudo-differential operator
$ \Phi_i D^{-1} \Psi_i$ is given by:
\be
\partder{}{t_k} \(\Phi_i D^{-1} \Psi_i \) = \llb B_k\, , \, \Phi_i
D^{-1} 
\Psi_i\rrb_{-} \qquad; \qquad B_k \equiv Q^k_{+}
\lab{tkppsi}
\ee
\label{lemma:eigenflo}
\end{lemma}
\begin{proof}
The proof is a consequence of the following technical observations
based on identity \rf{tkppsi-app} from Appendix \ref{section:appa} and 
eqs.\rf{eigenlax}:
\be
\(B_k \, \Phi_i D^{-1} \Psi_i\)_{-} = B_k \(\Phi_i\) D^{-1}
\Psi_i =
\partder{\Phi_i}{t_k} D^{-1} \Psi_i
\lab{bkppsi}
\ee
and
\be
-\(\Phi_i D^{-1} \Psi_i \, B_k \)_{-} = -  \Phi_i D^{-1} 
B_k^{\ast} \(\Psi_i\) =
\Phi_i D^{-1} \partder{\Psi_i}{t_k}
\lab{ppsibk}
\ee
\end{proof}
What we learn from the above lemma is that if we define a Lax operator
$L$, such that its purely pseudo-differential part is 
$L_{-} = \Phi D^{-1} \Psi$, then $L_{-} $ satisfies automatically 
the KP-type flow equations as in \rf{lax-eq}:
$\pa L_{-}/ \pa t_k = \lb B_k\, , \, L \rb_{-}$.

Following \ct{oevela,chengs} let $\pa_{\a_{i}}$ be a vector field, whose
action on the Lax operator $Q$ is induced by the (adjoint) eigenfunctions 
$\P_i, \Psi_i$ through:
\be
\pa_{\a_{i}} {Q} \equiv \lb Q \,  , \, \Phi_i D^{-1} \Psi_i \rb
\lab{ghostflo}
\ee
We find from the lemma \ref{lemma:eigenflo} the following proposition
\ct{oevela,chengs} :
\begin{proposition}
The vector fields (``ghost'' flows) $\pa_{{\a_{i}}}$ \rf{ghostflo} commute 
with the isospectral flows of the Lax operator $Q$:
\be
\sbr{\pa_{\a_{i}}}{\pa_n} Q = 0  \qquad n=1, 2, \ldots 
\lab{comm}
\ee
{\sl i.e.}, $\pa_{{\a_{i}}}$ define symmetries for the corresponding Lax
evolution eqs. $\pa_n Q = \Sbr{Q^n_{+}}{Q}$ .
\label{proposition:commflo}
\end{proposition}
\begin{definition}
The constrained KP hierarchy (denoted as ${\sf cKP}_{r,m}$) is obtained by
identifying the ``ghost'' flow $\sum_{i=1}^m \pa_{{\a_{i}}}$  with the 
isospectral flow $\pa_r$.
\label{definition:ckpdef}
\end{definition}
Comparing \rf{ghostflo} with the \rf{lax-eq} we find that for the
Lax operator
belonging to ${\sf cKP}_{r,m}$ hierarchy we have:
\be
Q^r_{-} = \sum_{i=1}^m \Phi_i D^{-1} \Psi_i 
\lab{qrpipsi}
\ee
Hence, we are led to the Lax operator $L = Q^r$ given by:
\be
L \equiv L_{r,m}  = D^r+ \sum_{l=0}^{r-2} u_l D^l + 
\sum_{i=1}^m \Phi_i D^{-1} \Psi_i
\lab{f-5} 
\ee
and subject to the Lax equation:
\be
\partder{ L}{t_k} = \lb \( L^{k/r} \)_{+} \, , \, L \rb
\lab{f-5a}
\ee
An equivalent useful form of \rf{f-5} is:
\be
L \equiv L_{r,m}  = D^r+ \sum_{l=0}^{r-2} u_l D^l + 
\sum_{i=1}^m a_i \( D - b_i\)^{-1}
\lab{f-5-1}
\ee
where
\be
\P_i = a_i e^{\int b_i} \quad , \quad \Psi_i = e^{-\int b_i}
\lab{f-5-2}
\ee

It will be convenient for us to parametrize the ${\sf cKP}_{r,m}$
hierarchy in terms of the Lax operator from \rf{f-5}.
The functions $\Phi_i$ and $ \Psi_i$ remain eigenfunctions of $L_{r,m}$
\rf{f-5} according to definition \ref{definition:eigen-def}, 
{\sl i.e.}, they satisfy eqs.\rf{eigenlax}.

We now recall results of \ct{avoda,rio}
which showed that the Lax operator \rf{f-5} has an equivalent
form as a ratio of two purely differential operators:
\begin{lemma}
The ${\sf cKP}_{r,m}$ Lax operator from \rf{f-5} can be
equivalently rewritten as a ratio 
\be
L_{r,m} = L_{m+r} L_{m}^{-1}  \qquad \; m , r \geq 1
\lab{laxmk} 
\ee
of two purely differential operators
\br
L_{m+r} &=& \( D + v_{m+r}\) \( D + v_{m+r-1} \) \cdots \( D + v_1 \)
\nonu \\
L_m &=& \( D + \tv_{m} \) \( D + \tv_{m-1} \) \cdots \( D + \tv_1 \)
\lab{i-lmk-a}
\er
with coefficients $v_i, \tv_i$ subject to a constraint:
\be
\sum_{j=1}^{m+r} v_j - \sum_{l=1}^{m}  \tv_l= 0
\lab{i-psimk1}
\ee
\label{lemma:ratio}
\end{lemma}
\begin{proof}
Let us start with the expression \rf{laxmk}. It can easily be
rewritten as
\be
L_{m+r} L_{m}^{-1}= \sum_{l=1}^{m} A_l  \prod_{i=l}^{m} \( D +
{\tv}_{i}\)^{-1}
+ \sum_{l=0}^{r-2} A_{l+m+1}  D^{l} + D^{r} \lab{1ff}
\ee  
taking into account condition \rf{i-psimk1}.
We therefore need only to concentrate on the pseudo-differential
part of
\rf{1ff}. We can rewrite it as
\br
\( L_{m+r} L_{m}^{-1}\)_{-} &= &\sum_{l=1}^m r_l \prod_{i=l}^m
D^{-1} q_i
\lab{iss-8aaa} \\
r_l= A_{l} e^{-\int \tv_{l}} \quad ;\;\; q_m= e^{\int \tv_m} \;\;
&,&
\;\; q_i= e^{\int \( \tv_{i} - \tv_{i+1}\)} \;\;, \;\;
i=1,\ldots ,m-1  \lab{4.5}
\er
Let us define the quantity:
\be
Q_{l,i} \equiv (-1)^{i-m} \int q_i \int q_{i-1} \int \ldots
\int q_l\, (dx^{\pr})^{i-l+1}
\quad \qquad 1 \leq l\leq i \leq m
\lab{qni}
\ee
Then, after using that $ D^{-1} Q_{1,i-1}  q_i = D^{-1} Q_{1,i} D- Q_{1,i}$, 
eq.\rf{iss-8aaa} acquires the form :
\be
\( L_{m+r} L_{m}^{-1}\)_{-}= \sum_{i=2}^{m}  r_i^{(1)} \prod_{l=i}^m
D^{-1} q_l
+ r_1 D^{-1} \( - Q_{1,m-1} q_m\)
\lab{rqlax-a}
\ee
where
\be
r_i^{(1)} \equiv r_i +r_1 Q_{1,i-1} \qquad i=2, \ldots , m
\lab{ronei}
\ee
The above process can be continued to yield the expression:
\be
\( L_{m+r} L_{m}^{-1}\)_{-}=  \sum_{i=1}^m \Phi_i D^{-1} \Psi_i
\lab{f-5b}
\ee
with
\br
\P_i \eq r_i + \sum_{m=1}^{i-1} r_{m}
\sum_{s_{i-m-1} =s_{i-m-2} +1}^{i-m} \cdots \sum_{s_2 =s_{1}
+1}^{i-m}
\sum_{s_{1} = 1}^{i-m} Q_{m,i-s_{i-m-1}-1}
Q_{i-s_{i-m-1},i-s_{i-m-2}-1} \cdots                    \nonu\\
&\cdots& Q_{i-s_2,i-s_{1}-1} Q_{i-s_{1},i-1}
\qquad \qquad\qquad \qquad 1 \leq i \leq m
\lab{piri} \\
\Psi_m \eq q_m \quad\; , \quad \;
\Psi_i = (-1)^{m-i} q_m \int q_{m-1} \int \ldots \int q_i
\, (dx^{\pr})^{m-i}  \qquad 1\leq i \leq m-1
\lab{psiqi}
\er

We now give an opposite construction starting from the
Lax operator \rf{f-5} with pseudo-differential part
$\(L\)_{-} =  \sum_{i=1}^m \P_i  D^{-1} \Psi_i$ .
First, let us consider the simplest nontrivial case $m=2$.
Consider the identity:
\br
\phi D^{-1} \psi &=& \phi \psi \( \chi^{-1} D^{-1} \chi \) -
\phi D^{-1} \llb \chi \(\pa_x \frac{\psi}{\chi}\)\rrb
\( \chi^{-1}D^{-1}\chi\) \phanta   \nonu \\
&=& \phi \psi \( \chi^{-1} D^{-1} \chi \) -
\phi \frac{W\llb \chi ,\psi\rrb}{W\llb \chi\rrb}
\( \frac{W\llb \chi\rrb}{W\llb \chi ,\psi\rrb} D^{-1}
\frac{W\llb \chi ,\psi\rrb}{W\llb \chi\rrb}\) \(
\chi^{-1}D^{-1}\chi\)
\lab{iss-4}
\er
for arbitrary functions $\phi,\psi,\chi$, where in deriving \rf{iss-4} we have
used the simple Wronskian identity \rf{iw} from Appendix.
Here and in what follows we shall use the notations:
\be
W_k \equiv W_k \lb \psi_1 ,\ldots ,\psi_k \rb =
\det {\Bigl\Vert} \pa^{i-1} \psi_j {\Bigr\Vert}_{i,j=1,\ldots ,k}
\quad ,\quad W_0 =1 
\lab{wronski-det}
\ee
where $W_k$ denotes the Wronskian determinant of $\{ \psi_1, \ldots ,\psi_k\}$.

We obtain from \rf{iss-4}:
\br
L &=& L_{+} + \Phi_1 D^{-1} \Psi_1 + \Phi_2 D^{-1} \Psi_2  \nonu \\
&=&L_{+} + A^{(2)}_2 \( D + B^{(2)}_2 \)^{-1} +
A^{(2)}_1 \( D + B^{(2)}_1 \)^{-1} \( D + B^{(2)}_2 \)^{-1} 
\lab{iss-1} \\
A^{(2)}_2 &=& \Phi_1 \Psi_1 + \Phi_2 \Psi_2  \quad , \quad
B^{(2)}_2 =  \pa_x \ln \Psi_2  \lab{iss-2}  \\
A^{(2)}_1 &=& -\Phi_1 \frac{W\llb \Psi_2 ,\Psi_1\rrb}{W\llb
\Psi_2\rrb}
\quad ,\quad
B^{(2)}_1 = \pa_x \ln \frac{W\llb \Psi_2 ,\Psi_1\rrb}{W\llb
\Psi_2\rrb}
\lab{iss-3}
\er
This construction can be extended by induction to yield for general $m$: 
\br
L &\equiv& L_{r,m} = L_{+} + \sum_{i=1}^m \P_i D^{-1} \Psi_i    \nonu \\
{} \eq  D^r + \sum_{l=0}^{r-2} u_l D^l + 
\sum_{i=1}^m A^{(m)}_i \( D + B^{(m)}_i \)^{-1}
\( D + B^{(m)}_{i+1}\)^{-1} \cdots \( D + B^{(m)}_m \)^{-1}  
\lab{iss-8b}
\er
with 
\br
A^{(m)}_k &=& (-1)^{m-k} \sum_{s=1}^k \Phi_s \frac{W_{m-k+1}\llb \Psi_m
,\ldots ,\Psi_{k+1},\Psi_s \rrb}{W_{m-k}\llb \Psi_m ,\ldots ,\Psi_{k+1}\rrb}
\lab{iss-8c}    \\
B^{(m)}_k &=& \pa_x \ln \frac{W_{m-k+1}\llb \Psi_m ,\ldots ,
\Psi_{k+1},\Psi_k \rrb}{W_{m-k}\llb \Psi_m ,\ldots ,\Psi_{k+1}\rrb}
\lab{iss-8d}
\er
{\sl i.e.}, the expression \rf{iss-8b} is of the same form as \rf{1ff}.
Comparing the latter two equations we see that ${\ti v}_i = B^{(m)}_i $.
\end{proof}

Let $\{ \psi_1, \ldots, \psi_m \}$ be a basis of linearly independent solutions
of the $m$-th order differential equation $K \psi = 0$,
where $K=\( D + v_m \) \( D + v_{m-1} \) \ldots \( D + v_1 \) $.
Then one can show that:  
\be
v_k = \pa \( \ln { W_{k-1} \over W_{k}} \) \qquad ; \quad 
W_k \equiv W_k \lb \psi_1 ,\ldots ,\psi_k \rb
\lab{wil}
\ee
Hence by comparing \rf{wil} and \rf{iss-8d} we obtain the following corollary:
\begin{corollary}
The adjoint eigenfunctions $\Psi_i$ are in the kernel of the
purely differential operator $\(L_m\)^{*}$ from \rf{laxmk}--\rf{i-lmk-a}. 
\label{corollary:kernel}
\end{corollary}

Similarly to the Miura-like transformation \rf{iss-8c}--\rf{iss-8d}, we can
express the Lax coefficients $u_l , A^{(m)}_i ,B^{(m)}_i\,$ of \rf{iss-8b} 
in terms of the Lax coefficients $v_j ,\tv_l\,$ of $L_{r,m}$ in the form
\rf{laxmk}--\rf{i-lmk-a} as follows \ct{office} :
\br
\sum_{n_s =1}^{m+r-s} \( \pa + v_{n_s +s} - \tv_{n_s}\)
\sum_{n_{s-1} =1}^{n_s} \( \pa + v_{n_{s-1} +s-1} - \tv_{n_{s-1}}\)
\times \cdots \times    \nonu  \\
\times \sum_{{n_2} =1}^{n_3} \( \pa + v_{n_2 +2} - \tv_{n_2}\)
\sum_{{n_1} =1}^{n_2} \( \pa + v_{n_1 +1} - \tv_{n_1}\)
\Bigl( \sum_{l= n_1 +1}^{m+r} \( \tv_l - v_l \) \Bigr)   \nonu  \\
= \left\{ \begin{array}{ll}
u_{r-1-s} \;\; , & \quad  {\rm for} \;\; s=1, \ldots ,r-1 \\
A^{(m)}_{m+r-s}\;\; , & \quad  {\rm for} \;\; s=r, \ldots ,m+r-1 \end{array} 
\right.
\lab{Miura-A-v}
\er
(here it is understood that $\tv_j \equiv 0\,$ for $j \geq m+1$ ).

We conclude this section with some remarks.

\mark
Let us point out that there is an alternative way
(w.r.t. definition \ref{definition:ckpdef}) to introduce the \cKPrm
hierarchies. Namely, one can view \cKPrm in either of their equivalent forms
\rf{f-5}, \rf{laxmk}--\rf{i-lmk-a} or \rf{iss-8b}, as consistent Hamiltonian
reductions \ct{multikp,no2rabn1,office} of the original full (unconstrained)
KP hierarchy \rf{lax-op} w.r.t. the bi-Hamiltonian Poisson structures
\rf{first-KP}--\rf{second-KP}. Moreover, $L \equiv L_{r,m}$ obeys the same 
fundamental Poisson bracket algebras \rf{first-KP}--\rf{second-KP} as the 
original unconstrained $L$ \rf{lax-op}. 

\mark
The first Hamiltonian structure \rf{first-KP} for $L \equiv L_{r,m}$
\rf{f-5-1} is equivalent to the following free field fundamental Poisson
brackets:
\be
\pbbr{\P_i (x)}{\Psi_j (x^\pr )}_1 = - \d_{ij} \d (x - x^\pr ) \quad , \quad
{\rm rest} = 0
\lab{fpb1-1}
\ee
or, equivalently, in terms of the fields $a_i ,b_i\,$ \rf{f-5-2} :
\be
\pbbr{a_i (x)}{b_j (x^\pr )}_1 = - \d_{ij} \pa_x \d (x - x^\pr ) \quad , \quad
{\rm rest} = 0
\lab{fpb1-2}
\ee
Expanding \rf{f-5-1} in a standard pseudo-differential operator series:
\br
L_{r,m} \eq L_{+} + \sum_{k=1}^\infty U_k (\P ,\Psi ) \, D^{-k} =
L_{+} + \sum_{k=1}^\infty U_k (a,b) \, D^{-k}
\lab{f-5-3} \\
U_k (\P ,\Psi ) \!\!&\equiv & \! \!\sum_{i=1}^m \P_i (-\pa_x )^{k-1} \Psi 
\; \, , \; \, 
U_k (a,b) \equiv  \sum_{i=1}^m a_i (-1)^{k-1} P_{k-1} (- b_i )
\lab{u-free-1}
\er
where $P_k (\cdot )$ are the familiar {\faa} polynomials, we obtain a series of
free-field realizations of the $\Win1$ algebra of the Lax coefficient fields
\rf{u-free-1} in terms of $\P_i ,\Psi_i\,$ \rf{fpb1-1} or $a_i ,b_i\,$
\rf{f-5-2}, respectively.

\mark
The second Hamiltonian structure \rf{second-KP} for $L \equiv L_{r,m}$
\rf{laxmk}--\rf{i-lmk-a} is equivalent to the following free-field fundamental
Poisson brackets:
\br
\pbbr{v_i (x)}{v_j (x^\pr )}_2 \eq \(\d_{ij} -{1\over r}\) \pa_x \d (x-x^\pr )
\; , \qquad i,j =1,\ldots , m+r  \nonu \\
\pbbr{{\ti v}_k (x)}{{\ti v}_l(x^\pr )}_2 \eq
- \( \d_{kl} + {1\over r}\) \pa_x \d (x-x^\pr ) \; ,\qquad k,l =1,\ldots ,m
\nonu \\
\pbbr{v_i (x)}{{\ti v}_l (x^\pr )}_2 \eq {1\over r} \pa_x \d (x-x^\pr )   
\lab{fpb2}
\er
which, as demonstrated in refs.\ct{Yu,office}, is precisely the Cartan
subalgebra of the graded $SL(m+r,m)$ Kac-Moody algebra. This latter property
justifies the alternative name of the ${\sf cKP}_{r,m}$
hierarchies -- $SL(m+r,m)$ KP-KdV hierarchies generalizing the usual
$SL(r)$ KdV hierarchies defined through purely differential Lax operators:
\be
L \equiv L_{r,0} = D^r + \sum_{l=0}^{r-2} u_l D^l =
\( D - v_r \) \cdots \( D - v_1\)
\lab{kdv-r}
\ee
Expanding \rf{laxmk}--\rf{i-lmk-a} in a standard pseudo-differential operator
series:
\br
L_{r,m} \eq D^r + \sum_{l=0}^{r-2} u_l (v,\tv ) D^l +
\sum_{j=1}^\infty U_j (v,\tv ) D^{-j}   
\lab{laxmk-a} \\
U_j (v,\tv ) \!\! &\equiv& \!\! \sum_{s=0}^{{\rm min}(m-1,j-1)} \!
(-1)^{j-1-s} A^{(m)}_{m-s} (v,\tv ) P^{(s+1)}_{j-1-s} \( \tv_{m-s}, 
\ldots ,\tv_m \)
\lab{u-free-2}
\er
where $u_l (v,\tv ) ,\, A^{(m)}_{s} (v,\tv )$ are given by \rf{Miura-A-v} and 
\be
P^{(s)}_n \( \p_1 ,\ldots ,\p_s \) =  \sum_{l_1 + \cdots + l_s = n} 
\(\pa + \p_1\)^{l_1} \ldots \(\pa + \p_s\)^{l_s}
\lab{multi-faa}
\ee
are the multiple \faa polynomials, we obtain a series of
free-field realizations of the nonlinear ${\hat W}_{\infty}^{(r)}$ algebra of
the Lax coefficient fields $u_l (v,\tv ) , U_j (v,\tv )\,$  
\rf{Miura-A-v},\rf{u-free-2} in terms of $v_i ,\tv_l\,$ \rf{fpb2}.

\sect{Darboux-B\"{a}cklund Solutions of Con\-strained KP $\,\,$ Hierarchies}
\label{section:noak2-2}
\subsection{DB Transformation and Eigenfunctions of the Lax Ope\-rator}
We begin with reviewing the basic properties
of DB transformations for the general (unconstrained) KP hierarchy 
stressing its basic property of preserving the form of the pertinent
Lax evolution equation. Related material can be found in, {\sl e.g.}, 
refs.\ct{chau,oevela}. Next, we discuss in more detail the DB transformations 
and explicit solutions for the constrained \cKPrm hierarchies.

To describe a result of acting with the DB transformation in the setting of 
pseudo-differential operators we need first to make a simple technical
observation.
\begin{lemma}
For an arbitrary pseudo-differential operator $A$ we have the
following identity \ct{oevelr}:
\be
\( \chi D \chi^{-1} A \chi D^{-1} \chi^{-1} \)_{+} =
\chi D \chi^{-1} A_{+}  \chi D^{-1} \chi^{-1}
-  \chi \pa_x \(\chi^{-1}  A_{+} (\chi )\)  D^{-1} \chi^{-1}
\lab{aonchi}
\ee
\label{lemma:aplusdb}
\end{lemma}
Let $L$ satisfy the generalized KP Lax equation
$\pa_k  L = \sbr{L_{+}^{k\ov r}}{L}$.
Consider a ``gauge'' transformation:
\be
L \to {\ti L} \equiv T L T^{-1}
\lab{gauge-transf}
\ee
with $T$ being a pseudo-differential operator.
Then the transformed Lax operator will satisfy:
\be
\pa_k {\ti L} =
\sbr{T L_{+}^{k\ov r}T^{-1} + (\pa_k T )T^{-1}}{{\ti L}}
\lab{tisato}
\ee
\begin{definition}
The DB transformation is defined by \rf{gauge-transf} with the following
special form of $T$ :
\be
T = \Phi D \Phi^{-1} \qquad ; \quad
L_{+}^{k\ov r} \Phi = \pa_k \Phi
\lab{T-Phi}
\ee
{\sl i.e.}, with $\Phi$ being an eigenfunction of $L$.
\label{definition:DB}
\end{definition}
  
One easily verifies that for $A= L^{k\ov r}$ and $\chi = \Phi$ eq.\rf{aonchi} 
becomes:
\be
\( T L^{k\ov r}T^{-1} \)_{+} =     T L_{+}^{k\ov r}T^{-1} + (\pa_k T
)T^{-1}
\lab{lonphi}
\ee
Correspondingly, \rf{tisato} takes the form:
\be
\pa_k {\ti L} =
\sbr{{\ti L}_{+}^{k\ov r}}{{\ti L}}
\lab{tisato-a}
\ee
and we have, therefore, established:
\begin{proposition}
The DB transformation \rf{gauge-transf} with an eigenfunction $\Phi$
\rf{T-Phi} preserves the form of the Lax equation \rf{lax-eq}, {\sl i.e.}, 
the DB transformed Lax operator satisfies
the same evolution equation \rf{tisato-a} as the original Lax operator.
\label{proposition:dbcovariance}
\end{proposition}

Using simple identities valid for any pseudo-differential operator $A$ and
arbitrary function $f$ :
\be
{\rm Res}\,\( f A f^{-1}\) = {\rm Res}\, A \quad ,\quad
{\rm Res}\, \( D A D^{-1}\) = {\rm Res}\, A + \pa_x A_{(0)}
\lab{Res-id}
\ee
where the subscript ${\small (0)}$ indicates taking the zeroth order term, we
obtain:
\be
{\rm Res} {\wti L}^{l\over r} =
{\rm Res} \( \Phi D \Phi^{-1}{L}^{l\over r} \Phi D^{-1} \Phi^{-1} \) = 
{\rm Res} {L}^{l\over r} + \pa_x \partder{}{t_l} \ln \Phi
\lab{resa}
\ee
Combining \rf{resa} with eq.\rf{tau-L} for the $\t$-function, we get therefore
\ct{chau}:
\begin{proposition}
Under DB transformation \rf{gauge-transf} with an eigenfunction $\Phi$
\rf{T-Phi} the $\t$-function associated with the Lax operator $L$
transforms according to: $\t \to {\wti \t} = \P \t$.
\label{proposition:dbtau}
\end{proposition}

It is also useful to write down the DB transformation in terms of
the dressing operator and the BA function as well:
\br
W \to {\wti W} = \( \Phi D \Phi^{-1}\) W D^{-1}
\lab{DB-W} \\
\psi (t,\l ) \to {\wti \psi (t;\l )} = 
\l^{-1} \Phi (t) \pa \( \frac{\psi (t,\l )}{\Phi (t)}\)
\lab{DB-BA}
\er
Inserting \rf{DB-BA} and ${\wti \t}(t) = \P (t)\t (t)$ 
(from Prop.\ref{proposition:dbtau}) in the expression \rf{psi-main} for the
DB-transformed BA function:
\be
{\wti \psi}(t,\l) =
\frac{{\wti \t}\bigl( t - [\l^{-1}]\bigr)}{{\wti \t}(t)}
e^{\xi(t,\l)}
\ee
we deduce the following identity relating the $\t$-function with an
arbitrary eigenfunction $\Phi$ of $L$ :
\be
\frac{\P\bigl( t-[\l^{-1}]\bigr)}{\P (t)} - 1 + \l^{-1} \pa \ln \P (t) =
\l^{-1} \pa \ln \frac{\t \bigl( t-[\l^{-1}]\bigr)}{\t (t)}
\lab{tau-Phi-id}
\ee
Let us illustrate possible applications of \rf{tau-Phi-id} by the 
following example. Taking into consideration that:
\be
 \l  \frac{\P\bigl( t-[\l^{-1}]\bigr)}{\P (t)} = \sumi{n=1}  { p_n
( - \lb \pa \rb ) \P (t)\over \l^{n-1} \P (t)}
\lab{212a}
\ee
and expanding in $\l$ we find from \rf{tau-Phi-id} :
\be
\pa \frac{\t \bigl( t-[\l^{-1}]\bigr)}{\t (t)}
= { p_2 ( - \lb \pa \rb ) \P (t) \over \l \P (t)} + O ( \l^{-2})
\lab{212b}
\ee 
This implies that: 
\br
- \( \ln \t \)^{\pr \pr } \eq  { p_2 ( - \lb \pa \rb ) \P \over \P}
= \h { \( \pa_2 - \pa_1^2 \) \P \over \P } 
\lab{212c}\\
\eq \h { \( (L^{2/r})_{+} - \pa_1^2 \) \P \over \P } ={ 1\over r} u_{r-2} 
\lab{212d}
\er
Hence by  only using the basic fact that $\P$ was an arbitrary eigenfunction we 
obtained from \rf{tau-Phi-id} a very
general result $\( \ln \t \)^{\pr \pr} = {\rm Res} L^{1/r}$.

Analogously to \rf{tau-Phi-id} one also obtains an identity 
relating the $\t$-function with an arbitrary adjoint
eigenfunction $\Psi$ :
\be
\frac{\Psi\bigl( t+[\l^{-1}]\bigr)}{\Psi (t)} - 1 - \l^{-1} \pa \ln \Psi (t) =
- \l^{-1} \pa \ln \frac{\t \bigl( t+[\l^{-1}]\bigr)}{\t (t)}
\lab{tau-Psi-id}
\ee

The next step is to consider DB transformations within ${\sf cKP}_{r,m}$ 
hierarchy, {\sl i.e.}, such transformations 
\rf{gauge-transf},\rf{T-Phi} which preserve the constrained \cKPrm
form \rf{f-5} of the corresponding Lax operator.
Under an arbitrary DB transformation 
$ {\wti L} =\(\chi D \chi^{-1}\)\, L\, \(\chi D^{-1} \chi^{-1}\) $,
where $\chi$ is an eigenfunction of $L$ \rf{f-5}, the
transformed Lax operator reads:
\br
{\wti L} &=& \chi D \chi^{-1} \( L_{+} + \sum_{i=1}^{m} \Phi_i
D^{-1} \Psi_i \)
\chi D^{-1} \chi^{-1} \equiv {\wti L}_{+} + {\wti L}_{-} 
\lab{baker-1} \\
{\wti L}_{+} &=& {L}_{+} + \chi \( \pa_x \( \chi^{-1} L_{+} \chi
\)_{\geq 1} D^{-1}\) \chi^{-1}    
\lab{baker-2}  \\
{\wti L}_{-} &=& {\wti \Phi}_0 D^{-1} {\wti \Psi}_0 +
\sum_{i=1}^{m} {\wti \Phi}_i D^{-1} {\wti \Psi}_i   
\lab{baker-3} \\
{\wti \Phi}_0 &=& \chi \llb \pa_x \( \chi^{-1} L_{+} (\chi )\) +
\sum_{i=1}^m
\( \pa_x \( \chi^{-1}\Phi_i\) \pa_x^{-1} \( \Psi_i \chi\)
+ \Phi_i \Psi_i \) \rrb \nonu \\
&\equiv& \( \chi D \chi^{-1} L \)(\chi ) 
\lab{baker-4} \\
{\wti \Psi}_0 &=& \chi^{-1} \quad ,\quad
{\wti \Phi}_i = \chi \pa_x \( \chi^{-1} \Phi_i \) \quad ,\quad
{\wti \Psi}_i = - \chi^{-1} \pa_x^{-1} \( \Psi_i \chi \)  
\lab{baker-5}
\er
{}Let us recall, as already stressed in the introduction, that all functions
appearing in \rf{baker-2}--\rf{baker-5} are (adjoint) eigenfunctions of
$L$ \rf{f-5} or ${\wti L}$ \rf{baker-2}, but they are {\em not} (adjoint) BA
(wave) functions. In other words, they satisfy:
\be
\partder{}{t_k} f = L^{{k\over r}}_{+} (f)
\qquad f \equiv \chi ,\Phi_i \quad ; \quad
\partder{}{t_k} \Psi_i = - {L^\ast}^{{k\over r}}_{+} (\Psi_i )
\lab{baker-6}
\ee
and similarly for 
${\wti f} \equiv {\wti \Phi_0},{\wti \Phi_i}\; ,\; {\wti \Psi_i}$
with $L \to {\wti L}$.

In case when $\chi$ coincides with one of the original eigenfunctions of $L$,
{\sl e.g.}, $\chi = \Phi_1$, it follows that ${\wti \Phi}_1 =0$ and the DB
transformation \rf{baker-1} preserves the form \rf{f-5} of the Lax operators
involved, {\sl i.e.}, it becomes an {\em auto}-\Back ~transformation. Applying
successive DB transformations
\be 
L^{(k)} = T^{(k-1)} L^{(k-1)} \( T^{(k-1)}\)^{-1} \quad \;, \quad \;
T^{(k)} \equiv \Phi_1^{(k)} D \( \Phi_1^{(k)}\)^{-1}
\lab{shabes-03}
\ee
yields:
\br
L^{(k)} \eq  \( L^{(k)}\)_{+} + \sum_{i=1}^m \Phi_i^{(k)} D^{-1} \Psi_i^{(k)}
\lab{shabes-3} \\
\Phi_1^{(k+1)} \eq \( T^{(k)} L^{(k)}\) \(\Phi_1^{(k)}\)  \quad ,\quad
\Psi_1^{(k+1)} = \( \Phi_1^{(k)}\)^{-1}
\qquad, \qquad k =0,1,\ldots \lab{shabes-2} \\
\Phi_i^{(k+1)} \eq T^{(k)} \(\Phi_i^{(k)}\) \equiv
\Phi_1^{(k)} \pa_x \( \( \Phi_1^{(k)}\)^{-1} \Phi_i^{(k)}\) 
\qquad , \;\; i=2,\ldots , m
\lab{shabes-5} \\
\Psi_i^{(k+1)} \eq \( T^{(k)}\)^{\ast\, -1} \(\Psi_i^{(k)}\) =
- \( \Phi_1^{(k)}\)^{-1} \pa_x^{-1} \( \Psi_i^{(k)}\Phi_1^{(k)} \) \
\qquad , \;\; i=2,\ldots , m 
\lab{shabes-5a}
\er

Using the first equality from \rf{shabes-3}, {\sl i.e.}, 
$ L^{(k+1)} T^{(k)}=T^{(k)} L^{(k)}$ , one can rewrite \rf{shabes-2} 
in the form:
\be
\Phi_1^{(k)} = T^{(k-1)} T^{(k-2)} \cdots T^{(0)}
\( \( L^{(0)}\)^k \bigl(\Phi_1^{(0)}\bigr)\)
\lab{shabes-4}
\ee
whereas:
\be
\Phi_i^{(k)} = T^{(k-1)} T^{(k-2)} \cdots T^{(0)}\bigl(\Phi_i^{(0)}\bigr)
\qquad ,\;\;
\quad i=2,\ldots , m \lab{shabes-6}
\ee
Accordingly, for the BA functions we have from \rf{DB-BA} :
\be
\psi^{(k)} (t,\l ) = \l^{-1} T^{(k-1)} \bigl(\psi^{(k-1)}(t,\l )\bigr) =
\l^{-k} T^{(k-1)} T^{(k-2)} \cdots T^{(0)} \bigl(\psi^{(0)} (t,\l )\bigr) 
\lab{shabes-BA}
\ee
which together with \rf{shabes-4} and \rf{Phi-psi} implies the following
``spectral'' representations:
\br
\Phi_1^{(k)}(t) \eq \int d\l \, \l^{k(1+r)} \p_1^{(0)}(\l )
\psi^{(k)} (t,\l ) = \int d\l \, \l^{k(1+r)} \p_1^{(0)}(\l )
\frac{\t^{(k)} \bigl( t -[\l^{-1}]\bigr)}{\t^{(k)}(t)} e^{\xi (t,\l )}
\lab{spectr-1} \\
\Phi_i^{(k)}(t) &=& \int d\l \, \l^{k} \p_i^{(0)}(\l )
\psi^{(k)} (t,\l )
\quad , \;\; i=2, \ldots , m
\lab{spectr-i}
\er
where $\xi (t,\l )$ was defined in \rf{xidef}.

Finally, for the coefficient of the next-to-leading differential
term in \rf{f-5} ~$ u_{r-2} = r \, Res\, L^{{1\over r}} = r\, \pa_x^2 \ln\t$
(recall eq.\rf{tau-L}) we easily obtain from \rf{baker-2} (with $\chi =\Phi_1$)
its $k$-step DB-transformed expression:
\be
{1\over r} \( u_{r-2}^{(k)} - u_{r-2}^{(0)}\) =
\pa_x^2 \ln \frac{\t^{(k)}}{\t^{(0)}} =
\pa_x^2 \ln \( \Phi_1^{(k-1)} \cdots \Phi_1^{(0)}\)   \lab{sol-3-a}
\ee
which conforms with the result of multiple application of 
Prop.\ref{proposition:dbtau} :
\be
\frac{\t^{(k)}}{\t^{(0)}} = \P_1^{(k-1)} \cdots \P_1^{(0)}
\lab{sol-3-aa}
\ee

Let us consider momentarily the special case of ${\sf cKP}_{r,1}$ hierarchies
({\sl i.e.}, $m=1$).
Combining eqs.\rf{shabes-2},\rf{shabes-3} (for $m=1$) and \rf{sol-3-aa},
we deduce the following:
\begin{proposition}
Any \DB orbit of ${\sf cKP}_{r,1}$ hierarchy \rf{f-5} (for any $r\geq 2$ and 
$m=1$) defines a structure of two-dimensional Toda lattice model \ct{U-T}
(here $\P \equiv \P_1$) :
\br
\pa_x \partder{}{t_r} \ln \P^{(n)} &=&
\frac{\P^{(n+1)}}{\P^{(n)}} - \frac{\P^{(n)}}{\P^{(n-1)}}  \qquad ,\quad
n=0,1,2,\ldots
\lab{rec-rel-n}  \\
\P^{(-1)} &\equiv& \( \Psi^{(0)}\)^{-1}   \nonu  \\
\pa_x \partder{}{t_r} \ln \t^{(n)} &=&
\frac{\t^{(n+1)}\t^{(n-1)}}{\(\t^{(n)}\)^2}  \qquad ,\quad
n=1,2,\ldots
\lab{2d-toda}
\er
\label{prop:2d-toda}
\end{proposition}

Indeed, introducing variables $\psi_n$, such that
$\,\P^{(n)} = \exp \lcurl \psi_{n+1} - \psi_n \rcurl\,$ and
$\,\psi_n = 0\,$ for $n = -1,-2,\dots$ , the recurrence relation 
\rf{rec-rel-n} for the ${\sf cKP}_{r,1}$ eigenfunctions turns into the
familiar two-dimensional Toda lattice equations of motion, whereas the
recurrence relation eq.\rf{2d-toda} for the corresponding 
${\sf cKP}_{r,1}$ $\t$-function $\t^{(n)}(t)$ coincides with 
the equation for a (partial) two-dimensional Toda lattice $\t$-function.

For $r=1$ case (recall $t_1 \equiv x$) eqs.\rf{rec-rel-n},\rf{2d-toda} 
degenerate into one-dimensional Toda lattice structure (see 
eqs.\rf{pkplus},\rf{smallp},\rf{newtoda} below).

Returning to the general \cKPrm case,
we can represent the $k$-step DB transformation \rf{shabes-4}---\rf{sol-3-aa}
in terms of Wronskian determinants involving the coefficient functions of the 
``initial'' ${\sf cKP}_{r,m}$ Lax operator:
\be
L^{(0)} = D^r + \sum_{l=0}^{r-2} u_l^{(0)} D^l +
\sum_{i=1}^m \Phi_i^{(0)} D^{-1} \Psi_i^{(0)}    \lab{seq-b}
\ee
only.
Defining:
\be
\(L^{(0)}\)^k \P_1^{(0)} \equiv \chi^{(k)} \qquad \; k=1,2,\ldots
\lab{defchi}
\ee
we arrive at the following general result:
\begin{proposition}
The $k$-step DB-transformed eigenfunctions and the tau-function
\rf{shabes-4}---\rf{sol-3-aa} of the \cKPrm hierarchy for
arbitrary initial $L^{(0)}$ \rf{seq-b} are given by:
\br
\P_1^{(k)}&=& \frac{W_{k+1}\lb \P_1^{(0)},\chi^{(1)},\ldots,
\chi^{(k)}\rb}
{W_{k}\lb \P_1^{(0)}, \chi^{(1)},\ldots ,\chi^{(k-1)}\rb}  
\lab{pchi-a}  \\
\P_j^{(k)}&=&\frac{W_{k+1} \lb \P_1^{(0)},\chi^{(1)},\ldots ,\chi^{(k-1)},
\P_j^{(0)}\rb}{W_{k}\lb \P_1^{(0)}, \chi^{(1)}, \ldots, \chi^{(k-1)}\rb}
\qquad , \;\; j=2,\ldots ,m    
\lab{pchi-aa}  \\
\tau^{(k)} &=& W_{k} \lb  \P_1^{(0)}, \chi^{(1)},
\ldots,  \chi^{(k-1)}\rb \tau^{(0)}      
\lab{tauok}
\er
where $\tau^{(0)}, \tau^{(k)}$ are the $\tau$-functions of $L^{(0)},
L^{(k)}$, respectively, and $\chi^{(i)}$ is given by \rf{defchi}.
\label{proposition:kstep}
\end{proposition}
The proof of \rf{pchi-a}--\rf{tauok} is accomplished via induction w.r.t.
$k$ by using the important composition formula for Wronskians
\rf{iw}--\rf{W-def} from Appendix \ref{section:appa}.

Now, comparing expressions \rf{spectr-1} and \rf{pchi-a} together with
\rf{tauok}, we get the following recurrence relation for the $\t$-functions
of the DB-orbit on \cKPrm :
\br
\t^{(k+1)}(t) &=& \int d\l \, \l^{k(1+r)} \p_1^{(0)}(\l )\,
e^{\xi (t,\l )} \t^{(k)}\bigl( t - [\l^{-1}]\bigr) \nonu \\
&=& \int d\l \, \l^{k(1+r)} \p_1^{(0)}(\l )\, 
: e^{- {\hat \th} (\l )}: \t^{(k)} (t)
\lab{tau-recur}
\er
where in the second eq.\rf{tau-recur} we have employed the well-known vertex 
operator \ct{KP,ldickey,cortona,moerbeke} with:
\be
{\hat \th} (\l ) = - \sum_{l \geq 1} \l^l t_l +  
\sum_{l \geq 1} \frac{1}{l \l^l} \partder{}{t_l}
\lab{theta-field}
\ee
where $: \ldots :$ indicates standard normal ordering, meaning
$\partder{}{t_l}$ ``modes'' to the right of $t_l$ ``modes''.
Note, that eq.\rf{tau-recur} is noth\-ing but a spe\-cial case of 
Sato B\"{a}ck\-lund trans\-forma\-tion for the con\-strained 
\cKPrm hie\-rarchies.
Using the Wick-theorem identity:
\be
:e^{-{\hat \th} (\l_k )}: \ldots :e^{- {\hat \th} (\l_0 )}: =
:e^{-\sum_{j=0}^k {\hat \th} (\l_j )}: \,
\prod_{l=0}^k \(\l_l\)^{-l}\, \prod_{i>j} \(\l_i -\l_j\)
\lab{vertex-prod}
\ee
one can solve the recurrence relation \rf{tau-recur} expressing
$\t^{(k+1)} (t)$ directly in terms of the ``initial''
$\t$-function $\t^{(0)}$ of $L^{(0)}$ \rf{seq-b} :
\br
\t^{(k+1)} (t) = \int \prod_{j=0}^k d\l_j \,
W_{k+1} \llb f_0 (t,\l_0 ), \pa_r f_0 (t,\l_1 ), \ldots ,
\pa_r^k f_0 (t,\l_k ) \rrb 
\t^{(0)} \bigl( t - \sum_{j=0}^k [\l_j^{-1}]\bigr)
\lab{tau-recur-0} \\
{} = {1\over{(k+1)!}}
\int\prod_{j=0}^k d\l_j \, \Bigl(\prod_{i>j} \(\l_i -\l_j\)\Bigr)
\, \Bigl(\prod_{i>j} \(\l_i^r -\l_j^r\)\Bigr) \,\prod_{s=0}^k  f_0 (t,\l_s )\;
\t^{(0)} \bigl( t - \sum_{j=0}^k [\l_j^{-1}]\bigr)
\lab{tau-recur-r} \\
f_0 (t,\l ) \equiv \p_1^{(0)}(\l ) e^{\xi (t,\l )} \phantc
\lab{f-0}
\er
where we used the simple antisymmetrization identity:
\be
{\rm antisymm}_{\lcurl\m_1 ,\ldots ,\m_N\rcurl} 
\(\prod_{s=1}^N \(\m_s\)^{s-1}\) = {1\over {N!}} \prod_{i>j} \(\m_i - \m_j\)
\lab{anti-symm}
\ee
Now, comparing expressions \rf{tau-recur-0} and \rf{tauok} for the Wronskian
$\t$-functions we find that they agree provided the following identity
involving $\t^{(0)}$ is satisfied:
\br
&&\prod_{i>j} \(\l_i -\l_j\) 
\frac{\t^{(0)}\bigl( t -\sum_{j=0}^k [\l_j^{-1}] \bigr)}{\t^{(0)}(t)} =
\lab{Fay-gen}\\
&&\vareps_{j_0 ,j_1 ,\ldots ,j_k} 
\frac{\t^{(0)}\bigl( t -[\l_0^{-1}] \bigr)}{\t^{(0)}(t)}\, \(\pa +\l_{j_1}\)
\(\frac{\t^{(0)}\bigl( t-[\l_{j_1}^{-1}]\bigr)}{\t^{(0)}(t)}\) \cdots
\(\pa +\l_{j_k}\)^k
\(\frac{\t^{(0)}\bigl( t -[\l_{j_k}^{-1}] \bigr)}{\t^{(0)}(t)}\)
\nonu
\er
It is easy to see that eqs.\rf{Fay-gen} coincide with the well-known
differential Fay identity (for $k=1$) and its higher derivative corollaries
\ct{moerbeke}. 

One can generalize the successive DB transformations \rf{shabes-4}
on general \cKPrm Lax operators \rf{f-5} as follows.
Within each subset of $m$ successive steps we can perform the DB
transformations w.r.t. the $m$ different eigenfunctions of \rf{f-5}
unlike \rf{shabes-4} where all DB transformations are given by the
first $\Phi_1$ eigenfunction solely.
Repeated use of eqs.\rf{iw}--\rf{transf} from Appendix and employing 
short-hand notations:
\be
T^{(k)}_i \equiv \Phi^{(k)}_i D \(\Phi^{(k)}_i\)^{-1} \qquad ;\qquad
\chi^{(s)}_i \equiv \( L^{(0)}\)^s \Phi^{(0)}_i \quad ,\;\;
i=1,\ldots ,m
\lab{defchi-i}
\ee
yields the following generalization of \rf{pchi-a}--\rf{tauok} :
\begin{proposition}
The most general discrete Darboux-B\"{a}cklund orbit on the space 
of the \cKPrm Lax operators, starting from an arbitrary initial 
$L^{(0)}$ \rf{seq-b}, consists of the following elements:
\be
L^{(km+l)} = D^r + \sum_{j=0}^{r-2} u_j^{(km+l)} D^j +
\sum_{i=1}^m \Phi_i^{(km+l)} D^{-1} \Psi_i^{(km+l)}    
\lab{DB-orbit}
\ee
with $k$ arbitrary non-negative, $1 \leq l \leq m$, and where:
\br
\Phi_i^{(km+l)} = T^{(km+l-1)}_l \ldots T^{(km)}_1
T^{(km-1)}_m \ldots T^{((k-1)m)}_1 \ldots T^{(m-1)}_m \ldots
T^{(0)}_1 \chi^{(k_{\pm})}_i   \nonu  \\
= \frac{W\llb \Phi^{(0)}_1 ,\ldots ,\Phi^{(0)}_m ,
\chi^{(1)}_1,\ldots ,\chi^{(1)}_m ,\ldots ,\chi^{(k-1)}_1,\ldots
,\chi^{(k-1)}_m 
,\chi^{(k)}_1,\ldots ,\chi^{(k)}_l ,\chi^{(k_{\pm})}_i\rrb}{W\llb
\Phi^{(0)}_1 ,\ldots ,\Phi^{(0)}_m ,
\chi^{(1)}_1,\ldots ,\chi^{(1)}_m ,\ldots ,\chi^{(k-1)}_1,\ldots
,\chi^{(k-1)}_m ,\chi^{(k)}_1,\ldots ,\chi^{(k)}_l \rrb}
\lab{pchi-a-1}  \\
\chi^{(k_{+})}_i \equiv \chi^{(k+1)}_i \quad {\rm for} \;\; 1 \leq i
\leq l
\qquad ; \qquad
\chi^{(k_{-})}_i \equiv \chi^{(k)}_i \quad {\rm for} \;\; l+1 \leq i \leq m
\nonu
\er
The corresponding $\t$ functions read:
\br
\frac{\t^{(km+l)}}{\t^{(0)}} = \Phi^{(km+l-1)}_l \ldots
\Phi^{(km)}_1 \Phi^{(km-1)}_m \ldots \Phi^{((k-1)m}_1 \ldots
\Phi^{(m-1)}_m \ldots \Phi^{(0)}_1
\nonu  \\
= W\llb \Phi^{(0)}_1 ,\ldots ,\Phi^{(0)}_m ,
\chi^{(1)}_1,\ldots ,\chi^{(1)}_m ,\ldots ,\chi^{(k-1)}_1,\ldots ,
\chi^{(k-1)}_m ,\chi^{(k)}_1,\ldots ,\chi^{(k)}_l \rrb
\lab{tauok-1}
\er
\label{proposition:kml-step}
\end{proposition}
\subsection{Truncated KP Hierarchies}
An important particular class of \cKPrm DB orbits \rf{DB-orbit} are those 
with a purely differential initial Lax operator $L^{(0)}$ \rf{seq-b} :
\be
L^{(0)} = \( L^{(0)}\)_{+} = D^r + \sum_{j=0}^{r-2} u_j^{(0)} D^j
\quad , \quad i.e. \;\; \Psi_j^{(0)} = 0
\lab{seq-b-plus}
\ee
Such \cKPrm DB orbits are characterized by the fact that the $m$ 
adjoint eigenfunctions
$\Psi_i \equiv \Psi^{(k)}_i$ are not independent from the $m$ eigenfunctions
$\Phi_i \equiv \Phi^{(k)}_i$ since both are parametrized in terms of
$m$ independent initial eigenfunctions $\Phi^{(0)}_i$ only. 

Formulas \rf{pchi-a}--\rf{tauok}, \rf{pchi-a-1}--\rf{tauok-1}
simplify significantly in this case since now we have:
\be
\chi^{(s)}_i \equiv \( L^{(0)}\)^s \Phi^{(0)}_i = \pa_r^s \Phi^{(0)}_i
\quad ,\;\; i=1,\ldots ,m
\lab{defchi-i-0}
\ee
leading to ({\sl e.g.}, for \rf{pchi-a}--\rf{tauok}) :
\br
\P_1^{(k)}&=& \frac{W_{k+1}\lb \P_1^{(0)},\pa_r \P_1^{(0)},\ldots,
\pa_r^k \P_1^{(0)} \rb}{W_{k}\lb \P_1^{(0)}, \pa_r \P_1^{(0)},
\ldots ,\pa_r^{k-1}\P_1^{(0)}\rb}  
\lab{pchi-a-tr}  \\
\P_j^{(k)}&=& \frac{W_{k+1}\lb \P_1^{(0)},\pa_r \P_1^{(0)},\ldots,
\pa_r^{k-1}\P_1^{(0)},\P_j^{(0)}\rb}{W_{k}\lb \P_1^{(0)}, \pa_r
\P_1^{(0)},\ldots ,\pa_r^{k-1}\P_1^{(0)}\rb}
\qquad , \;\; j=2,\ldots ,m    
\lab{pchi-aa-tr}  \\
\frac{\t^{(k)}}{\t^{(0)}} &=& W_{k}\lb \P_1^{(0)}, \pa_r \P_1^{(0)},
\ldots, \pa_r^{k-1}\P_1^{(0)} \rb       
\lab{tauok-tr}
\er

A special feature of the subclass of \cKPrm DB orbits with a ``free''
initial $L^{(0)}= D^r$ is that their dressing operators
are truncated (having only finite number of terms in the pseudo-differential
expansion, cf. third ref. in \ct{addsym-models}) :
\br
W^{(km+l)} &=&  T^{(km+l-1)}_l \cdots T^{(km)}_1 T^{(km-1)}_m
\cdots T^{((k-1)m)}_1 \cdots T^{(m-1)}_m \cdots T^{(0)}_1 D^{-km-l} 
\nonu\\
&=& \sum_{j=0}^{km+l} w_j^{(km+l)} D^{-j}
\lab{W-trunc}
\er
where notations \rf{defchi-i} were used. Recalling \rf{W-main},
eq.\rf{W-trunc} implies:
\be
w_i^{(km+l)} = \frac{p_i (-[\pa])\t^{(km+l)} (t)}{\t^{(km+l)} (t)}  
\qquad ;\qquad 
p_j (-[\pa])\t^{(km+l)} (t) = 0  \quad ,\;\; j \geq km+l+1
\lab{tau-trunc}
\ee

On the other hand, one can consider {\em a priori} truncated $W$ dressing
operators
outside the context of \DB transformations.
\begin{definition}
KP hierarchies defined through Lax operators built out of $m$-{\em truncated}
dressing operators:
\br
W \equiv W^{(m)} =  \sum_{i=0}^{m} w_i (t) D^{-i}
\qquad ; \qquad L \equiv L^{(m)} = W^{(m)} D \( W^{(m)}\)^{-1}
\lab{dress-tr} \\
\partder{}{t_n} W^{(m)} = -\( W^{(m)} D\bigl( W^{(m)}\bigr)^{-1}\)_{-} W^{(m)}
\lab{sato-b}
\er
are called $m$-{\em truncated} KP hierarchies.
\label{definition:KP-tr}
\end{definition}
As in \rf{tau-trunc} above, the $\t$-function of the $m$-truncated KP 
hierarchy obeys the constraints:
\be
p_j (-[\pa])\t^{(m)} (t) = 0  \quad ,\;\; j \geq m+1
\lab{tau-trunc-m}
\ee

Let us set $m=1$ and recall the Hopf-Cole transformation, which describes
the dressing and Lax operators by a function
$\p$ such that $w_1 = - \pa_x \ln \p$ (see eq.\rf{hopf-cole} below).
Hence $\p$ is a solution of $ W^{(1)} \pa \p = 0$.
Similarly, for arbitrary $m$ we generalize the Hopf-Cole transformation,
and consider a set of $m$ functions $\p_k$ being solutions 
to the system of differential equations: 
\be
W^{(m)} \pa^m \p_k = 0 \qquad k =1, \ldots, m
\lab{hopf-co}
\ee
It is well-known, that while solutions of the general KP hierarchy
form the universal Grassmann manifold UGM, solutions of \rf{hopf-co}
defining the $m$-truncated KP hierarchy form the Grassmann manifold
$ GM (m, \infty ) = {\rm Mat } (\infty \times m ) / GL (m; \IC)$
where ${\rm Mat } (\infty \times m ) $ denotes $\infty \times m$
matrices of rank $m$ \ct{harada,ohta}.

In terms of the solutions of the $m$-th order differential equation
\rf{hopf-co} the Wilson-Sato equations \rf{sato-b} take the simple form:
\be
\pa_n \p_i = \pa_x^n \p_i \qquad i=1, \ldots, m
\lab{sato-lin}
\ee
Note that $\p_i$ satisfying \rf{sato-lin} can be regarded as a set of $m$
independent eigenfunctions of the ``free'' Lax operator $L^{(0)} = D$, and
their explicit form reads:
\be
\p_i (t) = \int d\l\, \p_i^{(0)}(\l )\exp\Bigl\{\sum_{l\geq 1}\l^l t_l\Bigr\}
\lab{p-i}
\ee
with arbitrary ``densities'' $\p_i^{(0)}(\l )$. Moreover, it is assumed that
in gen\-eral the set $\lcurl \p_1 ,\ldots ,\p_m\rcurl$ is 
{\em nondegenerate} in
a sense that no one $\p_i$ is a derivative of another $\p_j$.

In different words we have the following {\sl Lemma}:
\begin{lemma}
Eqs.\rf{sato-lin} are equivalent to the Wilson-Sato equations \rf{sato-b}
for the $m$-truncated dres\-sing operator \rf{dress-tr}.
\label{lemma:hopf-sato}
\end{lemma}
\begin{proof}
Define for convenience $\cF = W D^m= D^m + \ldots$, which is a
purely differential operator of order $m$. The Lax operator can then
be rewritten as $L = \cF D \cF^{-1}$ (here we will be suppressing for brevity
the superscripts ${}^{(m)}$ on $W$ and $L$). Assume that \rf{sato-lin} holds.
It follows:
\be
0= \pa_n \( \cF \p_k\) = (\pa_n  \cF) \p_k +   \cF \( \pa_x^n\p_k\) 
= \( \pa_n  \cF + L_{-}^n \cF \) \p_k 
\lab{cfpk}
\ee
for some arbitrary integer $n$.
In \rf{cfpk} use was made of the obvious identity $ L^n = \cF D^n
\cF^{-1}$ or
$ \cF D^n = L^n_{+} \cF + L^n_{-} \cF$ and the fact that
$L^n_{+} \cF \p_k =0$.
Expression $\( \pa_n  \cF + L_{-}^n \cF \)$ on the right hand side
of \rf{cfpk} is a purely differential operator of order smaller than
$m$. Since by assumption all $m$ functions $\p_k$ are independent we
conclude, therefore, that $\pa_n  \cF = -  L_{-}^n \cF $, which is
equivalent to the Wilson-Sato equation \rf{sato-a} for the dressing
operator $W$.

Let us now prove the inverse statement.  From assumption
$\pa_n \cF = L^n_{+} \cF -\cF D^n$ and \rf{hopf-co} we find
that $ \cF\( \pa_n \p_k  -\pa_x^n \p_k  \)= 0$. Since $\p_k$ span
the $m$-dimensional space of solutions of $\cF$ operator,
we must have $\pa_n \p_k  -\pa_x^n \p_k = \sum_j c^{(n)}_{kj} \p_j$,
with $c^{(n)}_{kj}\, j,k =1, \ldots,m$ being $x$-independent
constants. Hence, $ \nabla_n \p_k = \pa_x^n \p_k$ for the
``covariant'' derivative
$ \(\nabla_n\)_{jk} \equiv \pa_n \d_{jk} - c^{(n)}_{kj}$.
Compatibility gives then $\sbr{\nabla_n}{\nabla_l} \p_k =0$ leading to 
$ c^{(n)}_{ij} = c^{-1}_{ik} \pa_n c_{kj}$ for some $x$-independent matrix
$c_{ij}$. Define now ${\bar \p}_k = \p_l \(\exp c\)_{lk}$.
It follows that $\pa_n {\bar \p_k} = \pa_x^n {\bar \p_k}$.
\end{proof}

Let us return to \rf{hopf-co}
which factorizes as follows:
\be
\cF \p_k = \( D^m + w_1 D^{m-1} + \cdots + w_m \) \p_k = 
\( D + v_m \) \( D + v_{m-1} \) \cdots \( D + v_1 \) \p_k = 0
\lab{hopf-coa}
\ee
There is a relation between coefficients $v_i$ of \rf{hopf-coa}
and solutions $\p_k$ of the same form as \rf{wil} with $\p_k$
replacing $\psi_k$ :
\be
v_i = \pa \( \ln { W_{i-1} \lb \p_1, \ldots, \p_{i-1} \rb \over 
W_{i}\lb \p_1, \ldots, \p_{i} \rb } \) \qquad ; \qquad W_0 =1 
\lab{wila}
\ee
and, therefore:
\br
\cF = T_m \cdots T_1 \qquad ,\qquad T_i = 
\frac{W_{i}}{W_{i-1}} D \frac{W_{i-1}}{W_{i}} = D + v_i   \nonu  \\
L^{(m)} = W^{(m)} D \bigl( W^{(m)}\bigr)^{-1} = \cF D \cF^{-1} =
T_m \cdots T_1 D \, T_1^{-1} \cdots T_m^{-1}
\lab{lax-m}
\er
Comparing \rf{lax-m} with \rf{shabes-3} we identify the $m$-truncated KP
hierarchy \rf{dress-tr}--\rf{sato-b} as the $m$-th member of the DB orbit 
of ${\sf cKP}_{1,m}$ with initial $L^{(0)}=D$. Accordingly, we have the
following:
\begin{proposition}
The $\t$-function and the BA function of the $m$-truncated KP hierarchy 
take the following explicit form:
\br
\t^{(m)}&= &W_m \lb \p_1, \ldots ,\p_m \rb =
\det {\Bigl\Vert} \pa^{i-1} \p_j {\Bigr\Vert}_{i,j=1,\ldots ,m}
\lab{m-tau}  \\
p_s ( -[\pa]) W_m \lb \p_1, \ldots ,\p_m \rb \eq 
\left\{ \begin{array}{ll}
(-1)^s \sum_{i_1 < \ldots < i_s} W_m^{(i_1 \ldots i_s)} \; , &
\quad {\rm for} \quad 1 \leq s \leq m   \\
0 \; , &\quad {\rm for} \quad s \geq m +1
\end{array}   \right.
\lab{pmk} \\
\psi^{(m)}(t,\l) &=& e^{\sum_{l\geq 1}\l^l t_l}
\Bigl\{ 1 + \sum_{s=1}^m (-\l )^{-s} \sum_{i_1 < \ldots < i_s}
\frac{W_m^{(i_1 \ldots i_s)}}{W_m} \Bigr\}
\lab{bamtr} \\
W_m^{(i_1 \ldots i_s)} &\equiv&
W_m \llb \p_1 ,\ldots ,\pa \p_{i_1}, \ldots ,\pa \p_{i_s},\ldots , \p_m\rrb
\lab{Wronski-i-s}
\er
where the Wronskian \rf{Wronski-i-s} is obtained from
$W_m \equiv W_m \lb \p_1, \ldots ,\p_m \rb$ by putting an additional
derivative $\pa$ on rows with numbers $i_1,\ldots ,i_s$, respectively.
In particular, we have:
\be 
p_{k} ( -[\pa])  \p_i =0 \quad {\rm for} \;\; k >1 \; , \; i=1,\ldots, m 
\lab{pmk-1}
\ee
which is identical with \rf{sato-lin}.
\label{proposition:tau-BA-m}
\end{proposition}
Accordingly, eq.\rf{dress-tr} now reads:
\be
W^{(m)} = \sum_{j=0}^m 
\frac{p_j (-[\pa ]) W_m\lb\p_1,\ldots ,\p_m\rb}{W_m\lb\p_1,\ldots ,\p_m\rb}
D^{-j}
\lab{W-m-tr}
\ee
\begin{proof}
Due to the above identification of $m$-truncated KP hierarchies, eq.\rf{m-tau}
follows straightforwardly from \rf{tauok-tr} with $r=1$. Furthermore, we have:
\br
W_m \lb \p_1 (t- \lb \l^{-1} \rb), \ldots ,\p_m (t- \lb \l^{-1} \rb) \rb &=&
\sum_{s=0}^{\infty} \l^{-s} p_s (-[\pa])
W_m \lb \p_1 (t) ,\ldots ,\p_m (t) \rb \lab{Wronski-shif} \\
 &=& \sum_{s=0}^m (-1)^s \l^{-s} 
\sum_{i_1 < \ldots < i_s} W_m^{(i_1 \ldots i_s)}
\lab{Wronski-shift}
\er
where use is made, respectively, of the definition of Schur polynomials 
\rf{Schur} and of the simple identity (cf. \rf{p-i}) :
\be
\p_i \bigl( t-[\l^{-1}] \bigr) =  \int d\m\, \p^{(0)}_i (\m ) 
\exp \lcurl \sum_{l\geq 1}\m^l \Bigl( t_l - {1\over {l \l^l}}\Bigr) \rcurl
= \p_i (t) - \l^{-1} \pa \p_i (t)
\lab{phi-l-id}
\ee
which follows from $\, \exp \( - \sum_1^\infty \m^l /l \l^l\) = \( 1 -\m/\l \)$
.
Then, eq.\rf{pmk} directly follows upon comparing of \rf{Wronski-shif} with 
\rf{Wronski-shift}. Accordingly,
eq.\rf{bamtr} is easily verified upon substituting \rf{m-tau} in the 
expression for the BA function:
\be
\psi^{(m)}(t,\l) = \frac{\t^{(m)} \bigl( t-[\l^{-1}] \bigr)}{\t^{(m)}(t)}
e^{\xi (t,\l )}   \nonu
\ee
and using again \rf{Wronski-shift}.
\end{proof}
\subsection{Generalized \BH Hierarchies}
For the simplest case $m=1$ relation \rf{hopf-coa} takes the form of a 
classical Hopf-Cole transformation:
\be
\( 1 + w_1 D^{-1} \)D (\p) = 0 \quad \to \quad
\( \pa + w_1 \) \p = 0 \quad \to \quad w_1 = - \pa_x \ln \p
\lab{hopf-cole}
\ee
Correspondingly, $\cF = W^{(1)} D = D -\pa_x \(\ln \p \) = \p D \p^{-1}$ and,
thus, we are lead to the Lax operator:
\be
L^{(1)} = \(\p D \p^{-1}\)  \; D \; \(\p D^{-1} {\p}^{-1}\)
\, =\, D+ \llb \p \( \ln \p\)^{\pr \pr} \rrb D^{-1} \p^{-1}    
\lab{lone}
\ee
One finds from Lemma \ref{lemma:hopf-sato} that the Lax equation
\rf{lax-eq} for $L^{(1)}$ is equivalent to $\pa_n \p = \pa_x^n \p$,
which upon comparison of \rf{lone} with \rf{shabes-3} identifies \rf{lone} as
$k=1$ member of ${\sf cKP}_{1,1}$ DB orbit.
 
Equations $\pa_n \p = \pa_x^n \p$ can be rewritten in terms of the coefficient
$w \equiv -w_1$ of the dressing operator $W^{(1)}=1 - w D^{-1}$ as:
\br
\pa_n w \eq  \pa_x P_{n} ( w) \lab{wfaa} \\
P_{n+1} ( w) \eq (\pa + w) P_n ( w) \;\;\; n = 0,1,2 ,\ldots \qquad
P_0 (w) = 1 
\lab{faa}
\er
where $P_n (w)$ are Fa\'a di Bruno polynomials fully determined by
the recurrence relation in \rf{faa}.
The system of nonlinear differential equations \rf{wfaa}, which is an
alternative form of the equations of the $k=1$ member of ${\sf cKP}_{1,1}$
DB orbit, is called a Burgers-Hopf hierarchy.

Let us now consider the fully degenerated $m$-truncated KP hierarchy,
{\sl i.e.}, for which:
\be
\p_j = \pa^{j-1} \p \;\; ,\; j=1,\ldots ,m
\lab{m-tr-degen}
\ee
Substituting \rf{m-tr-degen} in \rf{m-tau} and comparing with \rf{tauok-tr}
allows us to
identify it with the $m$-th member ($m\geq 1$) of the ${\sf cKP}_{1,1}$ DB
orbit starting from $L^{(0)}=D$. In analogy with the above identification for
the (ordinary) \BH hierarchy, we shall introduce the following:
\begin{definition}
We call the fully degenerated $m$-truncated KP hierarchies {\em generalized}
Burgers-Hopf hierarchies.
\label{definition:gen-BH}
\end{definition}
Alternatively, the  generalized \BH hierarchy is described as a \DB orbit
of the \BH hierarchy defined in \rf{lone}.
Below we demonstrate that the set of generalized \BH hierarchies
is also equivalent to the semi-infinite one-dimensional Toda chain system.

The Lax structure obtained in the process of successive DB transformations
applied to \BH hierarchy \rf{lone} takes the following form of recursive
relations:
\br
L^{(k+1)} \eq \(\Phi^{(k)}  D {\Phi^{(k)} }^{-1}\)  \; L^{(k)}
 \; \(\Phi^{(k)}  D^{-1} {\Phi^{(k)} }^{-1}\)
= D + \Phi^{(k+1)} D^{-1} \Psi^{(k+1)}
\lab{lkplus} \\
\Phi^{(k+1)} \eq \Phi^{(k)} \( \ln \Phi^{(k)}\)^{\pr \pr} +
\(\Phi^{(k)}\)^2
\Psi^{(k)} \quad ,\quad \Psi^{(k+1)} = \(\Phi^{(k)}\)^{-1}
\lab{pkplus}
\er
with $\Phi^{(0)} = \p$.

Proposition \ref{proposition:tau-BA-m} specializes in the case of generalized
\BH hierarchies \rf{lkplus} to the following:
\begin{proposition}
The $\t$-function and the BA function of generalized \BH hierarchies
\rf{lkplus}  read: 
\br
\t^{(k)} \eq W_k \lb \p ,\pa \p ,\ldots ,\pa^{k-1}\p \rb
= {1\over{k!}}
\int\prod_{j=1}^k d\l_j \, \Bigl(\prod_{i>j} \(\l_i -\l_j\)\Bigr)^2
\prod_{s=1}^k \( \p^{(0)}(\l_s ) e^{\xi (t,\l_s )}\)
\lab{BH-tau}  \\
&& p_s ( -[\pa]) W_k \lb \p ,\pa \p ,\ldots ,\pa^{k-1}\p \rb=  \phantc \nonu \\
\eq  \left\{ \begin{array}{ll}
(-1)^s W_k \lb \p ,\pa \p ,\ldots ,\pa^{k-s-1} \p ,\pa^{k-s+1}\p ,\ldots ,
\pa^{k}\p \rb \; , & \quad {\rm for} \quad 1 \leq s \leq k  \\
0 \; , & \quad {\rm for} \quad s \geq k+1
\end{array}   \right.
\lab{BH-pmk} \\
\psi^{(k)}(t,\l) \eq  e^{\xi (t,\l )} \Bigl\{ 1 + \sum_{s=1}^k (-\l )^{-s} 
\frac{W_k \lb \p ,\pa \p ,\ldots ,\pa^{k-s-1} \p ,\pa^{k-s+1}\p ,\ldots ,
\pa^{k}\p \rb}{W_k \lb \p ,\pa \p ,\ldots ,\pa^{k-s-1} \p ,\pa^{k-s}\p ,
\ldots ,\pa^{k-1}\p \rb} \Bigr\}
\lab{BH-bamtr}  \\
\P^{(k)}(t) & =& \frac{W_{k+1} \lb \p ,\pa \p ,\ldots ,\pa^k \p \rb}{W_k \lb \p
,\pa \p ,\ldots ,\pa^{k-1}\p \rb}
\lab{BH-eigenf}
\er
\label{proposition:tau-BA-BH}
\end{proposition}
The second equality in \rf{BH-tau} is a special case of eq.\rf{tau-recur-r}
with $r=1$, $\t^{(0)}=const$ .

Using the second equality in \rf{Wronski-shift} in the fully degenerated case
of
\rf{m-tr-degen} 
we also find further useful identities involving Wronskians of the type
appearing in \rf{BH-bamtr} :
\br
&&\int d\l \, \l^{2k-j} \p^{(0)}(\l ) e^{\xi (t,\l )} 
W_k \lb \p (t- \lb \l^{-1} \rb) , \pa \p(t- \lb \l^{-1} \rb) , 
\ldots , \pa^{k-1} \p (t- \lb \l^{-1} \rb)   \rb \nonu \\
\eq \sum_{s=0}^k  (-1)^s \pa^{2k-s-j} \p \;
W_k\lb\p ,\pa\p ,\ldots ,\pa^{k-s-1}\p ,\pa^{k-s+1}\p ,\ldots ,\pa^{k-1}\p\rb
\nonu  \\
\eq  \left\{ \begin{array}{ll}
W_{k+1} \lb \p , \pa \p, \ldots , \pa^{k} \p \rb\; , 
& \quad {\rm for} \quad  j=0 \\
0 \; , & \quad {\rm for} \quad  1\leq j \leq k 
\end{array}   \right.
\lab{wronb}
\er
In particular, identity \rf{wronb} with $j=0$ provides explicit solution for
the $\t$-function recurrence relation \rf{tau-recur} (with $r=1$).

In order to exhibit the semi-infinite Toda chain structure in 
\rf{lkplus}--\rf{pkplus}, let us introduce the following change of variables: 
\be
\Phi^{(k)} \equiv e^{\psi_{k+1} - \psi_{k}}  \qquad k=0,1,\ldots
\lab{smallp}
\ee
Substituting \rf{smallp} in \rf{pkplus}, we find that
$\psi_n$ satisfy the standard form of the one-dimensional Toda lattice
equations:
\be
\pa^2 \psi_{n}\, =\, e^{\psi_{n+1}+\psi_{n-1}-2\psi_{n}  }
\lab{newtoda}
\ee
with $\psi_n=0$ for $n \leq 0$. Furthermore, comparing \rf{smallp} with
\rf{BH-eigenf} implies that eq.\rf{newtoda} has a solution in the form 
of a Wronskian:
\be
\psi_{n} = \ln  W_n \lb {\ti\p} , \pa {\ti\p}, \ldots , \pa^{n-1} {\ti\p} \rb
\quad{\rm with}\quad {\ti\p} \equiv e^{\psi_1}
\lab{newtwron}
\ee

Let us return to the constraint equation \rf{pmk-1} in the context of
generalized \BH hierarchies (fully degenerate $m$-truncated KP hierarchies) :
\be
p_{k} ( -[\pa])  \p =0 \quad {\rm for} \;\; k >1 
\lab{pmk-0}
\ee
which is identical with \rf{sato-lin}.
There exists a solution of this condition which can be written as:
\be
\p =  \sum_{r=1}^K c_r \zeta_r (t) \qquad ; \qquad 
\zeta_r (t) \equiv e^{\sumi{l=1} \l_r^l t_l}
\lab{eigenexpand}
\ee
with arbitrary constants $c_r,\, \l_r, \, r=1, \ldots, K$. Alternatively,
the solution \rf{eigenexpand} can be directly obtained from the ``spectral''
representation \rf{p-i} by making the particular choice \ct{CZ94} : 
$\p^{(0)}(\l ) = \sum_{r=1}^K c_r \d (\l - \l_r )$.
One can easily prove that:
\begin{lemma}
\be
W_K \lb \p , \pa \p, \ldots , \pa^{K-1} \p \rb = \prod_{r=1}^K c_r 
\prod_{i>j} (\l_i - \l_j ) W_K \lb \zeta_1 , \zeta_2, \ldots ,
\zeta_{K} \rb =  \Bigl(\prod_{i>j} \(\l_i -\l_j\)\Bigr)^2 \,
\prod_{r=1}^K c_r \zeta_r (t)
\lab{wron-soleq}
\ee
\label{lemma:wron-sol}
\end{lemma}
Eq.\rf{wron-soleq} establishes relation between, on one hand, the DB 
construction of the $\t$-function and, on the other hand,
the Wronskian type multi-soliton solutions of the nonlinear evolution
equations of the (constrained) KP hierarchy common in the literature.
Hence, the solution space for the semi-infinite one-dimensional
Toda chain formulated as a generalized \BH system 
contains the well-known multi-soliton solutions of the constrained KP systems.

\sect{Additional Symmetries of Constrained KP Hierar\-chies}
\label{section:addsym}
\subsection{Background on Additional Sym\-metries}
Let $L$ be again as in eq.\rf{lax-op} a pseudo-differential Lax operator of 
the full generalized KP hierarchy:
\be
L = D^r + \sum_{j=0}^{r-2} v_j D^j + \sum_{i \geq 1} u_i D^{-i}  \qquad ,\quad
\partder{}{t_l} L = \Sbr{L^{l\over r}_{+}}{L} 
\lab{gen-KP}
\ee
(recall $x \equiv t_1$) and let $M$ be a pseudo-differential operator 
``canonically conjugated'' to $L$ such that:
\be
\Sbr{L}{M} = \one \quad , \quad
\partder{}{t_l} M = \Sbr{L^{l\over r}_{+}}{M} 
\lab{L-M}
\ee

Within the Sato-Wilson dressing operator
formalism, the $M$-operator can be expressed in terms of dressing of the
``bare'' $M^{(0)}$ operator:
\be
M^{(0)} = \sum_{l \geq 1} \frac{l}{r} t_l D^{l-r} =
X_{(r)} + \sum_{l \geq 1} \frac{l+r}{r} t_{r+l} D^l \quad ; \quad
X_{(r)} \equiv \sum_{l=1}^{r} \frac{l}{r} t_l D^{l-r}
\lab{M-0}
\ee
conjugated to the ``bare'' Lax operator $L^{(0)} = D^r$. The dressing gives
\br
M \eq  W M^{(0)} W^{-1} =
W X_{(r)} W^{-1} + \sum_{l \geq 1} \frac{l+r}{r} t_{r+l} L^{l\over r} =
\sum_{l \geq 0} \frac{l+r}{r} t_{r+l} L^{l\over r}_{+} + M_{-}
\lab{M-dress}  \\
M_{-} \eq W X_{(r)} W^{-1} - t_r -
\sum_{l \geq 1} \frac{l+r}{r} t_{r+l} \partder{W}{t_l}  \, .\, W^{-1}
\lab{M--}
\er
where in \rf{M--} we used eqs.\rf{sato-a}.
Note that $X_{(r)}$ is a pseudo-\-differen\-tial oper\-ator sat\-isfying
$\Sbr{D^r}{X_{(r)}} = \one$ .

On BA functions \rf{linsys} and eigenfunctions \rf{eigenlax},\rf{Phi-psi}
of $L$ the action of $M$ is as follows:
\br
M \psi (t,\l ) &=& \( \frac{\l^{1-r}}{r}\partder{}{\l}+\a_r (\l )\)\psi (t,\l )
\lab{M-psi}  \\
M \Phi (t) &=& 
\int d\l \,\lcurl \( -\partder{}{\l} + r\l^{r-1}\a_r (\l )\) 
\frac{\l^{1-r}}{r} \p (\l ) \rcurl \psi (t,\l )
\lab{M-Phi}
\er
where $\a_r (\l )$ is a function of $\l$ only\foot{The appearance of
$\a_r (\l )$ can be traced back to the ambiguity in the definition of the  
dressing operator \rf{dress-1}: $ W \longrightarrow W W_0\,$ where
$W_0 = 1 + \sum_{i \geq 1} c_i D^{-i}\,$ with {\em constant} coefficients
$c_i$.}.

The so called {\em additional (non-isospectral) symmetries}
\ct{Orlovetal,cortona} are defined as vector fields on the space of
KP Lax operators \rf{gen-KP} or, alternatively, on the dressing
operator through their flows as follows:
\be
{\bar \pa}_{k,n} L = - \Sbr{\( M^n L^k\)_{-}}{L} =
\Sbr{\( M^n L^k\)_{+}}{L} + n M^{n-1} L^k  \;\,;  \;\;\;\,
{\bar \pa}_{k,n} W = - \( M^n L^k\)_{-} W
\lab{add-symm-L}
\ee
The additional symmetry flows commute with the usual KP hierarchy 
isospectral flows given in \rf{lax-eq}. 
But they do not commute among themselves,
instead they form a $\Win1$ algebra (see e.g. \ct{cortona}).
One finds that the Lie algebra of operators ${\bar \pa}_{k,n}$
is isomorphic to the Lie algebra generated by $- z^n (\pa/\pa z)^k$.
Especially for $n=1$ this becomes an isomorphism to the Virasoro
algebra ${\bar \pa}_{k,1} \sim - \cL_{k-1}$, with $\sbr{\cL_n}{\cL_k} = (n-k)
\cL_k$.

\subsection{Modification of Additional-Symmetry Flows for \cKPrm
Hie\-rar\-chies}
The main task of this section is the formulation of additional
non-isospectral Virasoro symmetry structure for the \cKPrm hierarchies, or in
different words, we will redefine the standard additional symmetry flows
\rf{add-symm-L} for the full KP hierarchy in such a way that they will 
preserve the subspace of \cKPrm hierarchies \rf{f-5}. 
To be more precise, we first show that only the
$sl(2)$ subalgebra (containing Galilean and scaling symmetries) of the
standard Virasoro additional symmetry algebra, with flows given by
${\bar \pa}_{k,1}$ for $k=0,1,2$, 
preserves the constrained form of \cKPrm Lax operators. Next, we show
how the rest of Virasoro additional symmetry (for the Virasoro generators
$\cL_k\, , \, k\geq 2$) acting on \cKPrm is recovered through an appropriate
modification of its generators. This is accomplished by adding a 
new structure consisting of ``ghost'' flows \rf{ghostflo} related to the 
plethora of (adjoint) eigenfunctions characteristic for the 
\cKPrm hierarchies.
This is natural since the ``ghost'' symmetry flows commute with the ordinary
isospectral flows (prop. \ref{proposition:commflo}).

Let us begin by applying the additional-symmetry flows \rf{add-symm-L} on 
$L$ \rf{f-5}. For $n=1$ we get:
\be
\( {\bar \pa}_{k,1} L\)_{-} =
{\sbr{\( M L^k \)_{+}}{L}}_{-} + \( L^k  \)_{-}
\lab{symm1}
\ee
Using the simple identities \rf{tkppsi} and \rf{lkminus} from Appendix
\ref{section:appa}
for the Lax operator \rf{f-5}, we are able to rewrite \rf{symm1} as:
\br
\( {\bar \pa}_{k,1} L\)_{-} &=& \sum_{i=1}^m \( M L^k \)_{+} (\Phi_i) D^{-1}
\Psi_i - \sum_{i=1}^m \Phi_i D^{-1} \( M L^k \)^{\ast}_{+} ( \Psi_i )  \nonu\\
&+& \sum_{i=1}^m \sum_{j=0}^{k-1} L^{k-j-1} (\Phi_i) D^{-1}
\( L^{\ast}\)^{j} ( \Psi_i )
\lab{pak1}
\er
Here
\be
L (\Phi_i ) \equiv
L_{+} (\Phi_i ) + \sum_{j=1}^m \Phi_j \pa_x^{-1} \( \Psi_j \Phi_i\)
\lab{applic}
\ee
(and similarly for the adjoint counterpart) denotes action of $L$ on
$\Phi_i$.  Notice that  
$L^{k-j-1} (\Phi_i) $,  $\( L^{\ast}\)^{j} ( \Psi_i) $ are (adjoint)
eigenfunctions of $L$ \rf{f-5}. Hence,  
whereas the original $L$ \rf{f-5} belongs to the class of ${\sf cKP}_{r,m}$ 
hierarchies, the transformed Lax operator given by
${\bar \pa}_{k,1} L$ (cf. eq.\rf{pak1}) belongs to a {\em different} class --
${\sf cKP}_{r,m(k-1)}$ (for $k \geq 3$), since the number of eigenfunctions 
has increased.

For $k=0,1,2$ the flow equations \rf{pak1} can still
be rewritten in the desired original ${\sf cKP}_{r,m}$ form:  
\be
\(\pa_{\t} L\)_{-} = \sum_{i=1}^m \(\pa_{\t} \Phi_i\) D^{-1} \Psi_i + \Phi_i
D^{-1} \( \pa_{\t} \Psi_i\)
\lab{addflo}
\ee
with
$\pa_\t \equiv {\bar \pa}_{k,1}\; (k=0,1,2)$, where:
\br
{\bar \pa}_{0,1} \Phi_i \eq \( M\)_{+}  (\Phi_i) \qquad;\qquad
{\bar \pa}_{0,1} \Psi_i = - \( M\)^{\ast}_{+} ( \Psi_i  ) \lab{papsiz} \\
{\bar \pa}_{1,1} \Phi_i \eq \( ML\)_{+} (\Phi_i) + \a \Phi_i
\quad;\quad
{\bar \pa}_{1,1} \Psi_i = - \( ML\)^{\ast}_{+} (\Psi_i) + \b \Psi_i
\quad \; \a+\b =1\lab{papsi1} \\
{\bar \pa}_{2,1} \Phi_i \eq \( M L^2 \)_{+} (\Phi_i) + L (\Phi_i)
\quad;\quad
{\bar \pa}_{2,1} \Psi_i = - \( M L^2 \)^{\ast}_{+} (\Psi_i) + L^{\ast}
(\Psi_i)
\lab{papsi2}
\er
The coefficients $\a,\b$ on the right hand sides of
\rf{papsi1} are fixed to $\a=\b=\h$ from the requirement of closure of the
algebra of the flows \rf{papsiz}--\rf{papsi2}.

Since the additional flows satisfy the algebra
$\sbr{{\bar \pa}_{l,1}}{{\bar \pa}_{k,1}} = -\( l-k \) {\bar \pa}_{l+k-1,1}$ ,
we have an isomorphism ${\bar \pa}_{k,1} \sim - \cL_{k-1}$ with the Virasoro
operators and equations \rf{papsiz}--\rf{papsi2} contain the
$sl(2) $ subalgebra generators $\cL_{-1}, \cL_0 , \cL_1$.

However, for $\pa_\t \equiv {\bar \pa}_{k,1}\; k \geq 3$, eq.\rf{addflo} does
not hold anymore due to absence of consistent definitions for
${\bar \pa}_{k,1} \Phi_i ,\, {\bar \pa}_{k,1} \Psi_i$ generalizing
\rf{papsiz}--\rf{papsi2} for higher $k$. Thus, it appears
that the symmetry constrains behind the \cKPrm hierarchies have broken the 
standard Orlov-Schulman  
additional Virasoro symmetry down to the $sl(2)$ subalgebra.

To recover the complete Virasoro symmetry, our strategy will be to redefine 
the additional-\-symmetry generators. We first describe our technique for $k=3$
in which case eq.\rf{symm1} contains a term:
\be
\( L^3 \)_{-} = \sum_{i=1}^m \Phi_i D^{-1} \(L^{\ast}\)^2 \( \Psi_i\)  +
\sum_{i=1}^m  L\(\Phi_i\) D^{-1} L^{\ast} \( \Psi_i\)  +
\sum_{i=1}^m  L^2\(\Phi_i\) D^{-1} \Psi_i
\lab{l3m}
\ee
Note that the middle term in \rf{l3m} is not of the form of required
by equation \rf{addflo}.
At this point we recall that for
\be
X \equiv \sum_{k=1}^I M_k D^{-1} N_k
\lab{xdef}
\ee
with definitions \rf{f-5} and \rf{xdef} we find using identity \rf{x1x2} from
Appendix \ref{section:appa}:
\be
{\sbr{X}{L}}_{-} = \sum_{k=1}^I \( - L (M_k) D^{-1} N_k + M_k D^{-1} L^{\ast}
(N_k)\) + \sum_{i=1}^m \(  X ( \Phi_i ) D^{-1} \Psi_i
- \Phi_i D^{-1} X^{\ast} (\Psi_i )  \)
\lab{xlbra}
\ee
According to Proposition \ref{proposition:commflo}
the flows generated by \rf{xlbra} will commute with the isospectral flows
\rf{lax-eq} provided $M_i,N_i$ are (adjoint) eigenfunctions,
which will be the case in what follows. Consider now as an example:
\be
Y_3 \equiv \h \sum_{i=1}^m  \Bigl( \Phi_i D^{-1} L^{\ast} \( \Psi_i\) -
L \( \Phi_i\) D^{-1}  \Psi_i \Bigr)
\lab{xdef3}
\ee
Using the above formulas we find that
\br
{\sbr{Y_3}{L}}_{-} \eq -\( L^3 \)_{-} + {3 \over 2} \sum_{i=1}^m
\( \Phi_i D^{-1} \(L^{\ast}\)^2 \( \Psi_i\)  + L^2\(\Phi_i\) D^{-1} \Psi_i \)
\nonu \\
&+& \sum_{i=1}^m \Bigl( Y_3 ( \Phi_i )  D^{-1} \Psi_i
- \Phi_i D^{-1} Y^{\ast}_3 (\Psi_i )  \Bigr)
\lab{l3xl}
\er
Hence ${\sbr{-\( ML^3 \)_{-}+Y_3}{L}}_{-}$ still has a form of \rf{addflo}.
The interpretation of this result is evident.
It turns out that it is possible to find additional symmetries for the \cKPrm
model by combining the original ${\bar \pa}_{k,1}$ flows
and the ghost flows \rf{ghostflo} associated with operators of $Y_3$ type.
This will work provided that the above construction  
yields the Virasoro generator  $\cL_2$  obeying
the correct algebra with the unbroken
$sl(2)$ generators found above in \rf{papsiz}--\rf{papsi2}.
Note that each of the two terms in \rf{xdef3} could have been used with an
appropriate factor to obtain the similar conclusion as in \rf{l3xl}.
The choice of
the coefficients in $Y_3$  \rf{xdef3},
apart from being the most symmetric
combination, has the advantage that it will lead below to the correct
Virasoro algebra commutators.

We now generalize the above manipulations to an arbitrary $k$.
We introduce the pseudo-differential operators:
\br
X^{(0)}_k  &\equiv&  \sum_{i=1}^m \sum_{j=0}^{k-1}  L^{k-1-j} (\Phi_i) D^{-1} 
\( L^{\ast}\)^{j} (\Psi_i) \quad ;\quad k \geq 1   \lab{xk0}\\
X^{(1)}_k  &\equiv&  \sum_{i=1}^m \sum_{j=0}^{k-1} \(j - \h (k-1)\) L^{k-1-j}
(\Phi_i) D^{-1} \( L^{\ast}\)^{j} (\Psi_i) \quad ;\quad k \geq 1 \lab{xkb} \\
X^{(2)}_k  &\equiv&  \sum_{i=1}^m \sum_{j=0}^{k-1} \(j^2 - j (k-1)+ {
(k-2)(k-1) \ov 6}\) 
L^{k-1-j} (\Phi_i) D^{-1} \( L^{\ast}\)^{j} (\Psi_i) \quad ;\quad k \geq 1 
\lab{xk2} 
\er
Note that $X^{(1)}_2 = Y_{3}$ as defined above in \rf{xdef3}.

Acting on \rf{xk0}-\rf{xk2} with $ {\bar \pa}_{\ell,1}$ 
for $\ell = 0,1,2$ and using:
\br
\pa^{\ast}_{\ell}\, L^k (\Phi_i) \eq \( M L^\ell \)_{+} \(L^k (\Phi_i)\) +
\( k+ \h \ell \) L^{k+\ell-1} (\Phi_i) \nonu \\
\pa^{\ast}_\ell (L^{\ast})^{k} (\Psi_i) \eq - \( M L^\ell \)^{\ast}_{+}
\((L^{\ast})^{k}\,(\Psi_i)\) + \( k + \h \ell \)
(L^{\ast})^{k+\ell-1} (\Psi_i)
\lab{usepaast}
\er
valid for $\ell = 0,1,2$ and $k \geq 0$, we get:
\br
\pa^{\ast}_{\ell}\, X^{(0)}_k &= &
{\sbr{\( M L^{\ell}\)_{+}}{X^{(0)}_k}}_{-} +  k X^{(0)}_{k+\ell-1} 
\lab{paxk0} \\
\pa^{\ast}_{\ell}\, X^{(1)}_k &= &
{\sbr{\( M L^{\ell}\)_{+}}{X^{(1)}_k}}_{-} +  \(k - \ell +1 \) 
X^{(1)}_{k+\ell-1} \lab{paxk1} \\
\pa^{\ast}_{\ell}\, X^{(2)}_k &= &
{\sbr{\( M L^{\ell}\)_{+}}{X^{(2)}_k}}_{-} +  \(k - 2(\ell -1) \) 
X^{(2)}_{k+\ell-1}   \nonu \\
&& - {1\ov 6} \( (\ell -1)^3 - (\ell -1) \) X^{(0)}_{k+\ell-1}
\lab{paxk2}
\er
Here we recognize a structure of $\Win1$ algebra
under substitution $ \ell \to \ell -1$.
Now we are ready to restrict attention to the part of the algebra involving
$X^{(1)}_{k-1}$
operator from \rf{xkb}. 
We note that \rf{xlbra} and
identity \rf{lkminus} from Appendix \ref{section:appa}
enable us to obtain:
\br
{\sbr{X^{(1)}_{k-1}}{L}}_{-} &=& {k \over 2} \sum_{i=1}^m \( \Phi_i D^{-1}
\( L^{\ast}\)^{k-1} (\Psi_i) +  L^{k-1} (\Phi_i) D^{-1} \Psi_i\)
\lab{lkminnn} \\
&-&\(  L^k \)_{-} + \sum_{i=1}^{m} \Bigl( -\Phi_i D^{-1}
\(X^{(1)}_{k-1}\)^{\ast} (\Psi_i) + X^{(1)}_{k-1} (\Phi_i) D^{-1} \Psi_i \Bigr)
\nonu
\er
Our main result is contained in the following:
\begin{proposition}
The correct additional-symmetry flows for the \cKPrm hierarchies \rf{f-5},
spanning the Virasoro algebra, are given by:
\be
\pa^{\ast}_k\, L  \;\equiv\;  \sbr{- \( M L^k \)_{-} + X^{(1)}_{k-1}}{L}
\lab{pasta}
\ee
{\sl i.e.}, with the following isomorphism
$\cL_{k-1} \sim -\( M L^k \)_{-}  + X^{(1)}_{k-1}$, where $X^{(1)}_{k-1}$ are
defined in \rf{xkb}.
Accordingly, on dressing operators and BA functions the flows \rf{pasta} read:
\be
\pa^{\ast}_k\, W = \( - \( M L^k \)_{-} + X^{(1)}_{k-1}\) W \; \;  ; \; \;
\pa^{\ast}_k\, \psi (t,\l ) = 
\( -\( M L^k \)_{-} + X^{(1)}_{k-1}\)\bigl(\psi (t,\l )\bigr) 
\lab{pasta-psi}
\ee
\label{proposition:mainprop}
\end{proposition}

Indeed, using \rf{lkminnn} we first find that $\(\pa^{\ast}_k L\)_{-}$ can be
cast in the form of \rf{addflo} with:
\br
\pa^{\ast}_k\, \Phi_i &=&\( M L^k \)_{+} (\Phi_i) + 
{k \over 2} L^{k-1} (\Phi_i) + X^{(1)}_{k-1} (\Phi_i)   \nonu \\
\pa^{\ast}_k\, \Psi_i &=& - \( M L^k \)^{\ast}_{+} (\Psi_i) + {k \over 2}
(L^{\ast})^{k-1} (\Psi_i) - \(X^{(1)}_{k-1}\)^{\ast}(\Psi_i)
\lab{paast}
\er
Taking into account that $X^{(1)}_{i-1} =0$ for $i=0,1,2$ we see that
eq.\rf{paast}
reproduces
\rf{papsiz}-\rf{papsi2} (with ambiguity on the right hand side of \rf{papsi1}
removed by fixing $\a=\b=1/2$).
Hence $\pa^{\ast}_{\ell} = {\bar \pa}_{\ell,1}$ for $\ell = 0,1,2$.

Secondly, we note that the modified additional symmetry 
flows defined by \rf{pasta}
commute with the isospectral flows \rf{lax-eq} and, due to \rf{paast}, they
preserve the form of the \cKPrm Lax operator \rf{f-5}.
The remaining question is whether they form a closed algebra.
Indeed, using identity \rf{paxk1} we arrive at the fundamental
commutation relations for $\ell = 0,1,2$ and any $k \geq 0$ :
\be
\llb\, \pa^{\ast}_{\ell}\, , \,   \pa^{\ast}_k\, \rrb\, L\,= \,
\( k-\ell \) \, \pa^{\ast}_{k+\ell-1}\, L
\lab{paipay}
\ee
This discussion shows that
$\; \sbr{\cL_i}{\cL_{k}} = (i-k) \cL_{i+k}\;$ for $i=-1,0,1 \;$ ($sl(2)$
generators) and arbitrary $k$, where $\,\cL_{k-1} \sim - \pa^{\ast}_{k}$ .
Since according to proposition \ref{proposition:mainprop} the generator
$\cL_2$ is associated with $\, X_2^{(1)} - \( M L^3 \)_{-}\,$,
all higher Virasoro operators can be obtained recursively from:
\be
\cL_{n+1} = { -1 \over (n-1)} \sbr{\cL_{n}}{\cL_{1}} \qquad ,\quad n \geq 2
\lab{recursa}
\ee
Then eq.\rf{paipay} implies that $\, \cL_n$ with $n \geq 3\,$ may differ from
the generators given by the flows
$\, \pa^{]ast}_{n+1} \sim  - \( ML^{n+1}\)_{-} + X_n^{(1)}$ defined in
\rf{pasta} at most by flows commuting with the $sl(2)$ additional symmetry
generators, {\sl i.e.}, by ordinary isospectral flows.
Therefore, we can now easily show by induction that $\cL_k\, , \, k\geq -1$,
obtained in the above way form a closed Virasoro
algebra up to irrelevant terms containing ordinary isospectral flows.

For illustration consider, {\sl e.g.}, $\sbr{\cL_2}{\cL_3}\equiv Z $. 
Commuting $\cL_{-1}$ with $Z$ we find $\sbr{\cL_{-1}}{Z} = 6 \cL_4$, which
fixes $Z$ to be $-\cL_5$ up to isospectral flows commuting with the $sl(2)$ 
subalgebra. It is easy to see how to extend this argument to cover the whole
Virasoro algebra.

\subsection{Additional Symmetries versus DB Transfor\-ma\-tions for
\cKPrm{}$\,\,$ Hierarchies. String Condition}
Let $\Phi$ be an eigenfunction of $L$ defining a \DB transformation,
{\sl i.e.} :
\be
\partder{}{t_l} \Phi = L^{l\over r}_{+} (\Phi )  \quad ,\quad
{\wti L} = \(\Phi D \Phi^{-1}\)\, L\, \(\Phi D^{-1} \Phi^{-1}\)
\quad ,\quad
{\wti W} =  \(\Phi D \Phi^{-1}\)\, W\, D^{-1}
\lab{gen-L-DB}
\ee
Then the DB-transformed $M$ operator (cf. \rf{M-dress}) acquires the form:
\br
{\wti M} \eq \(\Phi D \Phi^{-1}\)\, M\, \(\Phi D^{-1} \Phi^{-1}\) =
\sum_{l \geq 0} \frac{l+r}{r} t_{r+l} {\wti L}^{l\over r}_{+} + {\wti M}_{-}
\lab{gen-M-DB}  \\
{\wti M}_{-} \eq {\wti W} {\wti X}_{(r)} {\wti W}^{-1} - t_r -
\sum_{l \geq 1} \frac{l+r}{r} t_{r+l}\partder{}{t_l}{\wti W}\,.\,{\wti W}^{-1}
\lab{M--ti}
\er
where ${\wti X}_{(r)} = D X_{(r)} D^{-1}$ with $X_{(r)}$ as in \rf{M-0}.
Clearly ${\wti X}_{(r)}$, like $X_{(r)}$, is
also admissible as canonically conjugated to $D^r$.
The DB-transformed $M$-operator acts on the DB-transformed BA function
\rf{DB-BA} as:
\be
{\wti M} {\wti \psi} (\l ) = 
\( \frac{\l^{1-r}}{r} \partder{}{\l} + {\wti \a_r} (\l )\) {\wti \psi} (\l )  
\quad , \quad 
{\wti \a_r} (\l ) = \a_r (\l ) + {1\over r} \l^{-r} 
\lab{M-psi-DB}
\ee

Now, let again $L$ belong to a \cKPrm hierarchy \rf{f-5} and let us con\-sider
auto-B\"{a}cklund trans\-formations \rf{shabes-3}--\rf{shabes-5a} preserving the
constrained \cKPrm form of $L$ :
\br
{\wti L} \eq  T_a L T_a^{-1} =
{\wti L}_{+} + \sum_{i=1}^m {\wti \Phi}_i D^{-1} {\wti \Psi}_i
\qquad , \qquad T_a \equiv \Phi_a D \Phi_a^{-1}
\lab{cKP-L-DB} \\
{\wti \Phi}_a \eq T_a L (\Phi_a ) \qquad ,\qquad {\wti \Psi}_a = \Phi_a^{-1}
\lab{DB-1} \\
{\wti \Phi}_i \eq  T_a (\Phi_i ) \quad ,\quad
{\wti \Psi}_i = {T_a^{-1}}^\ast \Psi_i =
- \Phi_a^{-1} \pa_x^{-1} ( \Psi_i \Phi_a ) \qquad, \quad i \neq a
\lab{DB-2}
\er
where the DB-generating $\Phi \equiv \Phi_a$ coincides with one of the
eigenfunctions of the initial $L$ \rf{f-5}.

With the help of identities \rf{DB-like-1}--\rf{DB-like-4} from the Appendix
\ref{section:appa}
we find the following explicit form of the DB transformation of the operators
$X^{(1)}_{k-1}$ \rf{xkb} :  
\br
T_a X^{(1)}_{k-1} T_a^{-1} &=& {\wti X}^{(1)}_{k-1} - \( {\wti
L}^{(a)}\)_{-}^{k-1} +
\lcurl T_a \( X^{(1)}_{k-1} + {k\over 2} L^{k-1}\) (\Phi_a )\rcurl D^{-1}
\Phi_a^{-1}
\lab{DB-Yk}   \\
\( {\wti L}^{(a)}\)_{-}^{k-1} &\equiv&
\sum_{j=0}^{k-2} {\wti L}^{k-j-2} ({\wti \Phi}_a) D^{-1}
\( {\wti L}^{\ast}\)^{j} ({\wti \Psi}_a )
\lab{lkminus-ti-a}
\er
Here ${\wti L},\, T_a $ are as in \rf{cKP-L-DB}
and the DB-transformed ${\wti X}^{(1)}_{k-1}$ have the same form as
$X^{(1)}_{k-1}$ in \rf{xkb}
with all (adjoint) eigenfunctions substituted with their
DB-transformed counterparts as in \rf{cKP-L-DB}--\rf{DB-2}. Also notice that
in the particular case of ${\sf cKP}_{r,1}$ hierarchies
$\( {\wti L}^{(a)}\)_{-}^{k-1}$ \rf{lkminus-ti-a} coincides with the
(pseudo-differential part of the power of the) full ${\sf cKP}_{r,1}$ Lax
operator (cf. eq.\rf{f-5} for $m=1$ and \rf{lkminus}).

Taking into account \rf{cKP-L-DB}--\rf{DB-2} and \rf{DB-Yk}--\rf{lkminus-ti-a}
we arrive at the following important
\begin{proposition}
The additional-symmetry flows \rf{pasta} for ${\sf cKP}_{r,1}$ hierarchies
(eq.\rf{f-5} with $m=1$)
commute with the \DB transformations \rf{cKP-L-DB} preserving the form of
${\sf cKP}_{r,1}$, up to shifting of \rf{pasta} by ordinary isospectral flows.
Explicitly we have:  
\be
\pa^\ast_k {\wti L} =
- \Sbr{\( {\wti M} {\wti L}^k \)_{-} - {\wti X}_{k-1}^{(1)}}{{\wti L}} +
\partder{\wti L}{t_{k-1}}
\lab{AS-DB-compat}
\ee
\label{proposition:dbcommute}
\end{proposition}

Proposition \ref{proposition:dbcommute} shows that the additional-symmetry
flows \rf{pasta} are well-defined for all ${\sf cKP}_{r,1}$ Lax operators
belonging to a given DB orbit of successive DB transformations.
Notice that it is precisely the class of (truncated) ${\sf cKP}_{r,1}$
hierarchies which is relevant for the description of discrete (multi-)matrix
models (refs.\ct{office,avoda,pirin} and sections 
\ref{section:noak4},\ref{section:noak5} below). 

Motivated by applications to (multi-)matrix models (see section 
\ref{section:noak4}), one can require invariance of \cKPrm hierarchies 
under some of the additional-symmetry flows,
{\sl e.g.}, under the lowest one $\pa^\ast_0 \equiv{\bar \pa}_{0,1}$, known as
``string-equation'' constraint (string condition) in the context of the
(multi-)matrix models:
\br
\pa^\ast_0 L = 0  \qquad \to \quad  \Sbr{M_{+}}{L} = - \one
\qquad ; \qquad
\pa^\ast_0 \Phi = 0  \qquad \to \quad  M_{+} \Phi = 0
\lab{01-constr-L}
\er
Eqs.\rf{01-constr-L}, using second eq.\rf{L-M},\rf{M-dress}
and first eq.\rf{paast} for $k=0$, lead to the following constraints on $L$
\rf{f-5}, the BA function $\psi (t,\l )$ and its DB-generating eigenfunction 
$\P (t)$, respectively:
\br
\sum_{l \geq 1} \frac{l+r}{r} t_{r+l} \partder{}{t_l} L +
\Sbr{t_1}{L}\, \d_{r,1} &=& - \one
\lab{L-constr} \\
\( \sum_{l \geq 1} \frac{l+r}{r} t_{r+l} \partder{}{t_l} +
t_r - \a_r (\l )\) \psi (t,\l ) &=& \frac{\l^{1-r}}{r}\partder{}{\l}\psi (t,\l
)
\lab{psi-constr} \\
\( \sum_{l \geq 1} \frac{l+r}{r} t_{r+l} \partder{}{t_l} + t_r \) \P (t) &=&  0
\lab{P-constr}
\er

Recall now formula \rf{tauok-1} for the $\t$-function of the ${\sf cKP}_{r,m}$
hierarchy \rf{f-5}. Noticing that the
eigenfunctions $\Phi^{(k)}$ of the DB-transformed Lax operators $L^{(k)}$
satisfy the {\em same} constraint eq.\rf{P-constr} irrespective of the
DB-step $k$, we arrive at the following result (``string-equation''
constraint on the $\t$-functions) :
\begin{proposition}
The Wronskian $\t$-functions \rf{tauok-1} of ${\sf cKP}_{r,m}$
hierarchies \rf{f-5}, invariant under the lowest additional symmetry flow
\rf{01-constr-L}, satisfy the constraint equation: 
\br 
&&\( \sum_{l \geq 1} \frac{l+r}{r} t_{r+l} \partder{}{t_l} + nt_r \)
\frac{\t^{(n)}}{\t^{(0)}} \equiv   \nonu  \\
&&\( \sum_{l \geq 1} \frac{l+m}{m} t_{m+l} \partder{}{t_l} + nt_m \) 
W_n \bigl\lb \Phi^{(0)}, L^{(0)} \Phi^{(0)}, \ldots , 
\( L^{(0)}\)^{n-1} \Phi^{(0)} \bigr\rb = 0
\lab{tau-M-constr}
\er
\label{proposition:taum}
\end{proposition}

In particular, eq.\rf{P-constr} applied to \rf{spectr-1},\rf{spectr-i} leads 
to the following restriction on the \DB generating eigenfunction 
$\P_1^{(0)}(t)$ of the ``initial'' $L^{(0)}$ which uniquely fixes the form of
the members of the pertinent \cKPrm DB-orbit:
\begin{corollary}
The constraint \rf{P-constr} imposes the following relations on the
eigenfunctions defining the \cKPrm DB-orbit:
\br
\p_1^{(0)}(\l ) \eq {\rm const} 
\lab{p-1-0-constr} \\
\Phi_1^{(k)}(t) \eq \int d\l \, \l^{k(1+r)} \psi^{(k)} (t,\l ) = 
\int d\l \, \l^{k(1+r)} \frac{\t^{(k)} \bigl( t -[\l^{-1}]\bigr)}{\t^{(k)}(t)}
e^{\xi (t,\l )} 
\lab{spectr-constr} \\
\P_i^{(k)}(t) \eq \int d\l \, \l^{k} \psi^{(k)} (t,\l )
\quad , \;\; i=2, \ldots , m        \nonu
\er
\label{corollary:eigenfconstr}
\end{corollary}

\sect{One-Matrix model, DB-Wronskian Technique and 
Virasoro constraints}
\label{section:noak3}
The standard (Hermitian) one-matrix model has partition function \ct{BIPZ} :
\be
Z_N \lb t \rb = \int dM \exp \lcurl \sumi{r=1} t_r \Tr M^r \rcurl  
\lab{ZN-1M}
\ee
where $M$ is a Hermitian $N \times N$ matrix.
One uses the method of generalized orthogonal polynomials \ct{ortho-poly}
to evaluate partition function \rf{ZN-1M}. After angular integration in 
\rf{ZN-1M} we obtain (up to an overall constant) :
\be
Z_N \lb t \rb ={ 1 \over N !} \int \prod_{i=1}^N d \l_i 
 \exp \lcurl \( \sum_{i=1}^{N} \sumi{k=1} t_k (\l_i )^k\) \rcurl 
\prod_{i >j =1}^N \( \l_i - \l_j \)^2 
\lab{partl}
\ee
Comparing \rf{partl} with \rf{BH-tau}, we notice that $Z_N$ \rf{ZN-1M}
precisely
coincides with the $\t$-function of the generalized BH hierarchy, {\sl i.e.},
the \DB orbit of ${\sl cKP}_{1,1}$ hierarchy starting from a ``free'' Lax
$L^{(0)}=D$, with the special ``initial'' condition \rf{p-1-0-constr} : 
\be
\p^{(0)}(\l ) = const
\lab{p-0-constr}
\ee
which, as discussed in section \ref{section:addsym} reflects the imposition of
the
additional requirement of invariance of the ${\sl cKP}_{1,1}$ solutions w.r.t.
the lowest additional symmetry flow.

Reversing the argument leading from \rf{ZN-1M} to \rf{partl}, we find that
the $\t$-function of a generic ${\sl cKP}_{1,1}$ \DB orbit (eq.\rf{tau-recur-r}
for $r=1$), {\sl i.e.}, a \DB orbit starting from a {\em non}-free
initial Lax $L^{(0)}= D + \P^{(0)} D^{-1}\Psi^{(0)}$ and with $\P^{(0)}$ having
an arbitrary ``density'' $\p^{(0)}(\l )$ not subject to the constraint
\rf{p-0-constr}, can be identified with the partition function of the following
generalized one-matrix model:  
\br
&&{\wti Z}_N \lb t \rb = \int dM \exp \lcurl \Tr \ln \p^{(0)}(M) +
\sumi{r=1} t_r \Tr M^r + \ln \t^{(0)} \bigl( t - \Tr [M^{-1}]\bigr) \rcurl  
\lab{ZN-1M-wti} \\
&&\t^{(0)} \bigl( t - \Tr [M^{-1}]\bigr) \equiv
\t^{(0)} \bigl( t_1 - \Tr M^{-1}, t_2 - \h \Tr M^{-2}, 
t_3 - {1\over 3} \Tr M^{-3}, \ldots \bigr)   \nonu \\
&& \pa_x^2 \ln \t^{(0)} = \P^{(0)} \Psi^{(0)}   \nonu
\er

Below we will exhibit in more detail the relevant integrability structure
leading to the \DB Wronskian solution \rf{BH-tau} for $Z_N$ \rf{ZN-1M} and its
generalizations \rf{ZN-1M-wti}.

One deals with the integral \rf{partl} (or the corresponding generalization
from \rf{ZN-1M-wti}) 
using orthogonal polynomials:
\be
P_n (\l) = \l^n + O \(\l^{n-1}\)\quad,\quad  n= 0,1,\ldots
\lab{opoly}
\ee
which enter the orthogonality relation:
\be
\int d \l   P_n (\l)  P_m (\l) \exp \( \sumi{k=1} t_k \l^k  \) = 
h_m (t) \d_{nm}
\lab{orthorel}
\ee
{}From \rf{opoly} and \rf{orthorel} follow the recursion relations for the 
orthogonal polynomials: 
\be
\l P_n (\l) = \sumi{l=0} Q_{nl} P_l (\l) 
\lab{recurpo}
\ee
The matrix elements of $Q$ can be found by comparing \rf{recurpo} with the 
definition of orthogonal polynomials, yielding:
\be
Q_{n,n+1} =1 \; ; \; Q_{n,l} = 0 \;\;\;\; l \geq n+2 \; \; ; \;\;
Q_{n,n-1} = {h_n \ov h_{n-1}} \;\; n \geq 1
\lab{qone}
\ee
The orthogonal polynomials approach leads to expression for the partition
function in terms of $h_n$:
\be
Z_N = const \, h_0 h_1 \cdots h_{N-2} h_{N-1}
\lab{partfctz}
\ee
{}From the orthogonal relation \rf{orthorel} and definitions \rf{qone}
we also obtain:
\be
\partder{\ln h_n}{t_r} = \(Q^r\)_{nn} 
\lab{hnttit}
\ee
as well as:
\be
\partder{P_n}{t_r} = - \sum_{l=0}^{n-1} Q^r_{nl} P_l (\l)
\lab{pntr}
\ee
Defining now a wave function:
\be
\Psi_n \( t,\l\) = P_n (\l) e^{V(t,\l)} \qquad ,\quad
V(t,\l) \equiv \sum_{k\geq 1} \l^k t_k
\lab{psipnv}
\ee
we obtain from \rf{recurpo} and \rf{pntr} the following eigenvalue problem:
\be
\l \Psi = Q \Psi \qquad ;\qquad  
\partder{\Psi}{t_r} = Q^r_{(+)} \Psi 
\lab{lpsiq}
\ee
where $\Psi$ is a semi-infinite column
$\(\Psi_0, \ldots, \Psi_n, \Psi_{n+1}, \ldots \)^{T}$ and
$Q$ is a semi-infinite matrix {\sl i.e.}, with indices running from $0$ to 
$\infty$.
Furthermore we adhere to the following notation:
the subscripts $-/+$ denote lower/upper triangular parts of the matrix,
whereas $(+)/(-)$ denote upper/lower triangular plus diagonal parts.
The compatibility of the eigenvalue problem \rf{lpsiq} 
gives rise to a discrete linear system:
\be
\partder{}{t_r} Q = \llb Q^r_{(+)} , Q \rrb    
\lab{L-3-1M} 
\ee
Choosing parametrization 
$Q_{nn} = a_0 (n)$ and $Q_{n,n-1} = a_1 (n)\equiv \frac{h_n}{h_{n-1}}$ 
for the matrix $Q$, we can rewrite \rf{lpsiq} as:
\br
\l \Psi_n \eq \Psi_{n+1}  + a_0 (n) \Psi_n
+ a_1 (n) \Psi_{n-1} 
\lab{specb} \\
\pa \Psi_n \eq \Psi_{n+1} + a_0 (n) \Psi_n \qquad ({\rm for}\;\; r=1)
\lab{speca}
\er
Consistency of \rf{speca} and \rf{specb} yields the following evolution
equations having form of the Toda lattice equations of motion:
\br
\pa a_0 (n) \eq a_1 (n+1) - a_1 (n)  \lab{t1eqa} \\           
\pa a_1 (n) \eq a_1 (n) \( a_0 (n) - a_0 (n-1) \)  \lab{t1eqb}
\er

Using an inverted version of \rf{speca}:
\be
\Psi_n = {1 \ov \pa - a_0 (n)} \Psi_{n+1}
\lab{psinminus}
\ee
one can re-express the linear problem \rf{specb}--\rf{speca} entirely in 
terms of $\Psi_n$ at a {\em fixed} lattice site $n$ \ct{BX93-94,discrete} :
\br
\l \Psi_n \eq L_n \Psi_n \qquad n \geq 0
\lab{specb-n} \\
\partder{}{t_r} \Psi_n \eq \( L_n\)^r_{+} \Psi_n
\lab{speca-n}
\er
where now $L_n$ is a continuum pseudo-differential Lax operator associated 
with the site $n$ : 
\be
L_n = \pa + a_1 (n) {1 \ov \pa - a_0 (n-1)}
\lab{lnn}
\ee

Note that from \rf{speca}--\rf{specb} we also have
~$\l \Psi_{n+1} = L_{n+1} \Psi_{n+1} $, {\sl i.e.}, the lattice shift 
$n \to n+1$ does not change the eigenvalue $\l$.
In fact, the lattice shift $n \to n+1$ can be given a meaning of a
\DB transformation within the generalized \BH hierarchy introduced in 
section \ref{section:noak2-2}. To see this, we rewrite \rf{psinminus} 
as follows:
\be
\Psi_{n+1}= e^{\int a_0 (n)} \pa\; e^{-\int a_0 (n)} \Psi_{n}
= \P (n) \pa \P^{-1} (n) \Psi_{n} = T (n) \Psi_{n}
\lab{spectodc}
\ee
where $\P (n) = e^{\int a_0 (n)} $, and $T (n)= \P (n) \pa \P^{-1} (n) $
plays a r\^{o}le of the DB transformation operator generating the
lattice translation $n \to n+1$. 
Combining \rf{spectodc} with \rf{specb}--\rf{specb-n} we find:
\br
L_{n+1} \Psi_{n+1} &=& \l \( \pa - a_0 (n) \) \Psi_n 
= \( \pa - a_0 (n) \) L_n \Psi_n \nonu\\
&=& \( \pa - a_0 (n) \) L_n \( \pa - a_0 (n) \)^{-1} \Psi_{n+1}
\lab{lntoln1}
\er
Thus, the Lax operators at different sites are related by a DB transformation:
\be
L_{n+1} = \( \pa - a_0 (n) \) L_n \( \pa - a_0 (n) \)^{-1}=
T (n) L_n T^{-1} (n)
\lab{transf3}
\ee
with the Lax operator \rf{lnn} taking the form:
\br
L_n &\equiv& L (n) = \pa + \P (n) \pa^{-1}  \Psi (n)  
\lab{specln} \\
\P (n) &\equiv& e^{\int a_0 (n)} = a_1 (n) e^{\int a_0 (n-1)}  \quad ,\quad
\Psi (n) \equiv  e^{-\int a_0 (n-1)} = \( \P (n-1)\)^{-1}
\lab{Phi-Psi-n}
\er
This is precisely the form of the Lax operators belonging to the
${\sl cKP}_{1,1}$ DB orbit starting from a ``free'' Lax $L_0 = \pa$ 
~\rf{lkplus}--\rf{pkplus} or, according to subsection 3.3, $L_n$ \rf{specln} 
describes the generalized \BH system.

We see that we proved:
\begin{proposition}
The one-matrix model problem \rf{ZN-1M} is equivalent to solving the 
generalized \BH system. 
\label{proposition:1matbh}
\end{proposition}

In more detail, realizing from \rf{hnttit} that $a_0 (n) = \pa \ln h_n$ and 
accounting for \rf{psipnv}, we have the following identification between
discrete one-matrix model objects $h_n,\, P_n (\l ),\, Z_N\,$, on one hand, and
eigenfunction $\P (n) \equiv \P^{(n)}(t)$, BA function $\psi^{(n)}(t,\l )$ and
$\t$-function of continuum $n$-truncated KP hierarchy 
(Prop.\ref{proposition:tau-BA-BH}), on the other hand:
\br
h_n \eq \P (n)   
\lab{h-n-P-n} \\
\Psi_n \( t,\l\) \eq P_n (\l) e^{V(t,\l)} = \l^n \psi^{(n)}(t,\l ) =
\Bigl(\sum_{j=0}^n w^{(n)}_j(t) \l^{n-j} \Bigr) e^{\sum_{l\geq 1} \l^l t_l} 
\lab{Toda-BA} \\
Z_N \eq h_0 h_1 \cdots h_{N-2} h_{N-1} = \P (0) \cdots \P (N-1) =
W_N \lb h_0, \pa h_0, \ldots , \pa^{N-1} h_0 \rb
\lab{partfctw}
\er
In particular, from \rf{Toda-BA} and \rf{partfctw} one obtains an explicit
expression for the coefficients of the orthogonal polynomials 
(cf. refs.\ct{shaw,kostov,avoda}):  
\br
P_n (\l ) &=& \sum_{j=0}^n \l^{n-j}
\frac{p_j (-[\pa ]) W_n \lb h_0,\pa h_0 ,\ldots ,\pa^{n-1}h_0\rb}{W_n \lb h_0,
\ldots ,\pa^{n-1}h_0\rb}   \nonu  \\
{}\eq \l^n \frac{W_n \lb h_0 (t - [\l^{-1}]), \pa h_0 (t - [\l^{-1}]),\ldots ,
\pa^{n-1} h_0 (t - [\l^{-1}]) \rb}{W_n \lb h_0 (t), \pa h_0 (t), \ldots ,
\pa^{n-1} h_0 (t) \rb}
\lab{ortho-wronski}
\er

Further, using an identity:
\be
0 =\int d \l \partder{}{\l} \lcurl\exp \(  \sumi{k=1} t_k \l^k \) 
P_n (\l) P_m (\l)\rcurl 
\lab{dortho}
\ee
we derive:
\be
\partder{\Psi}{\l}= P \Psi \lab{lpsibarq} \quad;\quad P= \sumi{k=2} k t_k
Q^{k-1}_{+}
\ee
Consistency with \rf{lpsiq} requires the so-called {\em discrete} string
equation:
\be
\sbr{Q}{P} = \one \quad  \to \quad  
\sumi{k=2} k t_k \partder{Q}{t_{k-1}} = - \one
\lab{1streq}
\ee
which using \rf{hnttit} can be rewritten as:
\be 
\partder{}{t_1} \( \sumi{k=2} k t_k \partder{}{t_{k-1}}\ln h_n \) = 0  
\qquad n \geq 0
\lab{strnga}
\ee
According to \rf{P-constr} and the identification \rf{h-n-P-n},
eq.\rf{strnga} is the discrete one-matrix model counterpart of the
``initial-condition'' constraint \rf{p-0-constr} on the associated
\BH hierarchy due to invariance under the lowest additional symmetry flow.

In refs.\ct{Morozov}, 
using Ward identities' techniques for the matrix integral \rf{ZN-1M}, it was
shown that the partition function $Z_N$ in fact satisfies an infinite number of
constraints spanning the Borel subalgebra of the Virasoro algebra:
\br
\cL^{(N)}_s Z_N &=& 0 \qquad ,\quad s \geq -1
\lab{Vir-Z-N} \\
\cL_s^{(N)} &=& \sumi{k=1} k t_k \partder{}{t_{k+s}} + 2 N \partder{}{t_s} +
\sum_{k=1}^{s-1} \partder{}{t_k}\partder{}{t_{s-k}} \quad , \quad s \geq 1
\lab{lsn1}  \\
\cL_{0}^{(N)} &=& \sumi{k=1} k t_k \partder{}{t_k} + N^2 \quad \; ; \; \quad
\cL_{-1}^{(N)} = \sumi{k=2} k t_k \partder{}{t_{k-1}} + N t_1
\lab{lsn2}
\er
Here we provide a simple alternative proof that the Wronskian solution
\rf{partfctw} indeed satisfies all the Virasoro constraints:
\begin{proposition}
The Wronskian $\t$-function of the generalized Burgers-Hopf 
hierarchy \rf{BH-tau}
subject to the ``initial'' condition \rf{p-0-constr} satisfy the
Virasoro constraints:
\be 
\cL^{(N)}_s W_N \lb \p,\pa \p,\ldots ,\pa^{N-1}\p \rb = 0 
\qquad ,\quad s \geq -1
\lab{W-N-constr}
\ee
\label{proposition:Vir-constr}
\end{proposition}
\begin{proof}
The case $s=-1$ in \rf{W-N-constr} is a particular case of 
Prop.\ref{proposition:taum}. For $s \geq 0$ the proof will go by induction
w.r.t. $N$ and we shall employ the $\t$-function recurrence relation
\rf{tau-recur} which in the special case of \rf{BH-tau} takes the form:
\be
W_{N+1} \lb \p,\pa \p,\ldots ,\pa^N \p \rb =  \int d\l \, \l^{2N} 
:e^{- {\hat \th} (\l )}: W_N \lb \p,\pa \p,\ldots ,\pa^{N-1}\p \rb
\lab{tau-recur-BH}
\ee
Using the explicit form of the vertex operator with ${\hat \th}(\l )$ as in
\rf{theta-field}, it is easy to deduce the following commutation relation:
\br
\cL^{(N+1)}_s :e^{- {\hat \th} (\l )}: = :e^{- {\hat \th} (\l )}: \cL^{(N)}_s
+ \( \l^{s+1} \partder{}{\l} + (2N+s+1) \l^s\) :e^{- {\hat \th} (\l )}:
\nonu \\
+ :e^{- {\hat \th} (\l )}: \l^s \sum_{k=1}^s \l^{-k} \partder{}{t_k}
\lab{commut-rel}
\er
Applying \rf{commut-rel} on both sides of \rf{tau-recur-BH} and integrating by
parts we get:
\br
\cL^{(N+1)}_s W_{N+1} \lb \p,\pa \p,\ldots ,\pa^N \p \rb =
\int d\l \, \l^{2N} :e^{- {\hat \th} (\l )}:
\cL^{(N)}_s  W_N \lb \p,\pa \p,\ldots ,\pa^{N-1}\p \rb  
\nonu \\
+ \int d\l \, \l^{2N} :e^{- {\hat \th} (\l )}: 
\Bigl( \sum_{k=1}^s \l^{-k} \partder{}{t_k} \Bigr) 
W_N \lb \p,\pa \p,\ldots ,\pa^{N-1}\p \rb    
\lab{Vir-recur}
\er
Thus, to complete the proof we have to verify that the last term on the right
hand side of \rf{Vir-recur} does vanish identically.
Indeed, this term can be rewritten using second equality \rf{BH-tau} as:
\be
\int \prod_{j=0}^N d\l_j \, \sum_{k=1}^s \sum_{l=0}^{N-1}
\(\l_l\)^k \(\l_N\)^{s-k} \prod_{i>j} \(\l_i -\l_j\) 
\prod_{j=0}^N \(\l_j\)^j \prod_{k=0}^N \Bigl\{ \p^{(0)}(\l_k )
e^{\xi (t,\l_k )} \Bigr\}
\lab{bad-term-1}
\ee
valid even without the restriction \rf{p-0-constr}. The integral
\rf{bad-term-1} vanishes due to the fact that the expression in its integrand:
\be
\sum_{k=1}^s \sum_{l=0}^{N-1}
\(\l_l\)^k \(\l_N\)^{s-k} \prod_{i>j} \(\l_i -\l_j\) \prod_{j=0}^N \(\l_j\)^j 
\lab{lambda-antisym}
\ee
is {\em anti-symmetric} w.r.t. at least one pair of $\l$'s.
\end{proof}

\sect{Two-Matrix Model and ${\bf cKP}_{\bf r=p_2-1,m=1}$ Hierarchy}
\label{section:noak4}
\subsection{Two-Matrix Model and the Corres\-ponding Linear Toda-\-like System}
We consider the two-matrix model with a partition function \ct{ortho-poly} :
\be
Z_N \lb t,{\ti t},g \rb = \int dM_1 dM_2 \exp \lcurl
\sum_{r=1}^{p_1} t_r \Tr M_1^r +
\sum_{s=1}^{p_2} {\ti t}_s \Tr M_2^s + g \Tr M_1 M_2 \rcurl   
\lab{ZN-2M}
\ee
where $M_{1,2}$ are Hermitian $N \times N$ matrices,
and the orders of the matrix ``potentials'' $p_{1,2}$ may be
finite or infinite. After angular integration in \rf{ZN-2M} one obtains 
\ct{ortho-poly} 
(up to an overall constant) :
\be
Z_N \lb t,{\ti t},g \rb ={ 1 \ov N !} \int \prod_{i=1}^N d \l_i d {\ti \l}_i
 \exp \lcurl \sum_{i=1}^{N}\( V (\l_i) + {\ti V} ({\ti \l}_i) +
g \l_i {\ti \l}_i\) \rcurl \Delta (\l_i) \Delta ({\ti \l}_i)
\lab{partl-2M}
\ee
where 
$ V ( \l) = \sum_{k=1}^{p_1} t_k \l^k $, 
${\ti V} ({\ti \l}) = \sum_{s=1}^{p_2}  {\ti t}_s {\ti \l}^s$.
$\Delta (\l_i), \Delta ({\ti \l}_i)$ are standard Van der Monde determinants.

Similarly to the one-matrix case, we can write down an explicit Wronskian
expression for \rf{ZN-2M}. To this end we notice that the integral over one
set of eigenvalues, {\sl e.g.} ${\ti \l}_i$, can be represented as:
\br
\int \prod_{i=1}^N d {\ti \l}_i \, \prod_{i>j}\({\ti \l}_i - {\ti \l}_j\)
\prod_{k=1}^N \Bigl\{\rho (\l_k ,{\ti \l}_k ) 
e^{\xi ({\ti t},{\ti \l}_k )} \Bigr\} =
W_N \lb \P ({\ti t};\l_1 ), \ldots , \P ({\ti t};\l_N )\rb
\lab{int-ti-l-Wronski} \\
\P ({\ti t};\l ) \equiv \int d{\ti \l} \, \rho (\l ,{\ti \l} )
e^{\xi ({\ti t},{\ti \l})} \quad ,\quad
\rho (\l ,{\ti \l} ) \equiv e^{g \l {\ti \l}} \quad; \quad x \equiv {\ti t}_1
\lab{P-rho}
\er
Here and below $\,\xi ({\ti t},{\ti \l}) = \sum_{s=1}^{p_2} {\ti t}_s {\ti
\l}^s \,$ and $\, \xi (t,\l ) = \sum_{r=1}^{p_1} t_r \l^r\,$.
Inserting \rf{int-ti-l-Wronski} into \rf{partl-2M} and using the simple
identity \rf{anti-symm} we obtain:
\br
Z_N \eq {1\over {N!}} \int \prod_{i=1}^N d\l_i \, \prod_{i>j}\( \l_i -\l_j\)
W_N \bigl\lb e^{\xi (t,\l_1 )} \P ({\ti t};\l_1 ),\ldots ,
e^{\xi (t,\l_N )} \P ({\ti t};\l_N ) \bigr\rb  \nonu \\
{}\eq  \int \prod_{i=1}^N d\l_i \, \prod_{j=1}^N \(\l_j\)^{j-1}
W_N \bigl\lb e^{\xi (t,\l_1 )} \P ({\ti t};\l_1 ),\ldots ,
e^{\xi (t,\l_N )} \P ({\ti t};\l_N ) \bigr\rb   \nonu  \\
{}\eq W_N \bigl\lb h_0 ,\partder{}{t_1} h_0 ,\ldots ,
\partderh{}{t_1}{N-1} h_0 \bigr\rb
\lab{Wronski-2M-MM}
\er
where:
\be
h_0 \equiv h_0 (t,{\ti t}) = \int d\l\, d{\ti \l}\, \rho (\l ,{\ti \l})
e^{\xi (t,\l ) + \xi ({\ti t},{\ti \l} )}
\lab{h0-rho}
\ee
Since in the derivation of \rf{Wronski-2M-MM} the ``density''
$\rho (\l ,{\ti \l})$ was in fact arbitrary (not necessarily fixed by the
expression in \rf{P-rho}), eq.\rf{Wronski-2M-MM} yields simultaneously the
Wronskian solutions of all higher multi-matrix models \ct{Morozov} 
provided we choose
$\rho (\l ,{\ti \l})$ in the form of eq.\rf{rho-q} from next section.

In ref.\ct{BX93-94} it was shown that, by using the method
of generalized orthogonal polynomials \ct{ortho-poly}, the partition function
\rf{ZN-2M} and its derivatives w.r.t. the parameters $\( t_r,{\ti t}_s ,g \)$
can be explicitly expressed in terms of solutions to generalized Toda-like
lattice systems associated with \rf{ZN-2M}. Unlike the case of the full
generalized Toda lattice hierarchy \ct{U-T}, however, the
associated Toda matrices are (i) semiinfinite, and (ii) contain in general
{\em finite} number of non-zero diagonals. In the present section we present
in detail the identification of the correct continuum integrable system
underlying the explicit Wronskian solution \rf{Wronski-2M-MM} of discrete
two-matrix models in the case when at least one of the matrix potentials is 
of finite order, {\sl e.g.}, $p_2 = finite$.

As in one-matrix model one can deal with the integral \rf{partl-2M} 
using generalized orthogonal polynomials \ct{ortho-poly} :
\be
P_n (\l) = \l^n + O \(\l^{n-1}\)\quad;\quad
{\ti P}_m ({\ti \l}_2) = {\ti \l}^m+ O \({\ti \l}^{m-1}\)\quad,\quad
n,m= 0,1,\ldots
\lab{opoly-2M}
\ee
which are orthogonal to each other with respect to the weight
$\exp \( V ( \l) + {\ti V} ({\ti \l}) + g \l {\ti  \l} \)$:
\be
h_n \d_{nm} = \int_{} \int_{} d\l d{\ti \l} \,
P_n (\l ) \exp \lcurl \sum_{r=1}^{p_1} \l^r t_r +
\sum_{s=1}^{p_2} {\ti \l}^s \tit_s + g \l {\ti \l} \rcurl {\wti P}_m ({\ti \l})
\lab{h-n}
\ee
Here $h_n$ are the normalization factors.
In this setting  of generalized orthogonal polynomials the multiplication 
by $\l$, ${\ti \l}$ can be represented by the Jacobi matrices $Q$ and $\tQ$:
\be
\l P_n (\l) = \sum_{l=0}^{n+1} Q_{n,l} P_l (\l) \qquad  ; \qquad
{\ti \l} {\ti P}_n ({\ti \l} ) = \sum_{l=0}^{n+1}{\ti P}_l ({\ti \l}) 
\tQ_{l,n} (h_n/h_l)
\lab{recurpo-2M}
\ee
$Q_{nl}$, $\tQ_{nl}$ are matrix elements of $Q$ and $\tQ$, which define in 
what follows the underlying linear system.
Due to definitions \rf{opoly-2M} they satisfy the relations:
\be
Q_{n,n+1} =1 \; ; \; Q_{n,l} = 0 \;\;\;\; l \geq n+2 \quad \mbox{and} \quad
\tQ_{n+1,n} =h_{n+1}/h_n \; ; \; {\wti Q}_{l,n} = 0 \;\;\; \; l \geq n+2
\lab{qtiqone}
\ee

The orthogonal polynomial approach leads to expression for the partition
function in terms of $h_n$ which is similar to the one-matrix case
\rf{partfctz} :
\be
Z_N = const ~\prod_{n=0}^{N-1} h_n
\lab{ZN-h}
\ee
{}From the orthogonal relation \rf{h-n} and the recursive relations 
\rf{recurpo-2M} we obtain:
\be
\partder{\ln h_n}{t_r} = \(Q^r\)_{nn} \qquad ; \qquad
\partder{\ln h_n}{{\ti t}_s} = \({\wti Q}^s\)_{nn}
\lab{hnttit-2M}
\ee
and, accordingly:  
\be
\partder{}{t_r} \ln Z_N = \sum_{n=0}^{N-1} Q^r_{n,n}   \quad ,\quad
\partder{}{\tit_s} \ln Z_N = \sum_{n=0}^{N-1} {\wti Q}^s_{n,n}
\lab{ZN-ab}
\ee

In terms of matrices $Q$ and $\tQ$ the corresponding linear problem
for the lattice wave function $\psi_n = P_n (\l)$ and the
Lax (or ``zero-curvature'') representation of the latter read \ct{BX93-94} :
\br
Q_{n,m} \psi_m &= &\l \psi_n  \qquad \;\;, \qquad \;\;
-g\tQ_{n,m} \psi_m = \partder{}{\l} \psi_n   \lab{L-1} \\
\partder{}{t_r}\, \psi_n &= & - \( Q^r_{-}\)_{nm} \psi_m    \quad , \quad
\partder{}{{\ti t}_s}\, \psi_n = - \( \tQ^s_{-}\)_{nm} \psi_m   \lab{L-2} \\
\partder{}{t_r}\, Q &= & \llb Q^r_{(+)} , Q \rrb  \qquad , \qquad
\partder{}{{\ti t}_s}\, Q = \llb Q , \tQ^s_{-} \rrb  \lab{L-3} \\
\partder{}{t_r}\, \tQ &= & \llb Q^r_{(+)} , \tQ \rrb  \qquad , \qquad
\partder{}{{\ti t}_s}\, \tQ = \llb \tQ , \tQ^s_{-} \rrb \lab{L-4}
\er
where the Toda matrices are subject to the following additional constraints
(known as ``coupling conditions'' \ct{BX93-94} or discrete
``string equation'') :
\br
\one &= & -g \llb \, Q \, ,\,  \tQ\, \rrb
\lab{string-eq-a} \\
Q_{(-)} &= & - \sum_{s=1}^{p_2 -1} \frac{(s+1)}{g} {\ti t}_{s+1} \tQ^s_{(-)}
 - {1\over g} {\ti t}_1 \one
\lab{string-eq-b} \\
\tQ_{(+)} &= & - \sum_{r=1}^{p_1 -1} \frac{(r+1)}{g} {t}_{r+1} Q^r_{(+)}
 - {1\over g} {t}_1 \one
\lab{string-eq-c}
\er
The parametrization of the matrices $Q$ and $\tQ$ is as follows:
\br
Q_{n,n} = a_0 (n) \quad , \quad Q_{n,n+1} =1 \quad ,\quad
Q_{n,n-k} = a_k (n) \quad k=1,\ldots , p_2 -1   \nonu  \\
Q_{n,m} = 0 \quad {\rm for} \;\;\; m-n \geq 2 \;\; ,\;\; n-m \geq p_2 \phanta
\lab{param-1}  \\
\tQ_{n,n} = b_0 (n) \quad , \quad \tQ_{n,n-1} = R_n \quad , \quad
\tQ_{n,n+k} = b_k (n) R_{n+1}^{-1} \cdots R_{n+k}^{-1}
\quad k=1,\ldots ,p_1 -1    \nonu  \\
\tQ_{nm} = 0 \quad {\rm for} \;\;\; n-m \geq 2 \;\; ,\;\; m-n \geq p_1
\phanta    \lab{param-2}
\er

The $Q,\tQ$-matrix elements can in general be
expressed in terms of the normalization factors $h_n$ from \rf{h-n}.
For some of the elements the relation to $h_n$ and $Z_n$ takes a very simple
form \ct{BX93-94} :
\br
a_0 (n) &= &\partder{}{t_1} \ln h_n \quad ,\quad
b_0 (n) = \partder{}{\tit_1} \ln h_n \quad ,\quad
R_n = \frac{h_n}{h_{n-1}}  
\lab{abR-h} \\
\partder{}{t_1} \ln Z_N &=& \sum_{n=0}^{N-1} a_0 (n) \quad ,\quad
\partder{}{\tit_1} \ln Z_N = \sum_{n=0}^{N-1} b_0 (n)
\lab{ZN-abnn}
\er
We now exhibit the Toda structure behind the linear problem
\rf{L-1}-\rf{L-4} focusing on the lowest flows with $r,s=1$.
We start with the second lattice equation of motion \rf{L-4} for $s=1$.
Using parametrization \rf{param-2} we obtain:
\br
\partder{}{\tit_1} R_n \eq R_n \( b_0 (n) - b_0 (n-1) \)\quad,\quad
\partder{}{\tit_1} b_0 (n)= b_1 (n) - b_1 (n-1)  
\phantom{aaa} \lab{motionx-0} \\
\partder{}{\tit_1} \( \frac{b_k (n)}{R_{n+1} \cdots R_{n+k}} \) \eq
\frac{b_{k+1}(n) - b_{k+1}(n-1)}{R_{n+1} \cdots R_{n+k}}  \quad  \;\; ,
\quad \; \; k \geq 1   \lab{motionx-k}
\er
Similarly, the first lattice equation of motion \rf{L-4} for $r=1$ gives :
\be
\partder{}{{t}_1} b_0 (n) = R_{n+1} - R_n    \quad,\quad
\partder{}{{t}_1} b_k (n) = R_{n+1} b_{k-1}(n+1) - R_{n+k} b_{k-1} (n)
\lab{motiony-k}
\ee
for $k \geq 1$. In complete analogy, the lattice equations of motion \rf{L-3}
for $r=1, s=1$ read explicitly:
\be
\partder{}{\tit_1} a_0 (n) = R_{n+1} - R_n \quad,\quad
\partder{}{\tit_1} a_k (n) = R_{n-k+1} a_{k-1}(n) - R_n a_{k-1}(n-1)
\lab{motionxx-k}
\ee
(with $k \geq 1$) and:
\be
\partder{}{{t}_1} a_0 (n) = a_1 (n+1) - a_1 (n) \quad,\quad
\partder{}{{t}_1} \( \frac{a_k (n)}{R_n \cdots R_{n-k+1}}\) =
\frac{a_{k+1}(n+1) - a_{k+1}(n)}{R_n \cdots R_{n-k+1}}
\lab{motionyy-k}
\ee
with $k \geq 1 $.
Following \rf{motionx-0}, \rf{motiony-k}, \rf{motionxx-k} we obtain the
``duality'' relations:
\be
\partder{}{{t}_1} b_1 (n) = \partder{}{\tit_1} R_{n+1}
\quad , \quad\partder{}{\tit_1} a_0 (n) = \partder{}{{t}_1} b_0 (n)
\quad , \quad \partder{}{\tit_1} a_1 (n) = \partder{}{{t}_1} R_n
\lab{dual}
\ee
{}From the above one gets the two-dimensional Toda lattice equation:
\be
\partder{}{{t}_1} \ln R_n = a_0 (n) - a_0 (n-1)
\quad \to \quad
\partder{}{t_1} \partder{}{{\ti t}_1} \ln R_n = R_{n+1}  - 2 R_{n} + R_{n-1}
\lab{motion-00}
\ee
Also, using \rf{motionyy-k}, \rf{motionx-0} and \rf{motiony-k}, we get from
\rf{ZN-ab} :
\br
\partderd{}{t_1} \ln Z_N &=& a_1 (N)    \quad ,\quad
\partderd{}{\tit_1} \ln Z_N = b_1 (N-1)  \nonu \\
\partderm{}{t_1}{\tit_1} \ln Z_N &=& R_N = \frac{h_N}{h_{N-1}} =
\frac{Z_{N+1} Z_{N-1}}{Z_N^2}
\lab{ZN2-abR}
\er
We note the two-dimensional Toda-like lattice structure exhibited by 
equation \rf{ZN2-abR}.

For later convenience, let us also display the lattice equations of motion 
for powers of $Q$ and $\tQ$ :
\br
\partder{}{t_1} Q^r = \llb Q_{(+)} , Q^r \rrb  \quad , \quad
\partder{}{{\ti t}_1} Q^r = \llb Q^r , {\tQ}_{-} \rrb  \lab{L-3-r} \\
\partder{}{t_1} {\tQ}^s = \llb Q_{(+)} , {\tQ}^s \rrb  \quad , \quad
\partder{}{{\ti t}_1} {\tQ}^s = \llb {\tQ}^s , {\tQ}_{-} \rrb \lab{L-4-s}
\er
as well as for their matrix elements :
\br
\partder{}{{\ti t}_1} \tQ^s_{nn} &=& R_{n+1} \tQ^s_{n,n+1} - R_n \tQ^s_{n-1,n}
\nonu \\
\partder{}{{\ti t}_1} \tQ^s_{n,n-k} &=& R_{n-k+1} \tQ^s_{n,n-k+1} -
R_n \tQ^s_{n-1,n-k} \quad\; k=1,\ldots, s
\lab{motionxx-k-s}  \\
\partder{}{t_1} \tQ^s_{nn} &=& \tQ^s_{n+1,n} - \tQ^s_{n,n-1}  \nonu \\
\partder{}{t_1} \( \frac{\tQ^s_{n,n-k}}{R_n \cdots R_{n-k+1}} \) &=&
\frac{\tQ^s_{n+1,n-k} - \tQ^s_{n,n-k-1}}{R_n \cdots R_{n-k+1}}  \quad
\quad \;\; k=1,\ldots, s
\lab{motionyy-k-s}
\er
Let us note that, due to the upper Jacobian form of $\tQ$ having only one
non-zero lower diagonal, the matrices $\tQ^s$ have $s$ non-zero lower
diagonals, {\sl i.e.}
\br
\tQ^s_{n,n-k} &=& 0 \quad {\rm for} \;\; k \geq s+1    \nonu \\
\tQ^s_{n,n-s} &=& R_n \cdots R_{n-(s-1)}   \quad ,\quad 
\tQ^s_{n,n-(s-1)} = R_n \cdots R_{n-(s-2)}\, \sum_{j=0}^{s-1} b_0 (n-j)
\lab{s-Jacobi}  \\
\tQ^s_{n,n-(s-2)} &=& R_n \cdots R_{n-(s-3)} \( \sum_{j=0}^{s-1} b_1 (n-j) +
\sum_{0 \leq i \leq j \leq s-2} b_0 (n-i) b_0 (n-j) \)
\nonu
\er

For completeness, let us also add the explicit expressions for the higher flow
equations for $b_0 (n)$, $ b_1 (n)$, $R_{n+1}$ resulting from \rf{L-3} and 
\rf{L-4} :
\br
\partder{}{{\ti t}_s}b_0 (n) \eq \partder{}{\tit_1} \( \tQ^s\)_{nn} 
\; \; ,\; \; 
\partder{}{{\ti t}_s}b_1 (n) =
\partder{}{\tit_1} \( R_{n+1}\( \tQ^s\)_{n,n+1}\)    
\lab{t-s-eqs} \\
\partder{}{{\ti t}_s}R_{n+1} \eq \partder{}{\tit_1} \( \tQ^s\)_{n+1,n}
\nonu \\
\partder{}{t_r}b_0 (n) \eq \partder{}{t_1} \( Q^r\)_{nn} \; \; ,\;\;
\partder{}{t_r}b_1 (n) =
\partder{}{t_1} \( R_{n+1}\( Q^r\)_{n,n+1}\) 
\lab{t-r-eqs} \\ 
\partder{}{t_r}R_{n+1} \eq  \partder{}{t_1} \( Q^r\)_{n+1,n} 
\nonu \\
\er
which easily follow by comparing equations \rf{L-3}--\rf{L-4}
with equations \rf{L-3-r}--\rf{L-4-s}, respectively,
for the main diagonal and the first upper/lower diagonal matrix elements.
\subsection{Equivalent Hierarchies Method and the String Equa\-tion}
The notion of equivalent hierarchies was introduced to answer a question
of equivalence between different formulations of integrable models.
The key role in this context is played by the particular form of coordinate
transformations on the space of evolution parameters:
\be
t_n \to {\bar t}_n = t_n + O \(t_{n+1}, t_{n+2}, \ldots \)
\lab{equa}
\ee
This coordinate transformation appeared in \ct{Tak} as a tool used to
establish an equivalence between Zakharov-Shabat hierarchy associated 
to the zero-curvature equation :
\be
  \frac{\partial B_n}{\partial t_m} - \frac{\partial B_m}{\partial t_n}
    + \sbr{B_n}{B_m} = 0 ,\; \; \qquad n, m = 1, 2, \ldots
\lab{zs-eq}
\ee
and the standard KP hierarchy formulated in terms of pseudo-differential
operators.
Because the zero-curvature equation  \rf{zs-eq} is coordinate free, 
the differential operators ${\bar B}_m $ defined by the transformation
\rf{equa} :
\be
B_n (t) = \sum_m \frac{\pa {\bar t}_m}{\pa t_n} {\bar B}_m ({\bar t})
\lab{equb}
\ee
will continue to satisfy equation  \rf{zs-eq}.
As shown in \ct{Tak}, there exists a unique transformation of the form
\rf{equa} for which the differential operator ${\bar B}_m $ takes a standard 
form  of $(L)_{+}^m$ with the Lax operator $L = D + \sum u_{i} D^{-i}$ 
being a solution of the Lax equations of the KP hierarchy with respect 
to times $ {\bar t}_n,\, n \geq 1 $ :
\be 
\frac{\pa L }{\pa {\bar t}_m} \,=\, \sbr{{\bar B}_m}{L} \qquad ,\quad
{\bar B}_m = (L)_{+}^m
\lab{equac}
\ee
The general conclusion is that the invariance of the hierarchy under 
\rf{equa} can be used to cast the differential operator hierarchy defined 
by \rf{zs-eq} into the particularly convenient standard form of the 
KP hierarchy. Similar considerations have also been made in \ct{Tak} about 
the Toda lattice hierarchy.

The method of equivalent hierarchies appeared also in the setting of the
Generalized Kontsevich Model \ct{KMMM}.

In ref.\ct{office} a similar in spirit notion was proposed to deal
with the string equation constraint in the setting of the two-matrix model
as represented by the Toda system \rf{L-1}-\rf{L-4} augmented by 
\rf{string-eq-a}-\rf{string-eq-c}.
In this reference an equivalent Toda hierarchy was found for which 
the string condition could be partially implemented by a simple 
identification of two flows.
We now expand further on these ideas.

The first step is to introduce the fractional powers of $Q$. 
\begin{definition}
Matrix $\hQ$ is defined as the fractional power of the $Q$-matrix
according to:
\be
\hQ \equiv Q^{{1\over{p_2-1}}}
\lab{hQ-def}
\ee
\label{definition:Qhatm}
\end{definition}
Let us note that it is at this point where the finiteness of the order 
$p_2$ of the second matrix potential in \rf{ZN-2M} becomes crucial.

The matrix $\hQ$ is, unlike $Q$, an {\em upper} Jacobi matrix with
only one non-zero lower diagonal, therefore, $\hQ$ has the same matrix
form as $\tQ$ and it is natural to parametrize it in the analogous way
(cf. eq.\rf{param-2}) :
\br
\hQ_{nn} = \hb_0 (n) \quad , \quad \hQ_{n,n-1} = \hR_n \quad &,& \quad
\hQ_{n,n+k} = \hb_k (n) \hR_{n+1}^{-1} \cdots \hR_{n+k}^{-1}
\quad\;\, k=1,\ldots ,p_1 -1    \nonu  \\
\hQ_{nm} &=& 0 \quad {\rm for} \;\;\; n-m \geq 2
\lab{param-2-h}
\er
Correspondingly, the matrix elements $\hQ^s_{n,n-k}$ have the same
ex\-pres\-sions as given by \rf{s-Jacobi} with $b_0 (n), b_1 (n), R_n$ 
replaced by $\hb_0 (n), \hb_1 (n), \hR_n$.

Furthermore, the lattice equations of motion for powers of $\hQ$:
$\hQ^s \equiv Q^{{s\over{p_2-1}}}$, which follow from \rf{L-3}, take the form:
\br
\partder{}{{\ti t}_1} \hQ^s_{nn} &=& R_{n+1} \hQ^s_{n,n+1} - R_n \hQ^s_{n-1,n}
\nonu \\
\partder{}{{\ti t}_1} \hQ^s_{n,n-k} &=& R_{n-k+1} \hQ^s_{n,n-k+1} -
R_n \hQ^s_{n-1,n-k} \quad k=1,\ldots s
\lab{motionxx-k-ss}  \\
\partder{}{t_1} \hQ^s_{nn}&=& \hQ^s_{n+1,n} - \hQ^s_{n,n-1}  \nonu \\
\partder{}{t_1} \( \frac{\hQ^s_{n,n-k}}{R_n \cdots R_{n-k+1}} \) &=&
\frac{\hQ^s_{n+1,n-k} - \hQ^s_{n,n-k-1}}{R_n \cdots R_{n-k+1}}  \quad
\quad k=1,\ldots, s
\lab{motionyy-k-ss}
\er
The system \rf{motionxx-k-ss}--\rf{motionyy-k-ss} is exactly of the same form
as the corresponding equations for $\tQ$ \rf{motionxx-k-s}--\rf{motionyy-k-s}.

Starting with eqs.\rf{motionxx-k-ss}--\rf{motionyy-k-ss} for $k=s$ and using
properties \rf{s-Jacobi} (and their analogs for $\hQ$),
we arrive by induction w.r.t. $k=s,s-1,\ldots ,0$ at the following result:
\begin{proposition}
The matrix elements of $Q$ are completely expressed in terms of the matrix
elements of $\tQ$ through the relations:
\be
Q^{{s\over {p_2 -1}}}_{(-)} \, = \, \sum_{\s =0}^{s} \g_{s\s} \tQ^{\s}_{(-)}
 \quad , \quad s=0,1,\ldots ,p_2
\lab{tatko-s}
\ee
where the coefficients $\g_{s,0}$ are $t_1$-independent,
whereas the coefficients $\g_{s\s}$ with $\s \geq 1$ are independent of
$t_1, {\ti t}_1$. All $\g_{s\s}$ are simply expressed through the subset
thereof:
\be
\a_s \equiv \g_{p_2 -1,s} = - \frac{s+1}{g} {\ti t}_{s+1}
\lab{alpha-s}
\ee
\label{proposition:tatko-lem}
\end{proposition}
The equality \rf{alpha-s} follows from the coincidence of eq.\rf{tatko-s}
for $s= p_2 -1$ with the ``string equation'' \rf{string-eq-b}.

Let us note that, although the system of equations \rf{tatko-s} for $Q$
is overdetermined, it is compatible by construction. A calculation yields:
\br
\g_{ss} \eq \( \g_{11}\)^s \;\quad , \;\quad
\g_{s,s-1} = s \(\g_{11}\)^{s-1} \g_{10}  \quad\;, \;\quad
\g_{11} = \( -{{p_2}\over g} \tit_{p_2} \)^{1\over {p_2 -1}} 
\lab{tatko-2-a}  \\
\g_{s,s-2} \eq \(\g_{11}\)^{s-2} \llb \frac{s(s-1)}{2} \(\g_{10}\)^2 +
s \( \frac{\g_{31}}{3\g_{11}} - \g_{10}^2 \) \rrb
\; \; ,\;\;
\g_{10} = - {1\over g} \tit_{p_2 -1}
\( -{{p_2}\over g} \tit_{p_2} \)^{-{{p_2 -2}\over {p_2 -1}}}   \nonu  \\
\frac{\g_{31}}{3\g_{11}} &-& \g_{10}^2 \, = \, - \frac{p_2 -2}{g(p_2 -1)}
\frac{1}{\( -{{p_2}\over g} \tit_{p_2} \)^{{p_2 -3}\over {p_2 -1}} } \,
\llb \tit_{p_2 -2} - \frac{p_2 -1}{2p_2} \frac{\tit_{p_2 -1}^2}{\tit_{p_2}}\rrb
\lab{tatko-2}
\er
Relation \rf{tatko-s} implies the following expressions for the 
$\hQ \equiv Q^{{1\over {p_2 -1}}}$-matrix elements:
\be
{\hat R}_n = \g_{11} R_n \quad , \quad {\hat b}_0 (n) = \g_{11} b_0 (n) +
\g_{10} \quad ,\quad {\hat b}_1 (n) = \g_{11}^2 b_1 (n) +
\frac{\g_{31}}{3\g_{11}} - \g_{10}^2  \lab{tatko-1}
\ee
etc..

The preceding discussion can be rewritten in the dual form 
with the r\^{o}les of $Q$ and $\tQ$ interchanged.
In particular, the matrix elements of $\tQ$ are completely expressed in 
terms of the matrix elements of $Q$ through the relations:
\be
\tQ^{{r\over {p_1 -1}}}_{(+)} =
\sum_{\rho =0}^{r} \eta_{r\rho} Q^{\rho}_{(+)}
\quad , \quad  r=0,1,\ldots ,p_1
\lab{tatko-r}
\ee
Here, the coefficients $\eta_{r,0}$ are ${\ti t_1}$-independent,
whereas the coefficients $\eta_{r\rho}$ with $\rho \geq 1$ are independent of
${\ti t}_1 ,t_1$. All $\eta_{r\rho}$ are expressed through the subset
thereof (cf. eq.\rf{string-eq-c}) :
\be
\b_r \equiv \eta_{p_1 -1,r} = - \frac{r+1}{g} {t}_{r+1}
\lab{beta-r}
\ee
in the same form as \rf{tatko-2} with $\g_{s\s}, \a_s, s, p_2$ replaced by
$\eta_{r\rho}, \b_r, r, p_1$, respectively.

Combining relations \rf{tatko-s} and \rf{tatko-r} lead to the following
explicit expression of $\tQ$ in terms of $Q$ or vice versa:
\br
\tQ &=& \( \a_{p_2 -1}^{-1} Q \)_{-}^{{1\over {p_2 -1}}} \,+ \,
\sum_{r=0}^{p_1 -1} \b_r Q^r_{(+)}
\lab{freib-a} \\
Q &=& \( \b_{p_1 -1}^{-1} \tQ \)_{+}^{{1\over {p_1 -1}}} \, + \,
\sum_{s=0}^{p_2 -1} \a_s \tQ^s_{(-)}
\lab{freib-b}
\er
with $\a_s ,\b_r$ as in \rf{alpha-s},\rf{beta-r}.

The next step, natural from the viewpoint of equivalent hierarchies (see
equation \rf{equa}), involves introducing a new subset of evolution 
parameters instead of $\lcurl {\ti t}_s \rcurl$.
\begin{definition}
The new subset of evolution parameters $\lcurl {\hat t}_s \rcurl$ 
is defined through the relations:
\be
\partder{}{{\hat t}_s} \equiv \sum_{\s =1}^s \g_{s\s} \partder{}{{\ti t}_{\s}}
\quad ,\;\; s=1,\ldots, p_2      \lab{tatko-a}
\ee
which imply:
\be
{\tit}_s = {\tit}_s \( \htt_s ,\htt_{s+1}, \ldots ,\htt_{p_2}\)
\lab{tt}
\ee
{\sl i.e.}, ${\tit}_s$ does not depend on $\htt_1 ,\ldots ,\htt_{s-1}$.
\label{definition:hatdef}
\end{definition}
In particular, from \rf{tatko-2-a}--\rf{tatko-2} we have a number of
technical relations (to be used later on) :
\br
\partder{}{\htt_1} &=& \g_{11} \partder{}{\tit_1}   \quad , \quad
\g_{11}^{-1} = \( -{{p_2}\over g} \tit_{p_2}\)^{-{1\over(p_2 -1)}} =
\frac{p_2}{g(p_2 -1)} \htt_{p_2}
\lab{ti-hat-rel-1} \\
\g_{10} &= &- \frac{(p_2 -1)\htt_{p_2 -1}}{p_2\htt_{p_2}}  \quad ,\quad
\frac{\g_{31}}{3\g_{11}} -  \g_{10}^2 =
- \frac{(p_2 -2)\htt_{p_2 -2}}{p_2\htt_{p_2}}
\lab{ti-hat-rel-2}  \\
{\ti t}_{p_2} &=&
-{g\over {p_2}} \( \frac{p_2}{g(p_2 -1)} \htt_{p_2}\)^{-(p_2 -1)} \quad ,\quad
{\ti t}_{p_2 -1} =
\( \frac{p_2}{g(p_2 -1)} \htt_{p_2}\)^{-(p_2 -1)} \htt_{p_2 -1} \nonu  \\
{\ti t}_{p_2 -2} &=&
\( \frac{p_2}{g(p_2 -1)} \htt_{p_2}\)^{-(p_2 -2)} \htt_{p_2 -2}
- \( \frac{p_2}{g(p_2 -1)} \htt_{p_2}\)^{-(p_2 -1)}
\frac{p_2 -1}{2g}\htt_{p_2 -1}^2
\lab{ti-hat-rel-3}
\er

Taking into account equation \rf{tatko-s}, we arrive at the following result:
\begin{proposition}
All constrained Toda lattice eqs.\rf{L-3}--\rf{string-eq-c} can be re-expressed
as a
single set of flow equations for one independent matrix $Q$ only:
\br
\partder{}{{\htt}_s} Q = \llb \, Q \, ,\, Q^{{s\over {p_2 -1}}}_{-} \, \rrb 
\quad \;,\; \;\; s&=&1,\ldots,p_2 , 2(p_2 -1), 3(p_2 -1), \ldots, p_1 (p_2 -1)
\lab{L-3-h} \\
t_r &\equiv& {\hat t}_{r(p_2 -1)} \quad {\rm for} \;\; r=2,\ldots , p_1
\lab{identif}
\er
where the ``string equation'' constraints \rf{string-eq-a}--\rf{string-eq-b}
become, taking into account \rf{freib-a} :
\be
\partder{}{t_1} Q = \partder{}{\htt_{p_2 -1}} Q \quad , \quad
\(\sum_{r=1}^{p_1 -1} (r+1) {t}_{r+1} \partder{}{t_r} +
\frac{p_2}{(p_2 -1)} \htt_{p_2}\partder{}{\htt_1}\) Q = - \one
\lab{string-eq-hat}
\ee
\label{proposition:identil}
\end{proposition}
Similar (dual) equations to \rf{L-3-h}--\rf{string-eq-hat} can, of course, 
be obtained for $\tQ$ as an independent matrix.

Also, let us write down for the later use the $\htt_1$-lattice equations of
motion for $\hQ = Q^{1\over{p_2 -1}}$ resulting from \rf{L-3-h}
(cf. the analogous $\tQ$ lattice eqs.\rf{motionx-0}) :
\br
\partder{}{\htt_1} \hQ^{p_2 -1}_{nn} & = & \hR_{n+1} - \hR_n
= \partder{}{\htt_{p_2 -1}} \hb_0 (n)
\lab{motionx-0-h1}   \\
\partder{}{\htt_1} \hR_n &=& \hR_n \( \hb_0 (n) - \hb_0 (n-1) \)  \quad ,\quad
\partder{}{\htt_1} \hb_0 (n)= \hb_1 (n) - \hb_1 (n-1)    \lab{motionx-0-h2}
\er
Similarly, for the higher flows of $\hQ$ we get the complete analogs of
eqs.\rf{t-s-eqs}--\rf{t-r-eqs} :
\br
\partder{}{{\htt}_s}\hb_0 (n) \eq  \partder{}{\htt_1} \( \hQ^s\)_{nn} 
\quad ,\quad  \partder{}{{\htt}_s} \hb_1 (n) =
\partder{}{\htt_1} \( \hR_{n+1}\( \hQ^s\)_{n,n+1}\) 
\lab{t-s-eqs-h} \\
\partder{}{{\htt}_s} \hR_{n+1} \eq  \partder{}{\htt_1} \( \hQ^s\)_{n+1,n}
s= 1,\ldots,p_2 , 2(p_2 -1), 3(p_2 -1), \ldots, p_1 (p_2 -1)   \nonu
\er
The last equality in \rf{motionx-0-h1} follows from first equation
in \rf{t-s-eqs-h}.

Comparing the constrained Toda lattice representation
\rf{L-3-h}--\rf{string-eq-hat} for the two-matrix model with that for
the one-matrix model (eqs.\rf{L-3-1M},\rf{1streq}), we notice their formal
resemblance up to the following differences:

\begin{itemize}
\item
$Q$ in the two-matrix-model case has $p_2 -1 >1$ non-zero lower diagonals
(this is, of course, due to the nonlocal measure in the orthogonal polynomial
formalism, cf. \rf{h-n}).

\item
In the two-matrix-model case we have, besides $\lcurl t_r \rcurl$, an
additional subset of {\em fractional} flows $\lcurl \htt_s \rcurl$.

\item
The fractional flows contribute an additional term to the ``string equation''
(cf. \rf{1streq} {\sl versus} \rf{string-eq-hat}).
\end{itemize}

\subsection{Construction of the Correspon\-ding Con\-tinuum KP Hie\-rar\-chy
with the String Equa\-tion Constraint}
The linear auxiliary problem associated with \rf{L-3-h}--\rf{string-eq-hat} 
reads:
\br
{Q}_{nm} {\psi}_m &=& \l {\psi}_n  \quad , \quad
\partder{}{{\htt}_s} {\psi}_n = - \({Q}^{{s\over {p_2 -1}}}_{-}\)_{nm} {\psi}_m
\quad , \quad \partder{}{\htt_{p_2 -1}} \psi_n = \partder{}{t_1} \psi_n 
\lab{L-hat} \\ 
-{1\over g} \partder{}{\l} \psi_n ( \eq \tQ_{nm} \psi_m ) =
\( \( \a_{p_2 -1}^{-1} Q \)_{-}^{{1\over {p_2 -1}}} +
\sum_{r=0}^{p_1 -1} \b_r Q^r_{(+)}\)_{nm} \psi_m
\lab{M-hat}
\er
In components, using the parametrization of \rf{param-1} and 
\rf{param-2-h}, we have:
\br
\l \psi_n &=& \psi_{n+1} + a_0 (n) \psi_n + \sum_{k=1}^{p_2 -1}a_k
(n)\psi_{n-k}
\lab{3-1-2M} \\
\partder{}{\htt_1} \psi_n &=& - \hR_n \psi_{n-1}
\lab{3-3-2M}
\er

Applying again the Bonora-Xiong procedure to \rf{3-1-2M},\rf{3-3-2M},\rf{M-hat} 
as in derivation of eqs.\rf{specb-n}--\rf{speca-n} in the one-matrix-model
case,
we obtain from \rf{L-3-h}--\rf{M-hat} the following continuum Lax problem at a
fixed lattice site $n$ (with continuum ``space'' coordinate
$x \equiv \htt_1$) :
\br
\l {\psi}_n &=& {\hat L}(n) {\psi}_n \quad ,\quad
\partder{}{{\htt}_s} {\psi}_n = - {\hat \cL}_s (n) {\psi}_n \quad ,\quad
-{1\over g} \partder{}{\l} \psi_n = {\hat M}(n) \psi_n
\lab{L-cont-h} \\
\partder{}{{\hat t}_s} {\hat L}(n) \eq \Sbr{{\hat L}(n)}{{\hat \cL}_s (n)}
\quad , \;\; s=1,\ldots,p_2 , 2(p_2 -1), 3(p_2 -1), \ldots ,p_1 (p_2 -1)
\lab{L-cont-1-h}  \\
\partder{}{t_1} L(n) \eq \partder{}{\htt_{p_2 -1}} L(n)    \quad ,\quad
{1\over g}\one = \Sbr{{\hat M}(n)}{{\hat L}(n)}
\lab{string-eq-cont}
\er
where:
\br
{\hat L}(n) \!\! &\equiv&\! \! a_0 (n) + \sum_{k=1}^{p_2 -1}
\frac{(-1)^k a_k (n)}{{\hat R}_n\cdots {\hat R}_{n-k+1}}
\Bigl( D_x - \pa_x \ln \( {\hat R}_n \cdots {\hat R}_{n-k+2}\)\Bigr)
 \cdots  \Bigl( D_x - \pa_x \ln {\hat R}_n \Bigr)\, D_x
 \nonu\\
&-& D_x^{-1}\, {\hat R}_{n+1} 
\lab{Ln-x} \\
{\hat {\cL}}_s (n) \! &\equiv&\! \sum_{k=1}^s
\frac{(-1)^{k} {Q}_{n,n-k}^{s\over{p_2 -1}}}{{\hat R}_n\cdots {\hat R}_{n-k+1}}
\Bigl( D_x - \pa_x \ln \( {\hat R}_n \cdots {\hat R}_{n-k+2}\)\Bigr)
\cdots D_x
\lab{Ls-x}  \\
{\hat M}(n) \! &\equiv&\! - \g_{11}^{-1} D_x  -{1\over g}t_1 +
\sum_{r=1}^{p_1 -1} \b_r \({\hat L}^r (n) - {\hat \cL}_{r(p_2 -1)} (n)\)
\lab{Mn-x}
\er
Here all coefficients can simply be expressed in terms of the matrix
elements ${\hR}_{n+1},{\hb}_0 (n)$, $\hb_1 (n),\ldots $
of $Q^{1\over {p_2 -1}}$ \rf{param-2-h} at a fixed site $n$ through the
${\hat t}_1 \equiv x$ lattice equations of motion \rf{L-3-h}.

More transparent form for the constrained continuum hierarchy
\rf{L-cont-1-h}--\rf{string-eq-cont} is obtained
after performing a suitable gauge transformation and operator conjugation:
\br
L(n) \!\!&\equiv&\!\!
e^{\int {\hb}_0 (n)} \({\hat L}(n)\)^{\ast} e^{-\int {\hb}_0 (n)} =
D_x^{p_2 -1} + \( p_2 -1\) {\hb}_1 (n) D_x^{p_2 -3} + \cdots + \nonu \\
&+& {\hR}_{n+1} \( D_x - {\hb}_0 (n) \)^{-1} 
\lab{Ln}  \\
\cL_s (n) &\equiv&
e^{\int {\hb}_0 (n)} \({\hat \cL_s}(n)\)^{\ast} e^{-\int {\hb}_0 (n)}
+ \partder{}{\htt_s} \int \hb_0 (n) = \( L(n)\)^{s\over {p_2 -1}}_{(+)}
\lab{Ls} \\
M(n) &\equiv&
e^{\int {\hb}_0 (n)} \({\hat M}(n)\)^{\ast} e^{-\int {\hb}_0 (n)} =
{1\over g} \llb \htt_{p_2 -1} +
\frac{p_2}{(p_2 -1)} \htt_{p_2} \( L(n)\)^{1\over{p_2 -1}}_{(+)}+ \right. 
\lab{Mn}\\ 
&+& \left. \sum_{r=1}^{p_1 -1} (r+1) t_{r+1} \( L(n)\)^r_{(+)} \rrb  
+ O\bigl( L(n)\bigr) \nonu
\er
In derivation of the above equations the expressions
\rf{ti-hat-rel-1}--\rf{ti-hat-rel-2} were used.
The last term $O\bigl( L(n)\bigr)$ in \rf{Mn} may be discarded
without changing the formalism.

Correspondingly, \rf{L-cont-1-h}--\rf{string-eq-cont} acquire the form:
\br
\partder{}{{\hat t}_s} L(n) \eq
\Sbr{\( L^{s\over {p_2 -1}}(n)\)_{(+)}}{L (n)}
\;, \; s=1,\ldots,p_2 , 2(p_2 -1), 3(p_2 -1), \ldots ,p_1 (p_2 -1)
\phantom{aaa} \lab{Lax-2M}  \\
\partder{}{t_1} L(n) \eq  \partder{}{\htt_{p_2 -1}} L(n)    \quad ,\quad
\Sbr{L(n)}{\frac{p_2}{(p_2 -1)} \htt_{p_2}\partder{}{\htt_1} +
\sum_{r=1}^{p_1 -1} (r+1) t_{r+1} \( L(n)\)^r_{(+)}} = \one
\lab{string-eq-2M}
\er
The continuum ``string equation'' constraints \rf{string-eq-2M}, upon using
the flows from equations \rf{Lax-2M}, become
(cf. the discrete ``string equations'' \rf{string-eq-hat}) :
\be
\partder{}{t_1} L(n) = \partder{}{\htt_{p_2 -1}} L(n)    \quad ,\quad
\(\sum_{r=1}^{p_1 -1} (r+1) t_{r+1} \partder{}{t_r} +
\frac{p_2}{(p_2 -1)} \htt_{p_2}\partder{}{\htt_1}\) L(n) = - \one
\lab{string-eq-hat-cont}
\ee
or, in terms of the original evolution parameters using \rf{ti-hat-rel-1}
and eq.\rf{tatko-a} for $s=p_2-1$ together with \rf{alpha-s} :
\br
\(\sum_{s=1}^{p_2 -1} (s+1) \tit_{s+1} \partder{}{\tit_s} +
g\partder{}{t_1}\) L(n) &= &- \one       \nonu \\
\(\sum_{r=1}^{p_1 -1} (r+1) t_{r+1} \partder{}{t_r} +
g\partder{}{\tit_1}\) L(n) &= & - \one
\lab{string-eq-hat-cont-1}
\er

Recalling \rf{f-5} (or \rf{iss-8b}), we notice that \rf{Ln}
is nothing but the Lax operator for the ${\sf cKP}_{p_2-1,1}$ hierarchy
which now is subject to the additional ``$M$-constraints''
\rf{string-eq-hat-cont}. The latter, in turn, are special case of 
the condition for invariance under the lowest additional symmetry flow
\rf{L-constr} with
the order of $L$ being $m = p_2 -1$ and the index $l \equiv s$ running over
the set of non-zero evolution parameters as in eq.\rf{Lax-2M}
(recall the identification $t_r \equiv {\htt}_{r(p_2 -1)}$ \rf{identif}).

We shall denote in the sequel as ${\sf scKP}_{p_2-1,1}$ the
${\sf cKP}_{p_2-1,1}$ Lax system \rf{Lax-2M},\rf{Ln} with the additional 
``string equation'' constraint \rf{string-eq-2M}. 
As shown above, it is the exact continuum analog
of the constrained Toda lattice Lax system \rf{L-3}--\rf{string-eq-c}
describing the two-matrix model \rf{ZN-2M}.
\subsection{Partition Function, the Toda Lat\-tice Struc\-ture}
Before presenting the explicit formula for the partition function $Z_N$
\rf{ZN-2M} in terms of the $\t$-function for the 
{\sf scKP}$_{p_2-1,1}$ hierarchy, let us discuss the
r\^{o}le of the Toda lattice site label $n$ in the context of the continuum
{\sf scKP}$_{p_2-1,1}$ hierarchy. First, we observe that, by construction, the
coefficients of the Lax operator $L(n)$ \rf{Ln} satisfy the \scKP$_{p_2-1,1}$
system \rf{Lax-2M}--\rf{string-eq-2M}~ {\em for any}~ $n=0,1,2 \ldots$ with
the following ``boundary'' conditions due to the semi-infiniteness of the
underlying Toda matrices $Q,\tQ$ and accounting for their definition
\rf{tatko-1} :
\br
\hR_0 \eq 0 \quad ,\quad
\hb_0 (-1) = \g_{10} = - \frac{(p_2 -1) \htt_{p_2 -1}}{p_2 \htt_{p_2}}
\nonu \\
\hb_1 (-1) \eq \frac{\g_{31}}{3\g_{11}} - \g_{10}^2 =
- \frac{(p_2 -2)\htt_{p_2 -2}}{p_2\htt_{p_2}}
\lab{bound-cond}
\er
where the expressions \rf{ti-hat-rel-1}--\rf{ti-hat-rel-2} were used.
Therefore, Toda lattice shifts $n \to n+1$ are equivalent to mappings
of one solution of the \scKP$_{p_2-1,1}$ hierarchy into another one of the
same hierarchy, {\sl i.e.}, Toda lattice shifts must correspond to auto-\Back
transformations for the \scKP$_{p_2-1,1}$ hierarchy. Here we shall show that,
indeed, Toda lattice shifts within the discrete constrained integrable
hierarchy \rf{L-3-h}--\rf{string-eq-hat} generate special (constrained)
\DB transformations for the \scKP$_{p_2-1,1}$ system 
\rf{Lax-2M}--\rf{string-eq-2M}.

To this end let us rewrite $L(n)$ \rf{Ln} in the equivalent
``eigenfunction'' form:
\br
L(n) &=& D_x^{p_2 -1} + \( p_2 -1\) {\hb}_1 (n) D_x^{p_2 -3} + \cdots +
\Phi (n+1) D_x^{-1} \Psi (n+1)
\lab{Ln1} \\
\Phi (n+1) &\equiv& \hR_{n+1} \exp \Bigl\{ \int \hb_0 (n) \Bigr\} \quad ,\quad
\Psi (n+1) \equiv \exp \Bigl\{ -\int \hb_0 (n) \Bigr\}
\lab{Phi-n}
\er
The $\hQ$ lattice equations of motion \rf{motionx-0-h1}--\rf{motionx-0-h2}
yield the following recurrence relations for the eigenfunctions
$\Phi (n),\Psi (n)$  \rf{Phi-n} and $\t$-function (recall 
prop.\ref{proposition:dbtau}) of \scKP$_{p_2-1,1}$ :
\br
\partder{}{\htt_1} \partder{}{\htt_{p_2 -1}} \ln \Phi (n) \eq
\frac{\Phi (n+1)}{\Phi (n)} - \frac{\Phi (n)}{\Phi (n-1)} 
\lab{Phi-n-rec}  \\
\Psi (n+1) \eq \( \Phi (n)\)^{-1} \quad \to \quad 
\Phi (n) = \exp \Bigl\{ \int \hb_0 (n) \Bigr\}
\lab{Phi-n-1} \\
\frac{\t (n)}{\t (n-1)} = \exp \Bigl\{ \int \hb_0 (n) \Bigr\} \eq \Phi (n) 
\quad \to \quad
\frac{\t (n)}{\t (-1)} = \Phi (n) \Phi (n-1) \cdots \Phi (0)
\lab{tau-Phi}
\er
Here $\t (-1)$ denotes the $\t$-function of the ``initial''
Lax operator $L(-1)$,
{\sl i.e.}, $L(n)$ \rf{Ln1} for $n=-1$, which is {\em pure differential}
due to the ``boundary'' conditions \rf{bound-cond} :
\be
L(-1) = D_x^{p_2 -1}
- \frac{(p_2 -1)(p_2 -2)\htt_{p_2 -2}}{p_2\htt_{p_2}} D_x^{p_2 -3} + \cdots
\lab{L1-1}
\ee
\begin{lemma}
In terms of the original $\lcurl \tit_s \rcurl$ flow parameters 
the ``initial'' Lax operator $L(-1)$ becomes:
\be
L(-1) \,= \, \sum_{s=0}^{p_2 -1} \( -\frac{s+1}{g}\tit_{s+1}\)
\(\partder{}{\tit_1} - \frac{\tit_{p_2 -1}}{p_2 \tit_{p_2}}\)^s
\lab{L1-1-ti}
\ee
\label{lemma:L1lemma}
\end{lemma}
\begin{proof}
The lemma follows easily from \rf{Ln-x} and \rf{Ln} upon using the original
``coupling conditions'' \rf{string-eq-b} to express the coefficients of
${\hat L}(n)$ as (using the short-hand notations \rf{alpha-s},\rf{tatko-2}) :
\be
\frac{a_k (n)}{{\hat R}_n\cdots {\hat R}_{n-k+1}} = \g_{11}^{-k} \a_k +
\sum_{s=k+1}^{p_2 -1} \a_s
\frac{\tQ^s_{n,n-k}}{{\hat R}_n\cdots {\hat R}_{n-k+1}} =
\g_{11}^{-k} \a_k  \quad {\rm for} \;\; n=-1
\lab{limit-1}
\ee
and subsequently plugging it into (recall $x \equiv \htt_1$) :
\be
L(-1) = e^{\g_{10} \htt_1} \( \sum_{s=0}^{p_2 -1} \a_s \g_{11}^{-s} D^s \)
e^{-\g_{10} \htt_1}
\lab{L-1-2M-psi}
\ee
\end{proof}

Comparing \rf{Phi-n-rec}--\rf{tau-Phi} with the general formulas of \DB
transformations for generic ${\sl cKP}_{r,m=1}$ hierarchies
\rf{shabes-2},\rf{rec-rel-n},\rf{sol-3-aa} allows us to identify
constrained Toda lattice site shifts with \DB transformations for the
\scKP$_{p_2-1,1}$ systems as follows:
\br
L(n) \eq T(n) L(n-1) T^{-1}(n)  \quad ,\quad
T(n) = \Phi (n) D_x \Phi^{-1} (n) = D_x - \hb_0 (n)
\lab{DB-2M} \\
\Phi (n+1) \eq \( T(n) L(n-1)\) \Phi (n) =
T(n) \cdots T(0) \( L(-1)^{n+1} \Phi (0) \)  \; \,,\;\,
\Psi (n+1) = \( \Phi (n)\)^{-1} \nonu 
\er

According to the general formulas \rf{pchi-a-tr},\rf{tauok-tr},
the solutions for the $\t$-function and (adjoint-)eigenfunctions of $L(n)$ 
\rf{Ln1} and, therefore, for the matrix elements of
$\hQ \equiv Q^{1\over {p_2 -1}}$ (cf. \rf{Phi-n}) are expressed
entirely in terms of the ``initial'' eigenfunction $\Phi (0)$ of the
``initial'' Lax operator $L(-1)$ \rf{L1-1}
\br
\Phi (n)&=&\frac{W_{n+1} \llb \Phi (0), \partder{}{t_1}\Phi (0), \ldots, 
\partderh{}{t_1}{n} \Phi (0) \rrb}{W_n \llb \Phi (0),\partder{}{t_1}\Phi (0),
\ldots, \partderh{}{t_1}{n-1} \Phi (0) \rrb}
\lab{Phi-W-n} \\
\frac{\t (n)}{\t (-1)} &=&  \prod_{j=0}^n \Phi (j) =
W_{n+1} \llb \Phi (0), \partder{}{t_1}\Phi (0), \ldots , 
\partderh{}{t_1}{n}\Phi (0) \rrb \nonu \\
 &=& \det {\Bigl\Vert}
\partderM{\Phi (0)}{i+j -2}{\htt_1}{i-1}{t_1}{j-1} {\Bigr\Vert}
\lab{tau-W-n}   \\
\partder{}{\htt_s} \Phi (0)\!\! &=&\!\! \( L(-1)\)^{s\over{p_2 -1}}_{(+)} 
\Phi (0) \lab{P0-eigen} \\
 s&=&1,\ldots,p_2 ,2(p_2 -1),3(p_2 -1),\ldots , p_1 (p_2 -1) \nonu
\er
where we used the constraint
$\partder{}{t_1} \Phi (0) = \partder{}{\htt_{p_2 -1}} \Phi (0)\,$ following
from the first constraint eq.\rf{string-eq-hat-cont}.
Furthermore, the ``string equation'' constraints \rf{string-eq-2M}
(or \rf{string-eq-hat-cont}) imply the following additional conditions
for the Lax eigenfunctions $\Phi (n)$, which generate the \DB transformations
\rf{DB-2M} :
\br
&&\(\sum_{r=1}^{p_1 -1} (r+1) t_{r+1} \partder{}{t_r} + t_1 + \htt_{p_2 -1} +
\frac{p_2}{(p_2 -1)} \htt_{p_2}\partder{}{\htt_1}\) \Phi (n) = 0
\nonu \\
&&\partder{}{t_1} \Phi (n) =  \partder{}{\htt_{p_2 -1}} \Phi (n)
\lab{DB-constr}
\er
where the second equality accounts for the identification
$t_r \equiv \htt_{r(p_2 -1)}$ \rf{identif}.
The above equations are nothing but a special case (for the current 
set of evolution parameters \rf{Lax-2M}--\rf{P0-eigen}) of the 
condition \rf{P-constr} for compatibility between additional symmetries 
and \DB transformations of general KP hierarchies.

To write down more explicit expression for $\Phi (0)$ \rf{P0-eigen} ,
we shall use the representation \rf{spectr-constr} of
constrained eigenfunctions in terms of constrained BA functions (satisfying
\rf{psi-constr}) which in the case under consideration reads (using the
short-hand notations \rf{alpha-s},\rf{tatko-2}) :
\br
\Phi (0) &=&  \int_{} d\l \, \psi^{(0)} (\l )
\lab{Phi-psi-0}  \\
L(-1) \psi^{(0)} (\l ) \!\!&=& \!\! \l \psi^{(0)} (\l ) \;\;,\;\, i.e. \;\;
e^{\g_{10} \htt_1} \( \sum_{s=0}^{p_2 -1} \a_s \g_{11}^{-s} D^s \)
e^{-\g_{10} \htt_1} \psi^{(0)} (\l ) = \l \psi^{(0)} (\l ) \phantom{aaa}
\lab{L-1-psi}  \\
\partder{}{\htt_s} \psi^{(0)} (\l )  &=& 
\( L(-1)\)^{s\over{p_2 -1}}_{(+)} \psi^{(0)} (\l ) \quad ,\;\;\; s=1,\ldots,p_2
\lab{L-s-psi}  \\
\partder{}{t_r} \psi^{(0)} (\l ) &=&  \( L(-1)\)^r \psi^{(0)} (\l ) =
\l^r \psi^{(0)} (\l )
\lab{L-r-psi}  \\
\partder{}{t_1} \psi^{(0)} (\l ) &=& \partder{}{\htt_{p_2 -1}} \psi^{(0)} (\l )
\nonu \\
\partder{}{\l} \psi^{(0)} (\l )&=&
\(\sum_{r=1}^{p_1 -1} (r+1) t_{r+1} \partder{}{t_r} + t_1 + \htt_{p_2 -1} +
\frac{p_2}{(p_2 -1)} \htt_{p_2}\partder{}{\htt_1}\) \psi^{(0)} (\l ) 
\lab{psi-0-constr}
\er
The solution for $\psi^{(0)} (\l )$ can be written in the form:
\br
\psi^{(0)} (\l ) &=& e^{\sum_{r \geq 1} \l^r t_r + \l \htt_{p_2 -1}}
\int_{} d{\ti \m} \,
\exp \lcurl {\ti \m} \( \htt_1 + \frac{p_2}{p_2 -1}\htt_{p_2}\l \)\rcurl
{\ti f}^{(0)}({\ti \m}; \{ \htt^{\pr}\} )   \nonu  \\
&=& \g_{11} e^{\sum_{r \geq 1} \l^r t_r + \g_{10} \htt_1} \int_{} d\m \,
e^{\( g\l + \g_{11} \htt_1\)\m} f^{(0)} (\m ; \{ \htt^{\pr}\} )
\lab{psi-0-sol-0}
\er
where $\{ \htt^{\pr}\} \equiv (\htt_2 ,\ldots ,\htt_{p_2})$,
and we performed change of integration variable
${\ti \m} = \g_{11} \m + \g_{10}$ to pass from the first to the second
equality in \rf{psi-0-sol-0}. Now, inserting the last expression
\rf{psi-0-sol-0} into \rf{L-1-psi} reduces the latter to an equation for
$f^{(0)} (\m ; \htt^{\pr})$ :
\be
\partder{}{\m} f^{(0)} (\m ; \{ \htt^{\pr}\} ) =
\llb \sum_{s=1}^{p_2 -1} (s+1) \tit_s \m^s + (\tit_1 - \g_{11} \htt_1 )\rrb
f^{(0)} (\m ; \{ \htt^{\pr}\} )
\lab{f-0-eq}
\ee
Notice, that due to $x \equiv \htt_1$-independence of 
$\g_{11} = \partder{}{\htt_1} \tit_1$ as given in \rf{ti-hat-rel-1},
the last term in the square brackets does not depend on $x \equiv \htt_1$ .
Substituting the solution of eq.\rf{f-0-eq} into \rf{psi-0-sol-0} we obtain:
\be
\psi^{(0)} (\l ) = \,e^{\vareps (\{\htt^{\pr}\} ) + \g_{10} \htt_1}
\, e^{\sum_{r \geq 1} \l^r t_r } \int_{} d \m\,
e^{g \l\m + \sum_{s=1}^{p_2} \m^s \tit_s }
\lab{psi-sol-1}
\ee
where the ``integration constant'' $\vareps (\{\htt^{\pr}\} )$ from \rf{f-0-eq}
is determined through the $\htt_s$-flow eqs.\rf{L-s-psi} :
\be
\partder{}{\htt_s} \vareps (\{\htt^{\pr}\} ) =
\( \sum_{\s =1}^{p_2 -1} \Bigl( - \frac{\s +1}{g} \tit_{\s +1}\Bigr)
\Bigl( - \frac{p_2}{g} \tit_{p_2}\Bigr)^{-{\s \over {p_2 -1}}} D^\s
- {1\over g} \( \tit_1 - \g_{11} \htt_1\) \)^{s\over {p_2 -1}}_{(0)}
\lab{vareps-eq}
\ee
Here the subscript ${}_{(0)}$ denotes taking the zero-order term of the
corresponding pseudo-differential operator which, moreover, has constant
$x \equiv \htt_1$-independent coefficients. Then, for the eigenfunction
\rf{Phi-psi-0} we get:
\br
\Phi (0) &=& e^{\vareps (\{\htt^{\pr}\} ) + \g_{10} \htt_1}
\int_{} \int_{} d\l_1 d\l_2 \,
\exp \lcurl \sum_{r=1}^{p_1} \l_1^r t_r +
\sum_{s=1}^{p_2} \l_2^s \tit_s + g \l_1 \l_2 \rcurl \lab{h-Phi-0}\\
&=& \exp \lcurl -\htt_1 \frac{(p_2 -1) \htt_{p_2 -1}}{p_2 \htt_{p_2}}
+ \vareps (\htt^{\pr}) \rcurl\, h_0
\nonu
\er
where the expression \rf{h-n} was used.

In complete analogy to \rf{h-Phi-0} we can find the relation between
$h_n$ -- the normalization factors in the orthogonal
polynomial formalism \rf{h-n}, and the Lax eigenfunctions $\Phi (n)$ in the
\scKP$_{p_2-1,1}$ formalism for any $n$. Namely, consider
eq.\rf{Phi-n-1} and compare it with the second eq.\rf{abR-h} for $h_n$
which yields (accounting for \rf{tatko-1}) :
\be
h_n = \Phi (n) \exp \lcurl - \htt_1 \,\g_{10} - {\bar \vareps}_n (\htt^{\pr})
\rcurl
\lab{h-Phi}
\ee
where again ${\bar \vareps}_n (\htt^{\pr})$ are ``integration constants''.
Employing the lattice equations of motion for $h_n$ :
\be
\partderm{}{t_1}{\tit_1} \ln h_n = \frac{h_{n+1}}{h_n} - \frac{h_n}{h_{n-1}}
\lab{h-n-eq}
\ee
which follow from \rf{motiony-k} and \rf{abR-h},
fixes the ``integration constants''
${\bar \vareps}_n (\htt^{\pr}) = n \,\ln \g_{11} + \vareps (\htt^{\pr})\,$
up to an overall $n$-independent function $\vareps (\htt^{\pr})$ which,
therefore, must coincide with $\vareps (\htt^{\pr})$ from
eqs.\rf{vareps-eq}--\rf{h-Phi-0}. Thus, we get:
\be
\Phi (n) = h_n \,
\exp \lcurl -\htt_1 \frac{(p_2 -1) \htt_{p_2 -1}}{p_2 \htt_{p_2}}
+ \vareps (\htt^{\pr}) \rcurl \(\frac{p_2}{g(p_2 -1)} \htt_{p_2}\)^{-n}
\lab{h-Phi-n}
\ee

Now, substituting the expressions \rf{h-Phi-n} into eq.\rf{ZN-h} and taking
into account \rf{tau-W-n}, we arrive at the final expression
for $Z_N$ as anticipated in \rf{Wronski-2M-MM}--\rf{h0-rho} :
\br
Z_N &=& \frac{\t (N-1)}{\t (-1)} \,
\exp \lcurl N \(\htt_1 \frac{(p_2 -1) \htt_{p_2 -1}}{p_2 \htt_{p_2}}
- \vareps (\htt^{\pr}) \) \rcurl
\(\frac{p_2}{g(p_2 -1)} \htt_{p_2}\)^{-\frac{N(N-1)}{2}}
\lab{ZN-det-0} \\
\!\!\!\!\! &= &\! \det {\Bigl\Vert}
\partderM{\Phi (0)}{i+j -2}{\tit_1}{i-1}{t_1}{j-1} {\Bigr\Vert}
\exp \lcurl N \(\htt_1 \frac{(p_2 -1) \htt_{p_2 -1}}{p_2 \htt_{p_2}}
- \vareps (\htt^{\pr}) \) \rcurl \nonu \\
&=& \det {\Bigl\Vert}
\partderM{h_0}{i+j -2}{\tit_1}{i-1}{t_1}{j-1} {\Bigr\Vert} 
\lab{ZN-det}
\er
To get the last equality we used again \rf{h-Phi-0} as well as the chain of
simple identities (recall second eq.\rf{DB-constr},\rf{P0-eigen} and the
``similarity'' form of the initial Lax operator \rf{L-1-psi}) :
\br
e^{- \( \htt_1 \g_{10} + \vareps (\htt^{\pr})\)} \partderh{}{t_1}{k} \Phi (0)
&=& e^{- \(\htt_1 \g_{10} + \vareps (\htt^{\pr})\)} \( L(-1)\)^k \Phi (0) =
\( e^{- \htt_1 \g_{10}} L(-1) e^{\htt_1 \g_{10}}\)^k h_0  \nonu  \\
&=& \( - \sum_{s=1}^{p_2 -1} \frac{s+1}{g} \tit_{s+1} \partderh{}{\tit_1}{s}
- {1\over g} \tit_1 \)^k h_0 = \partderh{}{t_1}{k} h_0
\lab{t-k-Phi}
\er

The last equality in \rf{ZN-det} was previously obtained \ct{Morozov}
from different approaches.

Thus, we conclude from \rf{ZN-det-0}--\rf{ZN-det} that the solution of
the discrete two-matrix string model \rf{ZN-2M} is provided  
(up to an overall ``coupling constant'' dependent factor) by the $\t$-function
of simple DB orbits of  the integrable \scKP$_{p_2-1,1}$ hierarchy 
\rf{Ln},\rf{Lax-2M}--\rf{string-eq-2M} with finite number of flows
(when $p_1$ is finite).
Furthermore, based on arguments parallel to those leading to
eq.\rf{ZN-1M-wti} in the one-matrix model case, we can identify $\t (N-1)$
\rf{tau-W-n} for generic initial eigenfunctions $\P (0)$, corresponding to
more general than \rf{L1-1-ti} initial Lax operators and not subject to
``string equation'' constraint \rf{DB-constr}, as a partition function of a 
generalized two-matrix model with matrix potentials of non-polynomial type.
\lskip
\underbar{\sl Example: ${\bf p_2 =3}$ case.}
~Let us illustrate the above general formulas for the simplest nontrivial case
when the second two-matrix model potential in \rf{ZN-2M} is of cubic order,
{\sl i.e.}, $p_2 = 3$. The corresponding \scKP$_{2,1}$ Lax formulation
specializes as follows (recall $x \equiv \htt_1$) :
\br
L(n) &=& D^2 + 2 \hb_1 (n) + \hR_{n+1} \( D - \hb_0 (n)\)^{-1} \quad , \quad
L(-1) = D^2 - \frac{2\htt_1}{3\htt_3}
\lab{Ln1-3} \\
\tit_3 \! &=& \!\! -{g\over 3} \(\frac{3\htt_3}{2g}\)^{-2}      \;\;,\;\;
\tit_2 = \(\frac{3\htt_3}{2g}\)^{-2} \htt_2      \;\; ,\;\;
\tit_1 = \(\frac{3\htt_3}{2g}\)^{-1} \htt_1 -
\(\frac{3\htt_3}{2g}\)^{-2} {1\over g}\htt_2^2 \phantom{aa}
\lab{ti-hat-rel-3-3} \\
\htt_3\! &=& \! {{2g}\over 3} \(-\frac{3\tit_3}{g}\)^{-\h}     \;\;, \quad
\htt_2 = \(-\frac{3\tit_3}{g}\)^{-1} \tit_2         \;\;,\quad
\htt_1 = \(-\frac{3\tit_3}{g}\)^{-\h} \(\tit_1 - \frac{\tit_2^2}{3\tit_3}\)
\lab{ti-hat-rel-3-3i}
\er
Eqs.\rf{vareps-eq} specialize to:
\be
\partder{}{\htt_2} \vareps (\htt_2 ,\htt_3 ) =
\(\frac{2\htt_2}{3\htt_3}\)^2    \;\; , \quad
\partder{}{\htt_3} \vareps (\htt_2 ,\htt_3 ) =
- \(\frac{2\htt_2}{3\htt_3}\)^3    \;\;  \;\, \to \quad \;
\vareps (\htt_2 ,\htt_3 ) = {1\over 6} \frac{\( 2\htt_2\)^3}{\( 3\htt_3\)^2}
\lab{vareps-eq-3}
\ee
and, therefore, \rf{h-Phi-0} acquires the form:
\br
\Phi (0) &=& h_0 \exp \Bigl\{ - \htt_1 \frac{2\htt_2}{3\htt_3} +
{1\over 6} \frac{\( 2\htt_2\)^3}{\( 3\htt_3\)^2} \Bigr\}   \nonu \\
&=& \exp \Bigl\{ \tit_1 \frac{\tit_2}{3\tit_3} -
{2\over 3} \frac{\( \tit_2\)^3}{\( 3\tit_3\)^2} \Bigr\}
\int_{} \int_{}  d\l \,d\m \,\exp\lcurl \sum_{r=1}^{p_1} \l^r t_r
+ \sum_{s=1}^3 \m^s \tit_s \rcurl
\lab{h-Phi-0-3}
\er
\sect{Multi-Matrix Models as Con\-strained {\sf KP} Hie\-rar\-ch\-ies: 
Dar\-boux-\-B\"{a}ck\-lund Solutions}
\label{section:noak5}
Similarly to the two-matrix model \rf{ZN-2M}, the partition function of
the multi-matrix ($q$-matrix) model reads:
\be
Z_N \lb \{ t^{(1)}\},\ldots ,\{ t^{(q)}\},\{ g\} \rb =
\int dM_1 \cdots dM_q \exp \lcurl
\sum_{\a =1}^q \sum_{r_\a =1}^{p_\a} t^{(\a )}_{r_\a} \Tr M_\a^{r_\a} +
\sum_{\a =1}^{q-1} g_{\a ,\a +1} \Tr M_\a M_{\a +1} \rcurl
\lab{ZN-qM}
\ee
where $M_\a$ are Hermitian $N \times N$ matrices, and the orders of the
matrix ``potentials'' $p_{\a}$ may be finite or infinite. As in the
two-matrix model case, one associates \ct{BX93-94} to \rf{ZN-qM} 
generalized Toda-like lattice systems subject to specific constraints.
Correspondingly, $Z_N$ and its derivatives w.r.t. the coupling parameters can 
be expressed in terms of solutions of the underlying Toda-like discrete 
integrable hierarchy where $\{ t^{(1)}\},\ldots ,\{ t^{(q)}\}$ play the role 
of ``evolution'' parameters.

It turns out that, in order to identify the continuum \cKP integrable 
hierarchy which provides the exact solution for \rf{ZN-qM}, we need the 
following subset of the associated linear system and the corresponding Lax
(``zero-curvature'') representation from \ct{BX93-94} :
\br
{Q(1)}_{nm} \psi_m = \l \psi_n  \quad , \;\;
\partder{}{t^{(1)}_r} \psi_n \eq - \( {Q(1)}^r_{-}\)_{nm} \psi_m  \quad , \;\;
\partder{}{t^{(q)}_s} \psi_n = - \( {Q(q)}^s_{-}\)_{nm} \psi_m
\lab{L-q-1-2} \\
\partder{}{t^{(1)}_r} Q(1) \eq \llb {Q(1)}^r_{(+)} , Q(1) \rrb  \quad , \quad
\partder{}{t^{(q)}_s} Q(1) = \llb Q(1) , {Q(q)}^s_{-} \rrb
\lab{L-q-3} \\
\partder{}{t^{(1)}_r} Q(q) \eq \llb Q^r_{(+)} , Q(q) \rrb  \quad , \quad
\partder{}{t^{(q)}_s} Q(q) = \llb Q(q) , {Q(q)}^s_{-} \rrb
\lab{L-q-4}
\er
Introducing the notations:
\be
t_r \equiv t^{(1)}_r \quad ,\;\; r=1,\ldots ,p_1 \quad ; \quad
\tit_s \equiv t^{(q)}_s \quad ,\;\; s=1,\ldots ,p_q  \qquad ; \qquad
Q \equiv Q(1) \quad ,\quad \tQ \equiv Q(q)
\lab{t-q}
\ee
the system \rf{L-q-1-2}--\rf{L-q-4} becomes exactly the same as the one in the
two-matrix model case \rf{L-1}--\rf{L-4} (with the exception of the second
eq.\rf{L-1} which now has a more complicated form but will not be
needed in the sequel). Furthermore, there is a series of additional constraints
(``coupling conditions'') relating $Q \equiv Q(1)$ and $\tQ \equiv Q(q)$
which have much more intricate form (involving also the ``intermediate''
$Q(2),\ldots ,Q(q-1)$ matrices) than the two-matrix model ones
\rf{string-eq-a}--\rf{string-eq-c}. However, their explicit form will not be
needed in what follows since we will be able to extract the
relevant information only from the discrete Lax system \rf{L-q-3}--\rf{L-q-4}
and the relations \rf{abR-h} expressing $Q \equiv Q(1),\,\tQ \equiv Q(q)$ in
terms of orthogonal polynomial factors.

Here, we shall introduce the same parametrization
for the matrix elements of $Q \equiv Q(1)$ and $\tQ \equiv Q(q)$ as in
\rf{param-1}--\rf{param-2} (with the obvious changes in the sizes of the
Jacobi matrices):
\br
Q_{nn} &\equiv& {Q(1)}_{nn} = a_0 (n) \quad , \quad
{Q(1)}_{n,n+1} \equiv Q_{n,n+1} =1 \quad ,\quad
{Q(1)}_{n,n-k} \equiv Q_{n,n-k} = a_k (n)    \nonu  \\
k\eq 1,\ldots , m(1) \;\; ; \;\; m(1) = (p_q -1) \cdots (p_2 -1) 
\nonu  \\
{Q(1)}_{nm} \! \!& \equiv &\!\!  Q_{nm} = 0  \;\; {\rm for} \;\; 
m-n \geq 2 \;\, ,\;\, n-m \geq m(1) +1
\lab{param-q-1}  \\
{Q(q)}_{nn}\! \!&\equiv&\!\! \tQ_{nn} = b_0 (n)  \;\; , \;\;
{Q(q)}_{n,n-1} \equiv \tQ_{n,n-1} = R_n  \nonu \\
{Q(q)}_{n,n+k} &\equiv & \tQ_{n,n+k} = b_k (n) R_{n+1}^{-1} \cdots R_{n+k}^{-1}
\nonu \\
 k\eq 1,\ldots , m(q)\;\; , \;\; m(q) =  (p_{q-1} -1) \cdots (p_1 -1)  
 \nonu  \\
{Q(q)}_{nm} &\equiv& \tQ_{nm} = 0  \;\; {\rm for}  \;\; n-m \geq 2 \; ,\;
m-n \geq m(q) +1
\lab{param-q-2}
\er

Then, we have the same system of relations \rf{ZN-h}--\rf{ZN-ab} between the
$Q \equiv Q(1),\,\tQ \equiv Q(q)$ matrix elements and the normalization
factors in the nonlocally generalized orthogonal polynomial formalism
\ct{ortho-poly} (using notations \rf{t-q}) :
\br
h_n \d_{nm} \eq \int \int d\l d\m \,
P_n (\l ) \exp \Bigl\{ \sum_{r=1}^{p_1} \l^r t_r \Bigr\}
\rho (\l ,\m ; \{ t^{\pr\pr}\}, \{ g\})
\Bigl\{ \sum_{s=1}^{p_q} \m^s \tit_s \Bigr\} {\wti P}_m (\m ) \phantom{a}
\lab{h-q-n} \\
\rho (\l ,\m ; \{ t^{\pr\pr}\}, \{ g\}) & =& 
\int_{\Gamma} \prod_{\a =2}^{q-1} d\n_\a \, \exp \Bigl\{
\sum_{\a =2}^{q-1} \sum_{r_\a =1}^{p_\a} t^{(\a )}_{r_\a} \n_\a^{r_\a} +
\sum_{\a =2}^{q-2} g_{\a ,\a +1} \n_\a \n_{\a +1} 
\nonu\\
&+& g_{12} \l \n_2 + g_{q-1,q} \n_{q-1} \m \Bigr\}
\lab{rho-q} \\
\{ t^{\pr\pr}\}& \equiv& \( t^{(2)},\ldots ,t^{(q-1)}\)  \phanta \nonu
\er

Furthermore, with the notations \rf{t-q}--\rf{param-q-2} all equations
from the two-matrix model case \rf{motionx-0}--\rf{t-r-eqs} are literally
satisfied for the $q$-matrix model matrices $Q \equiv Q(1),\,\tQ \equiv Q(q)$.
Also, as in the two-matrix case it is natural to introduce the fractional
power:
\be
\hQ = Q^{1\over {m(1)}} \equiv {Q(1)}^{1\over {m(1)}}
\lab{Q-hat-q}
\ee
with exactly the same parametrization as in \rf{param-2-h} and satisfying the
same system of lattice eqs. of motion \rf{motionxx-k-ss}--\rf{motionyy-k-ss}.
Therefore, we have the following:

\begin{proposition}
The matrix elements of $Q \equiv Q(1)$ are completely expressed in terms
of the matrix elements of $\tQ \equiv Q(q)$ through the relations:
\be
Q_{(-)} = \sum_{s=0}^{m(1)} \a_s \tQ^s_{(-)} \qquad ,\qquad
Q^{s\over {m(1)}}_{(-)} = \sum_{\s =0}^{s} \g_{s\s} \tQ^{\s}_{(-)}
 \quad , \;\; s=0,1,\ldots ,m(1)
\lab{tatko-q-s}
\ee
where the coefficients $\a_0 ,\g_{s,0}$ are $t_1$-independent,
whereas the coefficients $\a_s ,\g_{s\s}$ with $\s \geq 1$ are independent of
$t_1, {\ti t}_1$. All $\g_{s\s}$ are expressed through $\a_s\equiv \g_{m(1),s}$
exactly in the same way as in the two-matrix model case:
\br
\g_{ss} &=& \( \g_{11}\)^s \;\quad, \quad\;
\g_{s,s-1} = s \(\g_{11}\)^{s-1} \g_{10}  \lab{tatko-q-2-a}  \\
\g_{s,s-2}  &=& \(\g_{11}\)^{s-2} \llb \frac{s(s-1)}{2} \(\g_{10}\)^2 +
s \( \frac{\g_{31}}{3\g_{11}} - \g_{10}^2 \) \rrb
\lab{tatko-q-2-b}  \\
\g_{11} &=& \( \a_{m(1)}\)^{1\over {m(1)}} \quad ,\quad
\g_{10} = \frac{\a_{m(1)-1}}{m(1) \( \a_{m(1)}\)^{{m(1)-1}\over {m(1)}}}
\nonu  \\
\frac{\g_{31}}{3\g_{11}} - \g_{10}^2 \eq {1\over {m(1)}}
\llb \frac{\a_{m(1)-2}}{\( \a_{m(1)}\)^{{m(1)-2}\over {m(1)}}} -
\frac{m(1)-1}{2m(1)}\,
\frac{\a^2_{m(1)-1}}{\( \a_{m(1)}\)^{{2(m(1)-1)}\over {m(1)}}} \rrb
\lab{tatko-q-2}
\er
\label{propositon:mma}
\end{proposition}

\mskp
Similarly, we have the exact analog of the dual statement \rf{tatko-r}
with the r\^{o}les of $Q \equiv Q(1)$ and $\tQ \equiv Q(q)$ interchanged.
The difference with the two-matrix case is that now the coefficients
$\a_s$ and $\b_r$ do not have any more the simple expressions \rf{alpha-s}
and \rf{beta-r}.

As an important consequence of \rf{tatko-q-s}, let us take
its diagonal $00$-part and use the last eq.\rf{abR-h} which yields:
\be
\partder{}{t_1} h_0 =
\( \sum_{s=1}^{m(1)} \a_s \partderh{}{\tit_1}{s} + \a_0 \) h_0
\lab{h-0-constr}
\ee
This equation is the only remnant of the constraints (``coupling conditions'')
on the multi-matrix model $Q$-matrices which will be used in the sequel.

Further on we follow the same steps as in the previous section to derive the
continuum \cKPrm hierarchy associated with the general $q$-matrix model. 
Namely, as a first step using \rf{tatko-q-s} we re-express the Toda-like 
lattice hierarchy \rf{L-q-1-2}--\rf{L-q-4} as a single set of flow equations 
for $\hQ \equiv {Q(1)}^{1\over {m(1)}}$ :
\br
\hQ^{m(1)}_{nm} {\psi}_m &=& \l {\psi}_n  \quad , \quad
\partder{}{{\htt}_s} {\psi}_n = - \({\hQ}^s_{(-)}\)_{nm} {\psi}_m
\lab{L-q-hat} \\
\partder{}{{\htt}_s} \hQ &=& \llb \hQ , \hQ^{s}_{-} \rrb \quad
\;,\; \;\; s=1,\ldots,p_q , 2m(1), 3m(1), \ldots, p_1 m(1)
\lab{L-3-q-h} \\
t_r &\equiv& {\hat t}_{rm(1)} \quad {\rm for} \;\; r=1,\ldots , p_1
\lab{identif-q}
\er
where, as in the two-matrix case, we have introduced a new subset of
evolution parameters $\lcurl {\hat t}_s \rcurl$ instead of
$\lcurl {\ti t}_s \equiv t^{(q)}_s \rcurl$ defined as:
\be
\partder{}{{\hat t}_s} \,= \,\sum_{\s =1}^s \g_{s\s} \partder{}{{\ti t}_{\s}}
\quad \;\; ,\quad\; s=1,\ldots, m(q)      \lab{tatko-q-a}
\ee
As a second step, one employs the Bonora-Xiong procedure to get from the
discrete Lax system \rf{L-q-hat}--\rf{L-3-q-h} an equivalent continuum Lax
system which, upon operator conjugation and similarity transformation as in
\rf{Ln}--\rf{Ls}, acquires the form (as before $x \equiv \htt_1$) :
\br
\partder{}{{\hat t}_s} L(n) \eq
\Sbr{\( L^{s\over {m(1)}}(n)\)_{(+)}}{L (n)}
\quad , \;\, s=1,\ldots,p_q , 2m(1), 3m(1), \ldots ,p_1 m(1)
\lab{Lax-q-2M}  \\
L(n) &=& D_x^{m(1)} + m(1) {\hb}_1 (n) D_x^{m(1)-2} + \cdots +
{\hR}_{n+1} \( D_x - {\hb}_0 (n) \)^{-1}
\lab{Ln-q}
\er
Exactly as in the two-matrix case, lattice shifts $n \to n+1$ in the
underlying discrete Toda lattice system, described by
\rf{L-q-1-2}--\rf{L-q-4}, generate the \DB transformations in the continuum
${\sl cKP}_{r=m(1)+1,m=1}$ hierarchy \rf{Lax-q-2M}--\rf{Ln-q} and the solutions
for the eigenfunctions and $\t$-functions at each successive step of the \DB
transformation is given explicitly by eqs.\rf{Phi-W-n}--\rf{tau-W-n} where
everything is expressed in terms of the eigenfunction of the ``initial'' Lax
operator $L(-1)$. The difference with the two-matrix case is only the explicit
form of the latter (cf. \rf{L-1-psi}) :
\be
L(-1) \;= \;e^{\g_{10} \htt_1} \( \sum_{s=0}^{m(1)} \a_s \g_{11}^{-s} D^s \)
e^{-\g_{10} \htt_1}
\lab{L-1-q-psi}
\ee
where the coefficients $\a_s ,\g_{10},\g_{11}$ have more complicated
dependence on $\{ \htt_s\}$ than in the two-matrix case.

Exactly as in the two-matrix case, we obtain the relation between the $n$-th 
step DB eigenfunction $\Phi (n)$ and the orthogonal polynomial
normalization factor $h_n$ \rf{h-q-n} which generalizes \rf{h-Phi-n} :
\be
\Phi (n) = h_n \, \g_{11}^n \,
\exp \lcurl \htt_1 \g_{10} + \vareps (\htt^{\pr}) \rcurl
\lab{h-Phi-n-q}
\ee
Substituting \rf{h-Phi-n-q} into \rf{ZN-h} and using the Wronskian formula
\rf{tau-W-n} we get:
\br
Z_N &=& \prod_{n=0}^{N-1} h_n = \det {\Bigl\Vert}
\partderh{}{\htt_1}{i-1} \( L(-1)\)^{j-1} \Phi (0) {\Bigr\Vert}
e^{-N\( \htt_1 \g_{10} + \vareps (\htt^{\pr})\)} \g_{11}^{-\frac{N(N-1)}{2}}
\lab{ZN-det-q-0}  \\
&=& \det {\Bigl\Vert}
\partderh{}{\tit_1}{i-1}
\( e^{- \htt_1 \g_{10}} L(-1) e^{\htt_1 \g_{10}}\)^{j-1} h_0  {\Bigr\Vert}
\lab{ZN-det-q-1}
\er
where we absorbed the $\g_{11}$-factors via changing
$\partder{}{\htt_1} \to \partder{}{\tit_1}$ by the definition \rf{tatko-q-a},
{\sl i.e.}, $\g_{11}^{-1} \partder{}{\htt_1} = \partder{}{\tit_1}$.
Now, in complete analogy with \rf{t-k-Phi} we find using \rf{L-1-q-psi} and
\rf{h-0-constr} :
\be
\( e^{- \htt_1 \g_{10}} L(-1) e^{\htt_1 \g_{10}}\)^{j_1} h_0 =
\( \sum_{s=0}^{m(1)} \a_s \partderh{}{\tit_1}{s} \)^{j-1} h_0 =
\partderh{}{t_1}{j-1} h_0
\lab{L-1-j-h}
\ee
Substituting \rf{L-1-j-h} into \rf{ZN-det-q-1} yields the final result for the
multi-matrix model partition function:
\be
Z_N =\det {\Bigl\Vert}
\partderM{h_0}{i+j -2}{\tit_1}{i-1}{t_1}{j-1} {\Bigr\Vert}
\lab{ZN-det-q}
\ee
which, as anticipated in \rf{Wronski-2M-MM}--\rf{h0-rho}, is functionally the
same as for the two-matrix model \rf{ZN-det}, however, with a more complicated
expression for $h_0$ \rf{h-q-n}.
\sect{Concluding Remarks}
\label{section:noak-out}
The intimate interrelation between discrete multi-matrix models
and specific \DB orbits of generalized constrained {\sf KP} integrable
hierarchies (\cKPrm hierarchies or, equivalently,  $SL(r+m,m)$ {\sf KP-KdV}
hierarchies -- see the remark at the end of section \ref{section:noak2-1})
looks now well understood. Let us briefly sketch, 
in conclusion, some directions for further developments connected with the 
present approach.

First, we will be aimed at extending our 
construction of the modified additional symmetry flows for the \cKPrm 
hierarchies to cover the whole $\Win1$ algebra. The relevant steps 
were already indicated in section \ref{section:addsym}. 
Let us recall that eqs.\rf{xk2},\rf{paxk2} bear direct resemblance 
to the $\Win1$ structure, which appeared in \ct{kac-rad-zemba}
where the higher-spin $W$-generators have been 
built up 
from bosonic currents through a generalized Sugawara construction.

Next, it is desirable to prove a theorem, analogous to the theorem of
refs.\ct{ASvM,Dickey-addsym}, explicitly expressing the action of the
modified additional symmetry flows \rf{pasta} (and their generalizations for
the full $\Win1$ algebra) directly in terms of differential operators acting
on the $\t$-functions (cf. \rf{Vir-Z-N}--\rf{lsn2}).

Recently there was an active interest in multi-matrix models in the context of
various random matrix problems in condensed matter physics 
(see refs.\ct{yshai-kamoto-other} and citations therein). 
The main object of physical interest there is the so called joint 
distribution function $\cP \( \l_1 ,\ldots ,\l_N \)$, with the partition
function being: 
\be
Z_N = {\rm const} \int \prod_{j=1}^N d\l_j \, \cP \( \l_1 ,\ldots ,\l_N \)
\lab{Z-N-joint}
\ee
Recall that linking the discrete
(multi-)matrix models with the continuum integrable hierarchies
(sections \ref{section:noak3}--\ref{section:noak5}) amounts 
to identifying, up to an overall factor, the (multi-)matrix model partition
functions with $\t$-functions of \DB orbits of \cKPrm hierarchies
$\, Z_N \simeq \t^{(N-1)}$ (cf. eq.\rf{ZN-det-0}).
In this spirit one associates $\cP \( \l_1 ,\ldots ,\l_N \)$
with the integrand in the ``eigenvalue'' ($\l$-integral) representation 
\rf{tau-recur-r}. The latter can equivalently be written as:

\be
\cP \( \l_1 ,\ldots ,\l_N \) = 
\exp \lcurl - \sum_{j=1}^N H_{{\rm one-body}} (\l_j ) -
\sum_{i>j} H_{{\rm two-body}} (\l_i ,\l_j ) -
H_{{\rm many-body}} (\l_1 ,\ldots ,\l_N ) \rcurl
\lab{joint-dist} 
\ee
\br
H_{{\rm one-body}}(\l ) &\equiv& -\ln \p_1^{(0)}(\l )-\sum_{l\geq 1} t_l \l^l 
\lab{one-body} \\
H_{{\rm two-body}} (\l_i ,\l_j ) &\equiv& - \ln \( \l_i - \l_j \)^2 
- \ln \Bigl( \sum_{s=0}^{r-1} \l_i^s \l_j^{r-1-s}\Bigr)
\lab{two-body} \\
H_{{\rm many-body}} (\l_1 ,\ldots ,\l_N ) &\equiv& - 
\ln \t^{(0)} \bigl( t - \sum_{j=1}^N [\l_j^{-1}]\bigr)
\lab{many-body}
\er   

Thus, we see that the generalized multi-matrix models 
corresponding to the generic
DB orbits of ${\sf cKP}_{r,1}$ hierarchies, which do not necessarily obey the
ordinary ``string'' equation, yield additional 
new type of contributions to the familiar 
effective potential in the pertinent joint distribution function.
These contributions are: the $-\ln \p_1^{(0)}(\l )$ term in the one-body 
potentials \rf{one-body},
the second attractive term in the two-body potentials \rf{two-body} 
dominating at very long distances over the ordinary repulsive first term, 
as well as an additional genuine many-body potential \rf{many-body}.

In the same spirit we can interpret the integrand in the $\l$-integral
representation \rf{tau-recur-r} of the $\t$-function for the most general
DB orbit of \cKPrm hierarchies with arbitrary $m \geq 1$. 
It will correspond to a joint
distribution function of a system of $m$ different types of ``particles''.

It would be interesting to study the physical implications of the emerging
new type of joint distribution functions \rf{joint-dist}--\rf{many-body},
especially regarding critical behavior of correlations. On the other hand,
from mathematical point of view, 
it is desirable to derive an explicit
multi-matrix model representation of the $\t$-functions corresponding to
generic DB orbits of \cKPrm hierarchies generalizing the representation
\rf{ZN-1M-wti} for the generic ${\sf cKP}_{1,1}$ $\t$-function. 

\appendix
\sect{Appendix: Technical Identities.}
\label{section:appa}
We list here for convenience a number of useful
technical identities, which have been used extensively throughout the text.

For an arbitrary pseudo-differential operator $A$ we have the
following identity:
\be
\( \chi D \chi^{-1} A \chi D^{-1} \chi^{-1} \)_{+} =
\chi D \chi^{-1} A_{+}  \chi D^{-1} \chi^{-1}
-  \chi \pa_x \(\chi^{-1}  A_{+} (\chi ) \)  D^{-1} \chi^{-1}
\lab{aonchi-app}
\ee
where $A_{+}$ is the differential part of
$A= A_{+} + A_{-} = \sumi{i=0} A_i D^i + \sum_{-\infty}^{-1} A_i D^i$.

For a purely differential operator $K$ and arbitrary functions
$f,g$ we have the identities:
\be 
\( K\, f D^{-1} g \)_{-} =  K (f) D^{-1} g  \quad , \quad
\( f D^{-1} g \, K \)_{-} = f D^{-1} K^{\ast} (g)
\lab{tkppsi-app}
\ee
Another useful technical identity involves a product of two
pseudo-differential operators
of the form:
$X_i = f_i D^{-1} g_i\;, \, i=1,2$ :
\be
X_1 X_2 = X_1 (f_2) D^{-1} g_2 + f_1 D^{-1} X^{\ast}_2 (g_1)
\lab{x1x2}
\ee
where:    
$X_1 (f_2) = f_1 \pa_x^{-1} (g_1 f_2)$, etc. .
{}From the above identity it follows the relation \ct{EOR95}:
\be
\(  L^k \)_{-} = \sum_{i=1}^m \sum_{j=0}^{k-1} L^{k-j-1} (\Phi_i) D^{-1}
\( L^{\ast}\)^{j} ( \Psi_i  )
\lab{lkminus}
\ee
for the \cKP Lax operator \rf{f-5}.

Let us also list some useful identities involving {\DB}-like transformation
of pseudo-differential operators of the $X_i$-form above:
\br
T_a \( \Phi_a D^{-1} N \) T_a^{-1} &=& \( \Phi_a^2 N\) D^{-1} \Phi_a^{-1}
\lab{DB-like-1} \\
T_a \( M D^{-1} \Psi_a \) T_a^{-1} &=&
{\wti M} D^{-1} \( {\wti L}^\ast ({\wti \Psi_a})\) +
\lcurl T_a \( M \pa_x^{-1} (\Psi_a \Phi_a )\) \rcurl  D^{-1} \Phi_a^{-1}
\lab{DB-like-2} \\
T_a \( M D^{-1} N \) T_a^{-1} &=& {\wti M} D^{-1} {\wti N} +
\lcurl T_a \( M \pa_x^{-1} (N \Phi_a )\) \rcurl  D^{-1} \Phi_a^{-1}
\lab{DB-like-3} \\
\({\wti L}^{\ast}\)^{ s} ({\wti \Psi_a}) &=&
- \Phi_a^{-1} \pa_x^{-1} \( \Phi_a \({L^\ast}\)^{s-1} (\Psi_a )\)
\lab{DB-like-4}
\er
where $\Phi_a$ is one of the eigenfunctions of a \cKP Lax operator $L$
\rf{f-5} and:
\br
T_a &\equiv& \Phi_a D \Phi_a^{-1} \qquad ,\qquad  {\wti \Psi_a} = \Phi_a^{-1}
\nonu \\
{\wti M} &\equiv& T_a (M) = \Phi_a \pa_x \( \Phi_a^{-1} M\) \quad ,\quad
{\wti N} \equiv {T_a^{-1}}^\ast (N) = - \Phi_a^{-1} \pa_x^{-1} \( \Phi_a N\)
\nonu
\er

Finally, let us recall the following important composition formula for
Wronskians \ct{Wronski} :
\be
T_k \, T_{k-1}\, \cdots\, T_1 (f ) \; =\; { W_{k} (f) \over W_k}
\lab{iw}
\ee
where:
\br
T_j = { W_{j} \over W_{j-1} } D { W_{j-1} \over W_{j} } =
\( D + \( \ln { W_{j-1} \over W_{j} } \)^{\pr} \) \quad;\quad W_{0}=1
\lab{transf} \\
W_k \equiv W_k \lb \psi_1, \ldots ,\psi_k \rb =
\det {\Bigl\Vert} \pa_x^{i-1} \psi_j {\Bigr\Vert}
\quad ,\quad
W_{k-1} \(f \)\equiv W_{k} \lb \psi_1, \ldots ,\psi_{k-1}, f\rb
\lab{W-def}
\er

\end{document}